\documentclass[twocolumn,4pt,longbibliograhy,preprint,superscriptaddress,preprint,preprintnumbers,nobibnotes,amsmath,amssymb,aps, pra,tightenlines,nobalancelastpage]{revtex4-2}

\usepackage{graphicx}
\usepackage{dcolumn}
\usepackage{bm}

\usepackage{subfig}

\usepackage{bbold}

\usepackage{hyperref}
\usepackage{xcolor}

\usepackage{enumerate}

\newcommand{\ket}[1]{\left|{#1}\right\rangle}
\newcommand{\bra}[1]{\left\langle{#1}\right|}

\newcommand{\Tr}[1]{\text{Tr} \left[ #1 \right]}
\newcommand{\Trp}[2]{\text{Tr}_{#1} \left[ #2 \right]}
\renewcommand{\det}[1]{\text{det} \left( #1 \right)}
\newcommand{\hdet}[1]{\text{Hdet} \left( #1 \right)}

\newcommand{\W}{\ket{\text{W}}}
\newcommand{\tW}{\text{W}}

\newcommand{\type}[1]{\ket{\mathcal{T}_{\text{#1}}}}

\newcommand{\sign}[1]{\text{sgn}\left( #1 \right)}

\newcommand{\tGHZ}{\text{GHZ}}
\newcommand{\GHZ}{\ket{\text{GHZ}}}

\newcommand{\Bell}[2]{\ket{\phi_{+}}_{#1} \ket{0}_{#2} }

\newcommand{\conc}[1]{\mathcal{C}_{#1} }

\newcommand{\ZeroClass}{\left[ \ket{000} \right] }

\newcommand{\textClass}[1]{\left[ \text{#1} \right] }

\newcommand{\multiComment}[1]{}

\newcommand{\tangle}[1]{\tau \left( #1 \right)}
\newcommand{\genNorm}{r_{I}}
\newcommand{\norm}[1]{r_{#1}}

\newcommand{\upperTau}{\tau_{\text{M}}\left( R\right)}

\newcommand{\middleTau}{\tau_{\star}\left( R\right)}

\newcommand{\argmin}{\mathop{\mathrm{argmin}}}

\newcommand{\detZero}{\det{T_0'}=0}

\newcommand{\wtrans}[1]{\ket{W_{k=#1}}}

\begin{document}


\title{Visualizing Three-Qubit Entanglement}

\author{Alfred Benedito}
\email{alfred.benedito@ift.csic.es}
\affiliation{\mbox{Instituto de F\'isica Te\'orica, UAM-CSIC, Universidad Aut\'onoma de Madrid, Spain.}}
\author{Germ\'an  Sierra}%
\email{german.sierra@csic.es}
\affiliation{\mbox{Instituto de F\'isica Te\'orica, UAM-CSIC, Universidad Aut\'onoma de Madrid, Spain.}}
\affiliation{\mbox{Kavli Institute for Theoretical Physics, University of California, Santa Barbara, CA 93106, USA.}}%


\multiComment{
\begin{abstract}

We present a graphical framework to represent entanglement in three-qubit states. The geometry associated with each \textit{entanglement class} and \textit{type} is analyzed, revealing distinct structural features. We explore the connection between this geometric perspective and the tangle, deriving bounds that depend on the entanglement class. Based on these insights, we conjecture a purely geometric expression for both the tangle and Cayley's hyperdeterminant. As an application, we analyze the energy eigenstates of physical Hamiltonians, identifying two distinct forms of \textit{genuine tripartite} entanglement. We further propose that, due to level repulsion effects, only one of these forms persists in typical physical systems.
\end{abstract}
}

\maketitle

\onecolumngrid 
\vspace{-3em}
\begin{center}
\begin{minipage}{0.9\linewidth}
\noindent
We present a graphical framework to represent entanglement in three-qubit states. The geometry associated with each \textit{entanglement class} and \textit{type} is analyzed, revealing distinct structural features. We explore the connection between this geometric perspective and the tangle, deriving bounds that depend on the entanglement class. Based on these insights, we conjecture a purely geometric expression for both the tangle and Cayley's hyperdeterminant for non-generic states. As an application, we analyze the energy eigenstates of physical Hamiltonians, identifying the sufficient conditions for \textit{genuine tripartite} entanglement to be robust under symmetry-breaking perturbations and level repulsion effects.
\end{minipage}
\end{center}

\twocolumngrid 



\section{Introduction}\label{section_Intro}


Entanglement is a consequence of the superposition principle, where quantum states cannot be written in product form on any local basis \cite{entangled_states__horodecki}. Although their existence was first pointed out by Einstein, Podolski and Rosen in 1935 \cite{EPR__paper}, it was not until the late 1990s and early 2000s that the study and classification of entanglement in systems of more than 2 qubits picked the interest of physicists due to the realization that they could be used as a resource in information processing and communication. This is because entanglement differs from classical correlations even if one uses local hidden variables \cite{Bell_theorem__paper}.\newline

\begin{figure}
	\centering
	\includegraphics[scale=0.35]{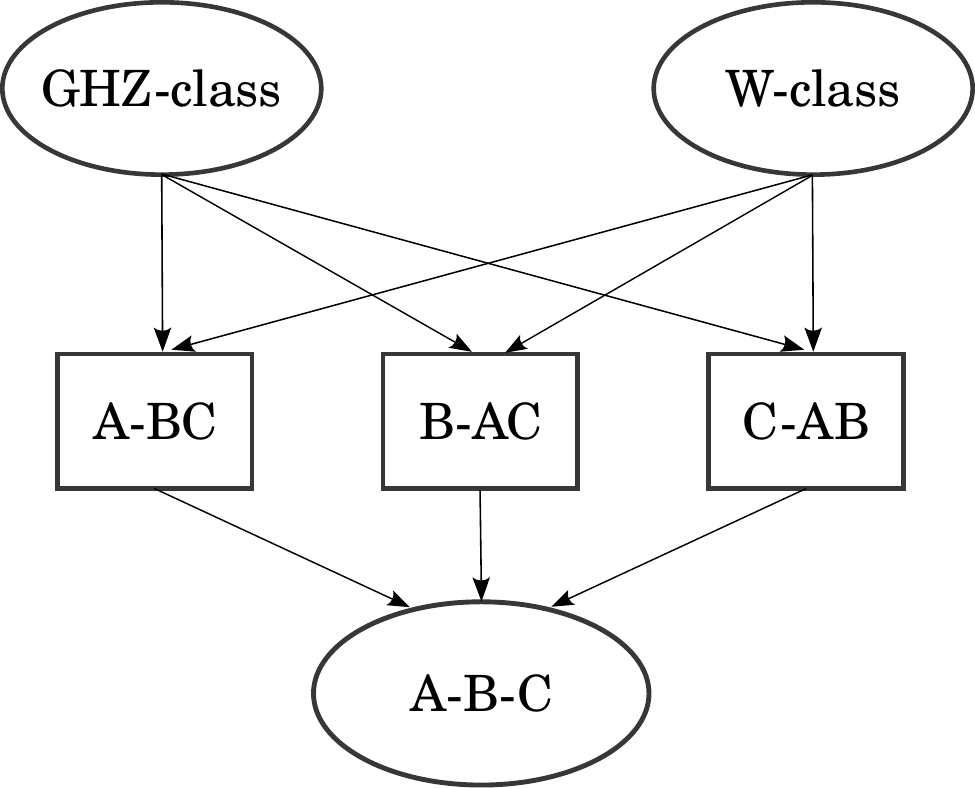}
	\caption{Entanglement classes (from \cite{Cirac_3qbit_classes})}
	\label{cirac_classes_diagram}
\end{figure}

The $\GHZ$ state \cite{GHZ_state__paper} sparked interest in the aforementioned classification\footnote{It also turned out to be the unique state that is maximally entangled in 3-qubit systems  \cite{acin_decomp_paper__cit_5}, which means that all its information is encoded purely on its entanglement. This can be concluded by observing that all its 1-qubit reduced density matrices are just the maximally mixed state.}, with early attempts revolving around the study of the orbits of $U(2)^{\otimes n}$ \cite{acin_decomp_paper__cit_4}, but later moving to the modern paradigm of \textit{entanglement measures} \cite{earliest_entang_measure__paper} such as Benett's \textit{entanglement of formation} \cite{00__bennet}. This measure was later extended to any 2-qubit mixed state \cite{01__1__wooters, 01__2__Wooters} through a quantity known as \textit{the concurrence} $\conc{}$. With it, it wash shown that, if one has a system of 3 qubits $A$, $B$ and $C$, then there is a trade-off between $A$'s entanglement with $B$ and with $C$ \cite{tangle_introduction}. In other words, the sharing of entanglement is restricted. This is a core difference between classical and quantum correlations, measured by \textit{the tangle} $\tau$ \footnote{Moreover, Wooters, Coffman and Kundu conjectured that such a relation should also exist for systems of more qubits (known as the CKW conjecture, later proven true by Osborne and Verstraete \cite{verstraete_nTangle}).}. This showed that there were two inequivalent ways of entangling all 3 qubits in a pure 3-qubit state, allowing the classification of pure 3-qubit states \cite{Cirac_3qbit_classes} based on defining entanglement classes as sets of states that map onto themselves under invertible SLOCC \footnote{The methods can be used to classify 4 qubit states as well \cite{verstraete_4qubits_classes}.}. They identified 6 different entanglement classes (see Fig. \ref{cirac_classes_diagram}), which can be fully characterized by 4 parameters (see Table \ref{cirac_table}).\newline

\begin{table}
	\centering
	\begin{tabular}{c  c  c  c  c}
		\hline
		\hline
		Class & $S_A$ & $S_B$ & $S_C$ & $\tau$ \\
		\hline
		A-B-C & 0 & 0 & 0 & 0 \\
		\hline
		A-BC & 0 & $>0$ & $>0$ & 0 \\
		\hline
		B-AC & $>0$ & 0 & $>0$ & 0 \\
		\hline
		C-AB & $>0$ & $>0$ & 0 & 0 \\
		\hline
		W & $>0$ & $>0$ & $>0$ & 0 \\
		\hline
		GHZ & $>0$ & $>0$ & $>0$ & $>0$ \\
		\hline
		\hline
	\end{tabular}
	\caption{Values of the local entropies and the tangle for the different classes (from \cite{Cirac_3qbit_classes}).}
	\label{cirac_table}
\end{table}

From the group orbit analysis \cite{acin_decomp_paper__cit_4} it was known that the number of entanglement invariant parameters had to be 5 for 3-qubit states but so far only 4 had been used. This pointed towards the existence of further structure yet to be found within these classes. This structure was understood by means of a \textit{Generalized Schmidt} / Canonical Decomposition (CD) for 3-qubit states \cite{acin_decomp_paper}. The CD showed there is a \textit{canonical form} unique for all states related by Local Unitaries (LUs) that uses only 5 of the 8 basis elements. Depending on which basis elements had null coefficients, one could identify different sub-classes (or \textit{Types}).\newline

This classification is quite difficult to visualize, because it depends on the 5 entanglement invariants $\{J_l\}_{l=1}^5$ (which are hard to relate to physical observables). For 1 qubit states, a typical graphical representation is the Bloch sphere $S^2$ representing the expectation values of spin observables. For 2 qubits, Mosseri and Dandoloff \cite{first_hopfQubit_paper} reinterpreted this Bloch sphere map as fibrating $S^3$ with $S^1$: a Hopf Fibration. This way, the generalization of the Bloch sphere follows immediately. For 3-qubits one can do a similar construction \cite{second_hopfQubit_paper}, but the fundamental difference (for our interests) is that the construction is then sensitive to the entanglement between $A$ and $BC$ but it says nothing about entanglement between $B$ and $C$. Mosseri proposed for this case to instead use the three Bloch-norms $(\norm{A}, \norm{B}, \norm{C})$ which are also entanglement invariants (they belong to a set first found by Sudbery \cite{Sudbery_2001_simpler_invariants, acin_decomp_paper}). It's these three invariants along with the tangle that will be the main focus of our work.\newline

The main aim of this paper is investigating how 3-qubit entanglement can be visualized with physical observables in a geometrical picture. We study how Mosseri's proposal results in a more physical and geometrical characterization of different entanglement classes and \textit{types} in what we call \textit{Bloch-norm representation}. This naturally leads to a series of bounds between the tangle and the norm of a vector. An important consequence of those results is that we can conjecture a formula for calculating Cayley's hyperdeterminant \cite{cayley_paper, adrian_latorre_paper, latorre_cited_book, tangle_introduction} which relies purely on geometrical characteristics. This is fairly striking, since contrary to the regular determinant (which can be understood geometrically as a measure of volume) the hyperdeterminant does not have a simple geometric interpretation. The second goal is the application of these geometrical tools to the study of the entanglement present in the energy eigenstates of physical Hamiltonians\textcolor{black}{, where we identify the sufficient conditions for \textit{genuine tripartite entanglement} to be robust under perturbations.}\newline

The paper is organized as follows. In Sec. \ref{section_Tetrahedrons} we briefly review the graphical aspect of Mosseri's proposal. The original work of this paper begins in Sec. \ref{section_Tangle} which is devoted to the study of the relation between the Bloch-norm representation and the tangle. We derive bounds for $\tau$ depending on the Bloch-norms and geometrically characterize it for states belonging to the GHZ class. Finally, in Sec. \ref{section_Chains} we consider physical systems of three qubits characterized by different Hamiltonians. We study the Bloch-norm properties of their energy eigenstates and characterize the source of their tangle. To do so, we developed a Python library to automate the analytic computations as much as possible. Finally, Appendix \ref{section_DetailsBounds} contains the details of the fibration procedure needed to obtain the bounds of Sec. \ref{section_Tangle}, and in Appendix \ref{section_DetailsChains} we provide the exact calculations for the chains in further detail.\newline

\section{Graphical representation of entanglement}\label{section_Tetrahedrons}

\multiComment{
\begin{align}
\frac{1}{2} \le Q_1 & := \Tr{\rho_A^2}  = 1-2\lambda_0^2 \left( 1-\lambda_0^2 - \lambda_1^2 \right) \le 1 \nonumber \\
\frac{1}{2} \le Q_2 & := \Tr{\rho_B^2}  = 1-2\lambda_0^2 \left( 1-\lambda_0^2 - \lambda_1^2 - \lambda_2 \right) - 2\Delta \le 1\nonumber  \\
\frac{1}{2} \le Q_3 & := \Tr{\rho_C^2}  = 1-2\lambda_0^2 \left( 1-\lambda_0^2 - \lambda_1^2 - \lambda_3^2 \right) - 2\Delta \le 1\nonumber  \\
\frac{1}{4} \le Q_4  & := \Tr{ \left( \rho_A \otimes \rho_B \right) \rho_{AB} } &  \nonumber \\
 & = 1+\lambda_0^2 \left( \lambda_2^2 \lambda_3^2 - \lambda_1^2 \lambda_4^2 - 2\lambda_2^2 - 3 \lambda_3^2 - 3 \lambda_4^2  \right) & \nonumber \\
& - \left(2 - \lambda_0^2 \right)\Delta \le 1 &  \nonumber \\
	0 \le Q_5 & := \left| \hdet{t_{ijk}} \right|^2  = \lambda_0^4 \lambda_4^4 \le \frac{1}{16}
\label{Q_invariants}
\end{align}

\begin{equation}
	\begin{matrix} \text{where } \Delta=J_1 \text{, } \rho_{I} = \Trp{\overline{I}}{\ket{\psi}\bra{\psi}} \text{, } \rho_{\overline{I}} =\Trp{I}{\rho} \\ \\  \text{ and } \ket{\psi} = \displaystyle\sum_{i,j,k \in \{0,1\}} \Big( t_{ijk} \ket{i}_A \ket{j}_B \ket{k}_C \Big) \end{matrix}
	\label{partial_traces_definitions}
\end{equation}
}

Let us first present a summary of the classification of 3-qubit states and their tangle. Starting by the concurrence, this is a quantity which measures the entanglement between two bipartitions of a state. In the simplest case, a 2-qubit state of components $\{v_{ij}\}_{i,j\in\{0,1\}}$, it reduces to $\propto | v_{00} v_{11} - v_{01} v_{10} |$ that is 0 if the state is separable and $>0$ if it is entangled. In 3-qubit states, there exist three possible bipartitions: $\conc{AB}$, $\conc{AC}$ and $\conc{A(BC)}$. By comparing them, one can find:

\begin{equation}
	\conc{AB}^2 + \conc{AC}^2 \le \conc{A(BC)}^2
	\label{concurrence_inequality}
\end{equation}

\noindent which motivates the definition of the \textit{tangle}:

\begin{equation}
	\tau_{ABC} := \conc{A(BC)}^2 - \left( \conc{AB}^2 + \conc{AC}^2 \right)
	\label{tangle_first_appearance}
\end{equation}



This tells us that $A$ can be entangled with $BC$ (measured by $\conc{A(BC)}$) in an essential way that cannot be described, in general, by a combination of entanglement of $A$ with $B$ (measured by $\conc{AB}$) and of $A$ with $C$ (measured by $\conc{AC}$). If that is the case, we say that this tripartite entanglement is \textit{genuine}. As stated in the introduction, this is what differentiates the entanglement present the state $\ket{\text{W}} = \frac{1}{\sqrt{3}} \left[ \ket{001} + \ket{010} + \ket{100} \right]$ and the state $\ket{\text{GHZ}} = \frac{1}{\sqrt{2}} \left[ \ket{000} + \ket{111} \right] $:

\begin{itemize}
	\item $\tau\left( \tW \right)=0$. This reflects that W can be written as a superposition of all three possible Bell pairs: $\W \propto \Bell{AB}{C} + \Bell{AC}{B} + \Bell{BC}{A}  $, so its entanglement is fully pair-wise generated. 
	
	\item $\tau\left( \tGHZ \right)=1$. This reflects that for $\GHZ$ no pair-wise decomposition exists. In fact, $\conc{IK}=0$, $\forall I,K \in \{A,B,C\}$ and $\conc{I(\overline{I})}=1$, $\forall I$, so its tripartite entanglement is \textit{genuine}.
\end{itemize}


We now present the CD:


\begin{eqnarray}
\ket{\psi}  &=&   \displaystyle\sum_{i,j,k \in \{0,1\}} \Big( t_{ijk} \ket{i}_A \ket{j}_B \ket{k}_C \Big) \nonumber \\
\ket{\psi} & \overset{\text{CD}}{\rightarrow} & \ket{\lambda_0, \vec{\lambda}, \lambda_4; \varphi}  := \Big[ \lambda_0 \ket{000} + \lambda_1 e^{i \varphi} \ket{100} \nonumber \\
& & + \lambda_2 \ket{101} + \lambda_3 \ket{110} + \lambda_4 \ket{111} \Big];
\label{cannonical_form}
\end{eqnarray}

\noindent where:
\begin{equation}
    \lambda_j \in [0,1] \forall j;  \text{ } \displaystyle\sum_{j=0}^4 \lambda_j^2 = 1; \text{ } \varphi \in [0,\pi].
    \label{cannonical_form_conditions}
\end{equation}

The $\lambda$ parameters can be used to calculate $\{J_l\}_{l=1}^5$ \cite{acin_decomp_paper}. Finally, we introduce Mosseri's Bloch-norms: given any $n$-qubit state one can compute $n$ different 1-qubit reduced density matrices \cite{fanoStuff}:

\begin{equation} 
	\rho = \frac{1}{2} \left( \mathbb{1} + \vec{r} \cdot \vec{\sigma} \right); \text{ } \mu_{\pm}\left( \rho\right)=\frac{1\pm r}{2};
	\label{1_qubit_density_matrix_canonical_form}
\end{equation}

\noindent where $\mu_{\pm}\left( \rho \right)$ are the eigenvalues of $\rho$. We call the Bloch-norm $r\equiv \left| \vec{r} \right|$, which fulfills $r=1$ if the reduced state is pure and $r<1$ if the state is mixed. Any 3-qubit state will have three different Bloch-norms. The resulting vector of Bloch-norms $(r_A, r_B, r_C)$ is restricted to a unit cube $[0,1]^3$, so it allows to graphically visualize the states. However, the full cube cannot be filled.\newline

\begin{figure}[hbpt!]
	\centering
	\subfloat[Representatives.]{\includegraphics[trim=16cm 2cm 14cm 0, clip, scale=0.30]{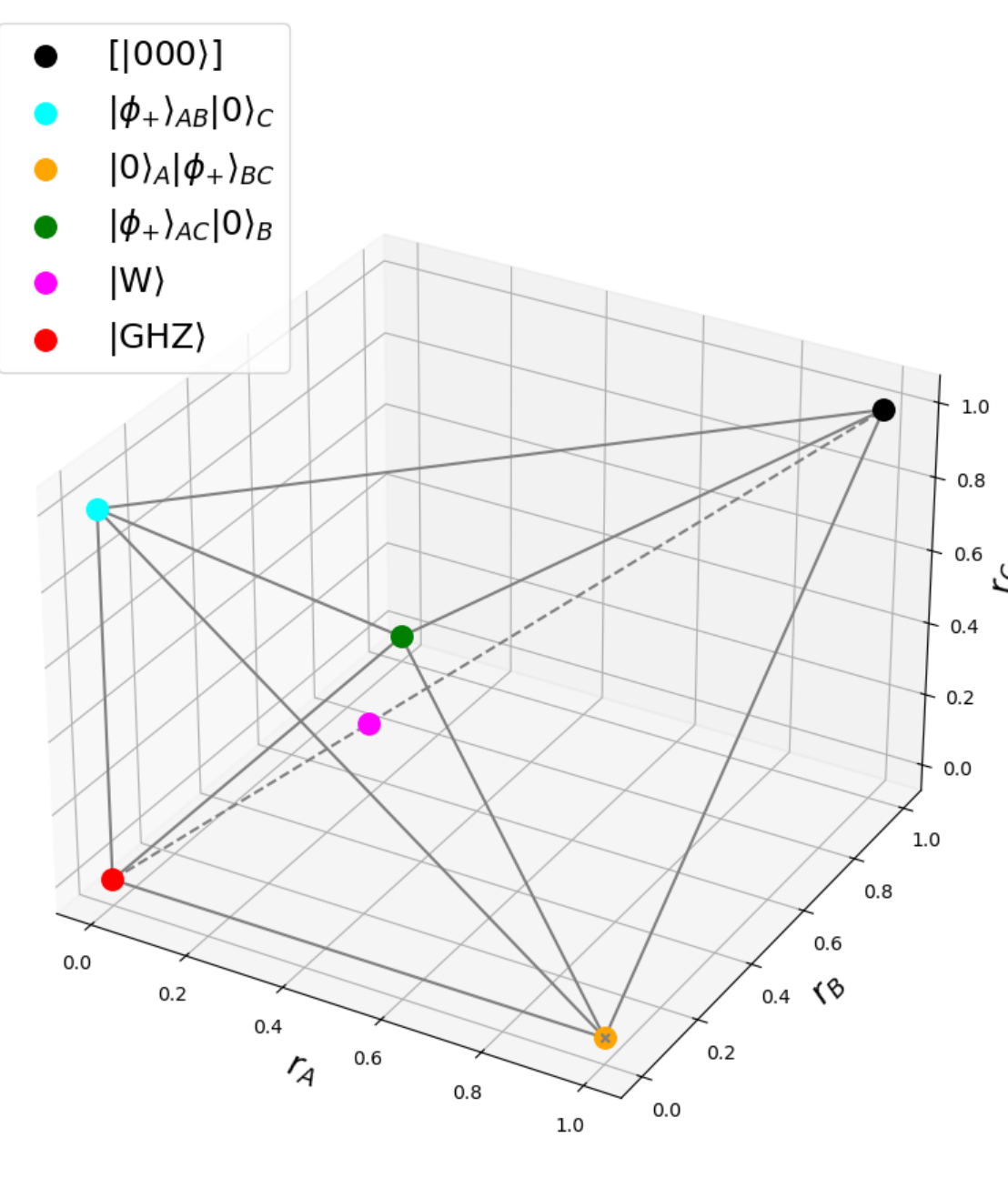}\label{tetra_01__packed}} \\
	\subfloat[Bipartite classes.]{\includegraphics[trim=16cm 2cm 14cm 0, clip, scale=0.30]{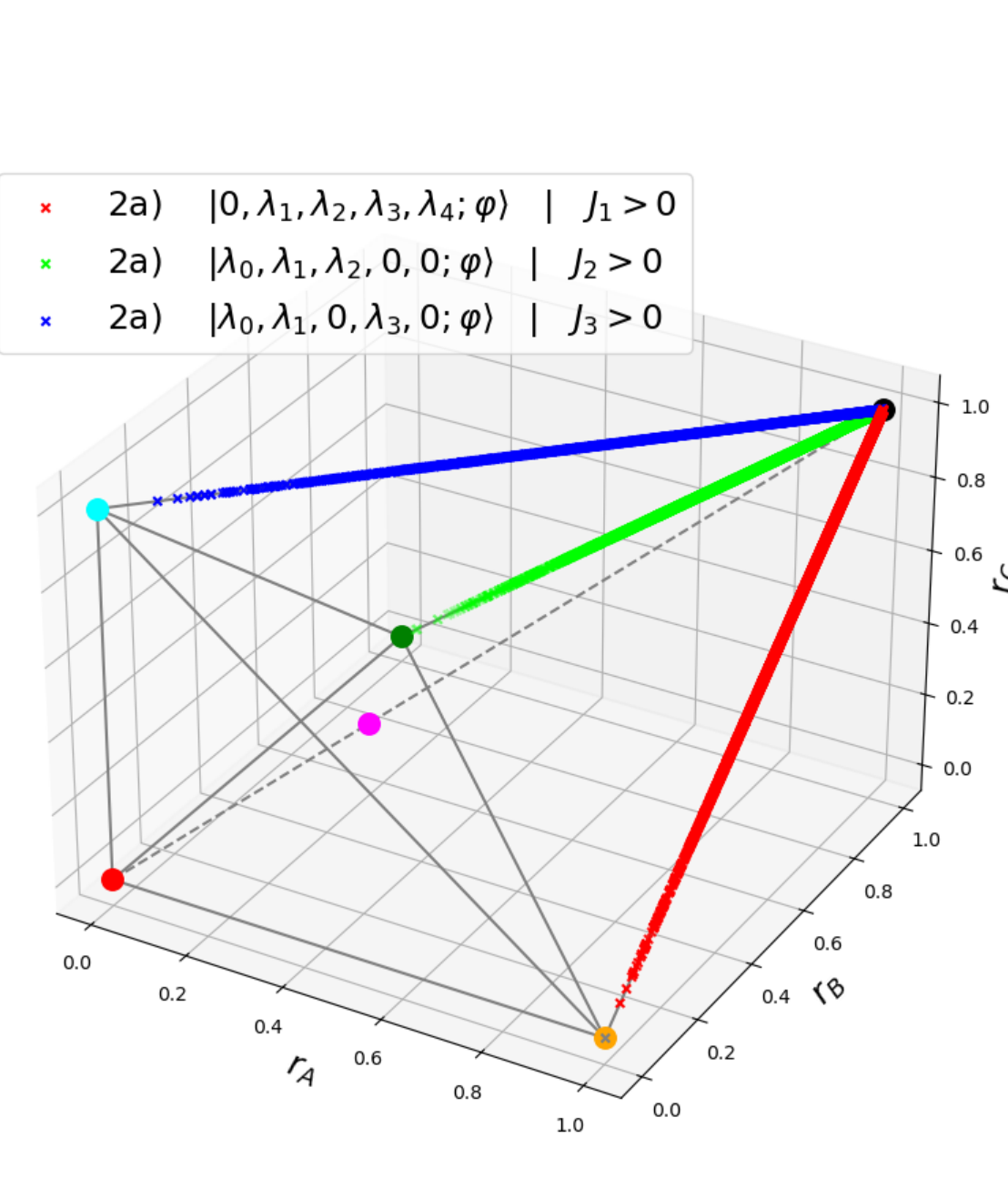}\label{tetra_2to1_class__packed}} \\
	\subfloat[W class.]{\includegraphics[trim=16cm 2cm 14cm 0, clip, scale=0.30]{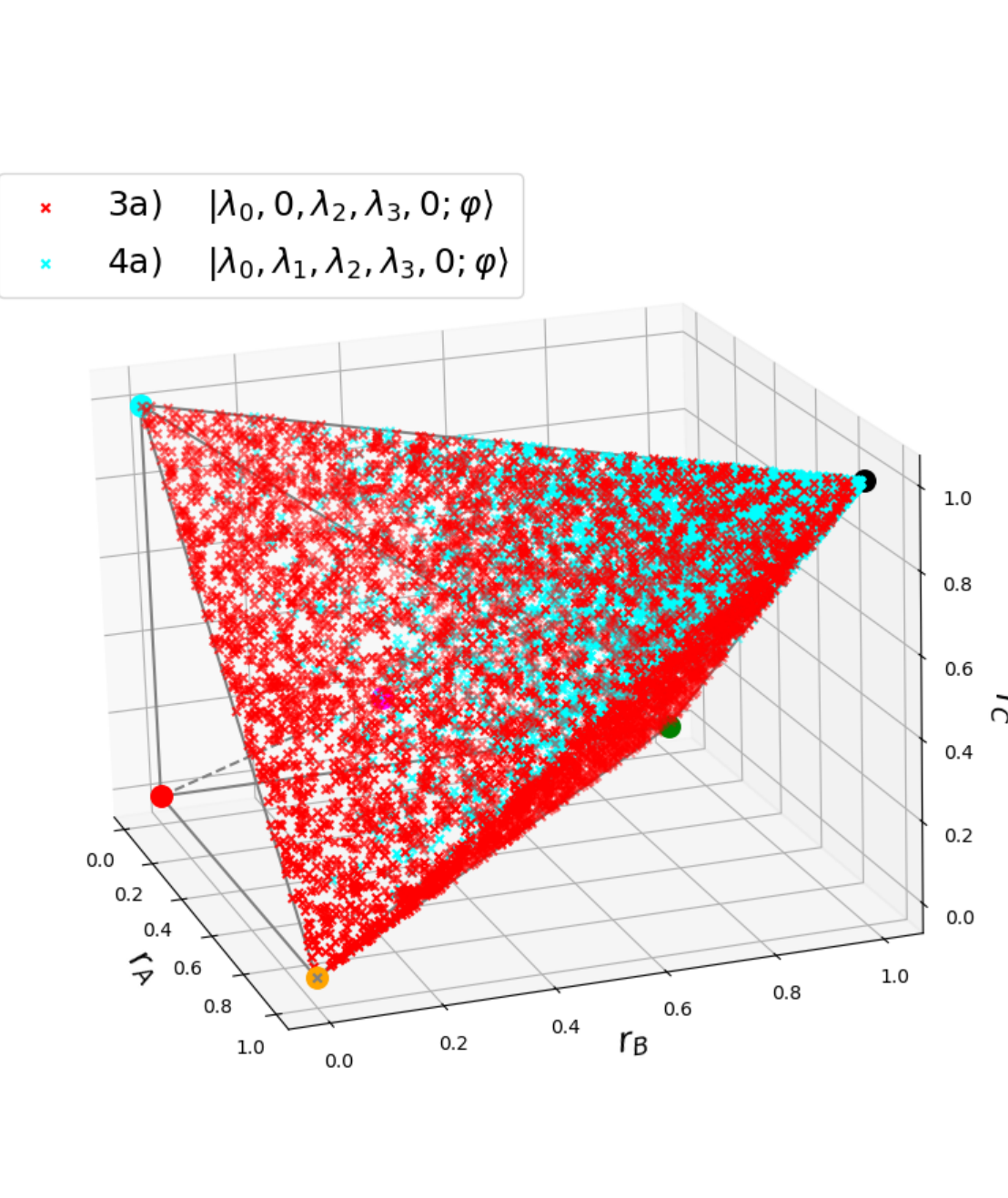}\label{tetra_W_class__packed}} \\
	\caption{Non GHZ classes.}
	\label{trial_classes_superFig}
\end{figure}

For any given entanglement class, there is a list of linear inequalities that the eigenvalues of the single-particle reduced-density matrices have to obey. These inequalities define a \textit{polytope} (a higher dimensional polygon) in which the states reside \cite{Polytope_paper}. If the eigenvalues violate the inequality, then the point lies outside the polytope and the state does not belong to the specified entanglement class. These inequalities apply as well to Mosseri's Bloch-norms. Furthermore, there exists a $1:1$ relation between the entanglement entropy of each individual qubit $S_{I}$ ($I\in \{A,B,C\}$) and its Bloch-norm $\genNorm$:


\begin{eqnarray}
    S(\rho_{I}) &=  \frac{1+\genNorm}{2} \log{\frac{2}{1+\genNorm} }   + \frac{1-\genNorm}{2} \log{\frac{2}{1-\genNorm} }
	\label{entropy_and_blochNorms}
\end{eqnarray}

\noindent so we can reproduce the table \ref{cirac_table} in terms the Bloch-norms and the tangle. This will enable us to provide a geometric viewpoint of the states as points inside the polytope. The 3-qubit polytope consists of two tetrahedron glued at the common base. This particular geometrical figure is known as \textit{triangular bipyramid} \cite{wikipedia_triangular_bipyramid}. The lower and upper tetrahedron have vertices $\{ (0,0,0), (1,0,0), (0,1,0), (0,0,1) \}$ and $\{ (1,1,1), (1,0,0), (0,1,0), (0,0,1) \}$ respectively. The representatives states of each entanglement class lie in one of the vertices of the polytope except the W state which lies at the center of the common base (see Fig. \ref{tetra_01__packed}).\newline

Finally, we list the detailed entanglement classification for 3-qubit states combining both the clases from \cite{Cirac_3qbit_classes} and the types from \cite{acin_decomp_paper}. We also include our observations on the different geometrical patterns:

\begin{enumerate}
	\item \textbf{Product state class/Type 1:} containing all 3-qubit states with no entanglement, denoted $\textClass{A-B-C}$. It is the equivalence class of $\ket{000}$ under LUs: $\ZeroClass$. All have $J_r=0$ $\forall r$ and $\genNorm=1 \text{ }\forall I$, so all are mapped to $(1,1,1)$ in the polytope.

    \item \textbf{Bipartite classes/Type 2a:} states of the form $ \ket{\varphi}_{I} \otimes \ket{\text{Entangled pair}}_{\overline{I}}$, that is $\genNorm = 1 \text{ and } \norm{I'} < 1$. These states have $J_l=0$ for all $l$ but one, which can be $J_1$, $J_2$ or $J_3$. States with $J_1 > 0$ correspond to class $\textClass{BC-A}$, $J_2 > 0$ to $\textClass{B-AC}$ and $J_3 > 0$ to $\textClass{C-AB}$. Each class covers one of the three edges of the upper tetrahedron connected to $(1,1,1)$ (see Fig. \ref{tetra_2to1_class__packed}).

	\item \textbf{W class:} includes all states with all three qubits entangled, without genuine tripartite entanglement. They can always be brought to the form:
    \begin{eqnarray}
        \sqrt{c}\ket{000} + \sqrt{d}\ket{100} \nonumber \\
        + \sqrt{a}\ket{101} + \sqrt{b}\ket{110}
    \label{general_W_class_form}
    \end{eqnarray}
\noindent with $a,b,c >0$ and $d\ge 0$ \cite{Cirac_3qbit_classes}. They can be of two types: \textbf{Type 3a} \textit{tri-Bell states} and \textbf{Type 4a}. \textbf{Type 3a} lie exclusively on the faces of the upper tetrahedron and have $\lambda_1=\lambda_4=0$ corresponding to the family with $d=0$ in \eqref{general_W_class_form}. The W state belongs to this type, having $\lambda_0 = \lambda_2 =\lambda_3 = 1/\sqrt{3}$ and $\genNorm= 1/3$, $\forall I$. \textbf{Type 4a} have $\lambda_4=0$ corresponding to the family with $d>0$ in \eqref{general_W_class_form}. They are located in the upper tetrahedron, and they accumulate near the $(1,1,1)$ point \cite{probDistributions_3qubits} (see Fig. \ref{tetra_W_class__packed}).

	\item \textbf{GHZ class:} contains states with genuine tripartite entanglement. There are 5 types:
\end{enumerate}
	\begin{enumerate}[]
		\item \textbf{Type 2b}  \textit{generalized GHZ states}. They have $J_l=0,\forall l$ except for $J_4=\tau/4 \implies \lambda_j=0$ for $j\in\{1,2,3\}$. The standard GHZ state corresponds to the values $\lambda_0=\lambda_4=1/\sqrt{2}$. They lie on the central diagonal connecting $(0,0,0)$ and $(1,1,1)$ (see Fig. \ref{tetra_2b_type__packed}). Notice that for $\lambda_0\in\{1/\sqrt{3}, \sqrt{2/3}\}$, they occupy the same point in the polytope as the $\tW$ state.

		\item \textbf{Type 3b} \textit{extended GHZ states}: They have $\lambda_i=\lambda_j=0$ for $j,k \in \{1,2,3\}$ with $j \ne k$, so either $\lambda_1=\lambda_2=0$ or $\lambda_1=\lambda_3=0$ or $\lambda_2=\lambda_3=0$. Each one spans a different triangle connecting the main diagonal with any of the three vertices of the face $\{ (1,0,0),(0,1,0),(0,0,1) \}$ (see Fig. \ref{tetra_3b_type__packed}).

		\item \textbf{Type 4b} have either $\lambda_2 = 0$ or $\lambda_3=0$. They lie in the space between two of the three triangles defined by type 3b. If $\lambda_2=0$, they lie between the triangles of kinds \textit{1-2} and \textit{2-3}, while if $\lambda_3=0$ then between \textit{2-3} and \textit{1-3}. No states of type 4b lie between \textit{1-3} and \textit{kind 2-3} (see Fig. \ref{tetra_4b_type__packed}).

		\item \textbf{Type 4c} have $\lambda_1 = 0$. These populate the polytope without any clear pattern.

            \item \textbf{Type 5} \textit{generic GHZ states:} these have $\lambda_j\ne 0$ and $J_k\ne 0$, $\forall j,k$. They may lie anywhere in the polytope.

	\end{enumerate}

\multiComment{
\begin{figure}
	\centering
	\subfloat[\textbf{Type 2b}.]{\includegraphics[trim=16cm 2cm 14cm 0, clip, scale=0.29]{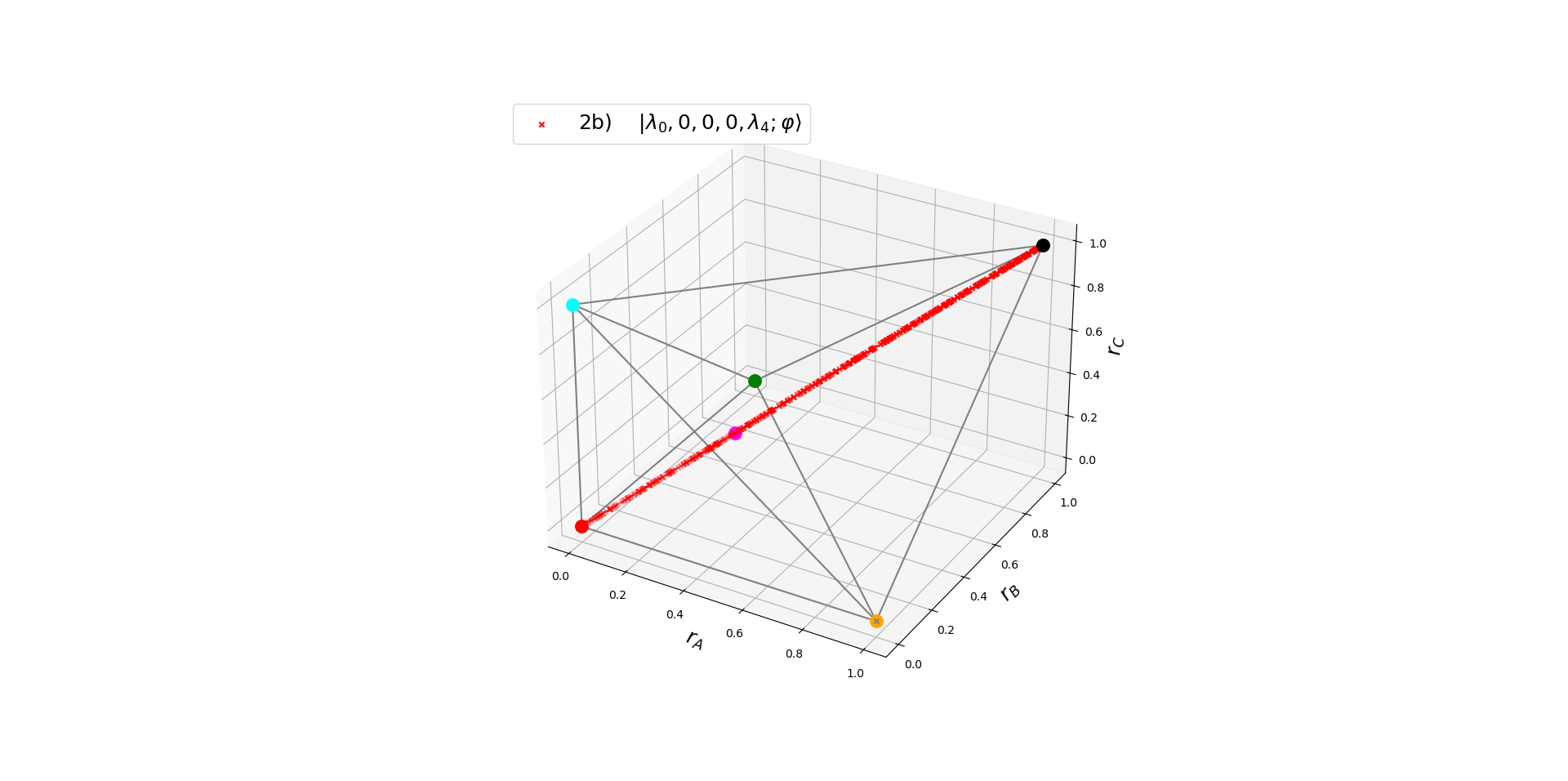}\label{tetra_2b_type__packed}} \\
	\subfloat[\textbf{Type 3b}.]{\includegraphics[trim=16cm 2cm 14cm 0, clip, scale=0.29]{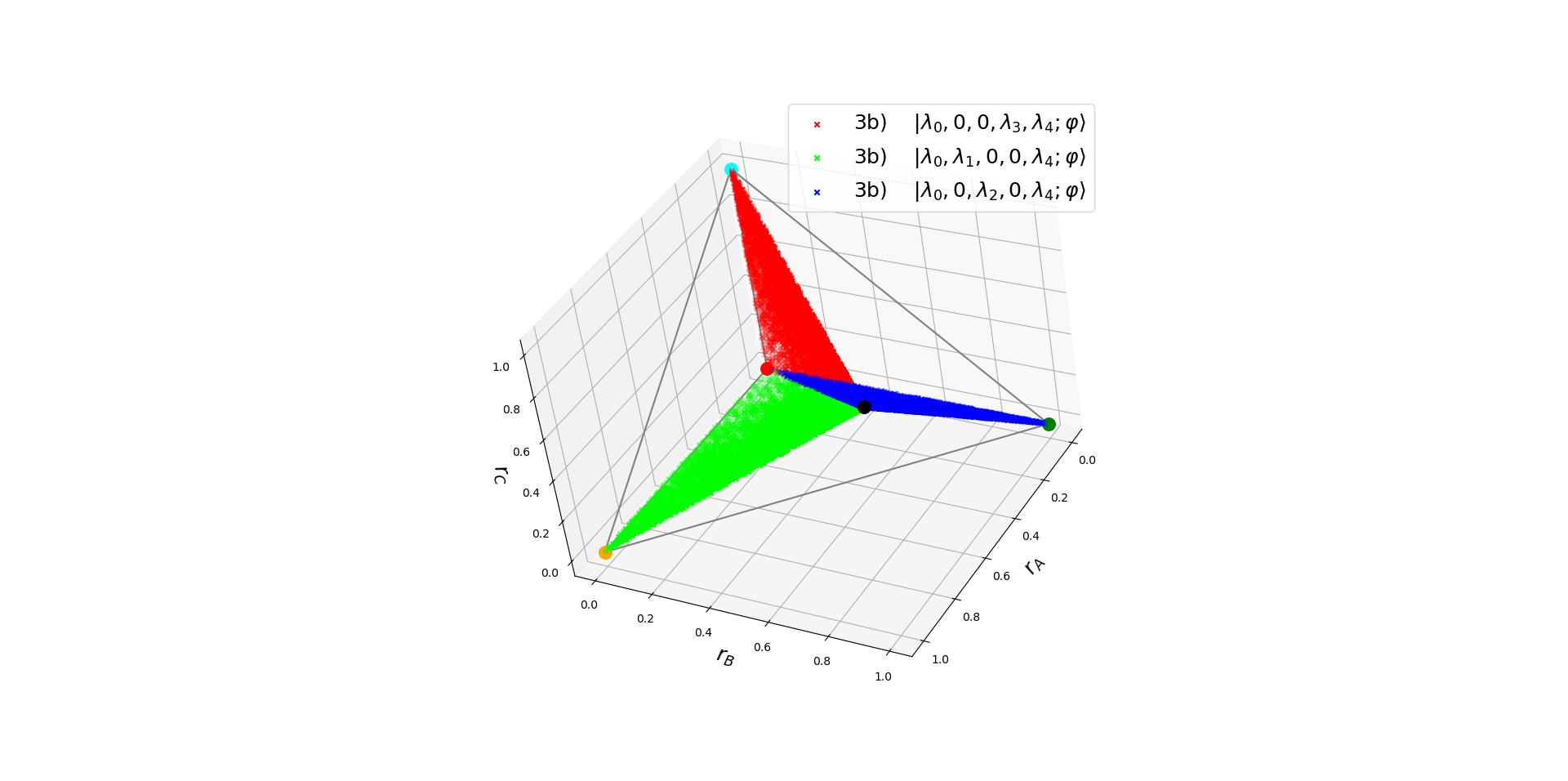}\label{tetra_3b_type__packed}} \\
	\subfloat[\textbf{Type 4b}.]{\includegraphics[trim=15cm 2cm 14cm 0, clip, scale=0.29]{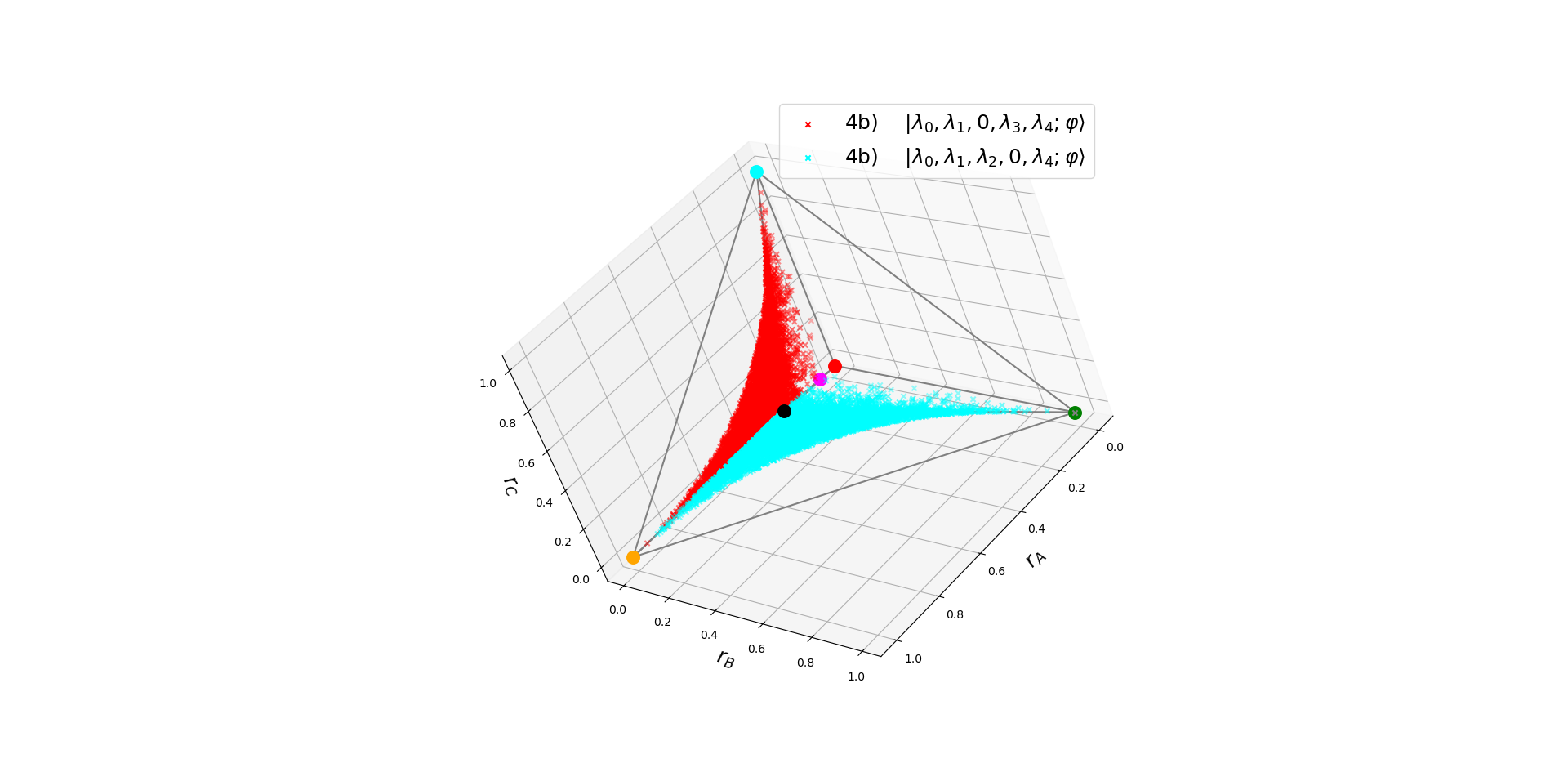}\label{tetra_4b_type__packed}} \\
	\caption{GHZ class.}
	\label{trial_classes_superFig}
\end{figure}
}

\begin{figure}[hbpt!]
	\centering
	\subfloat[\textbf{Type 2b}.]{\includegraphics[trim=16cm 2cm 14cm 0, clip, scale=0.3]{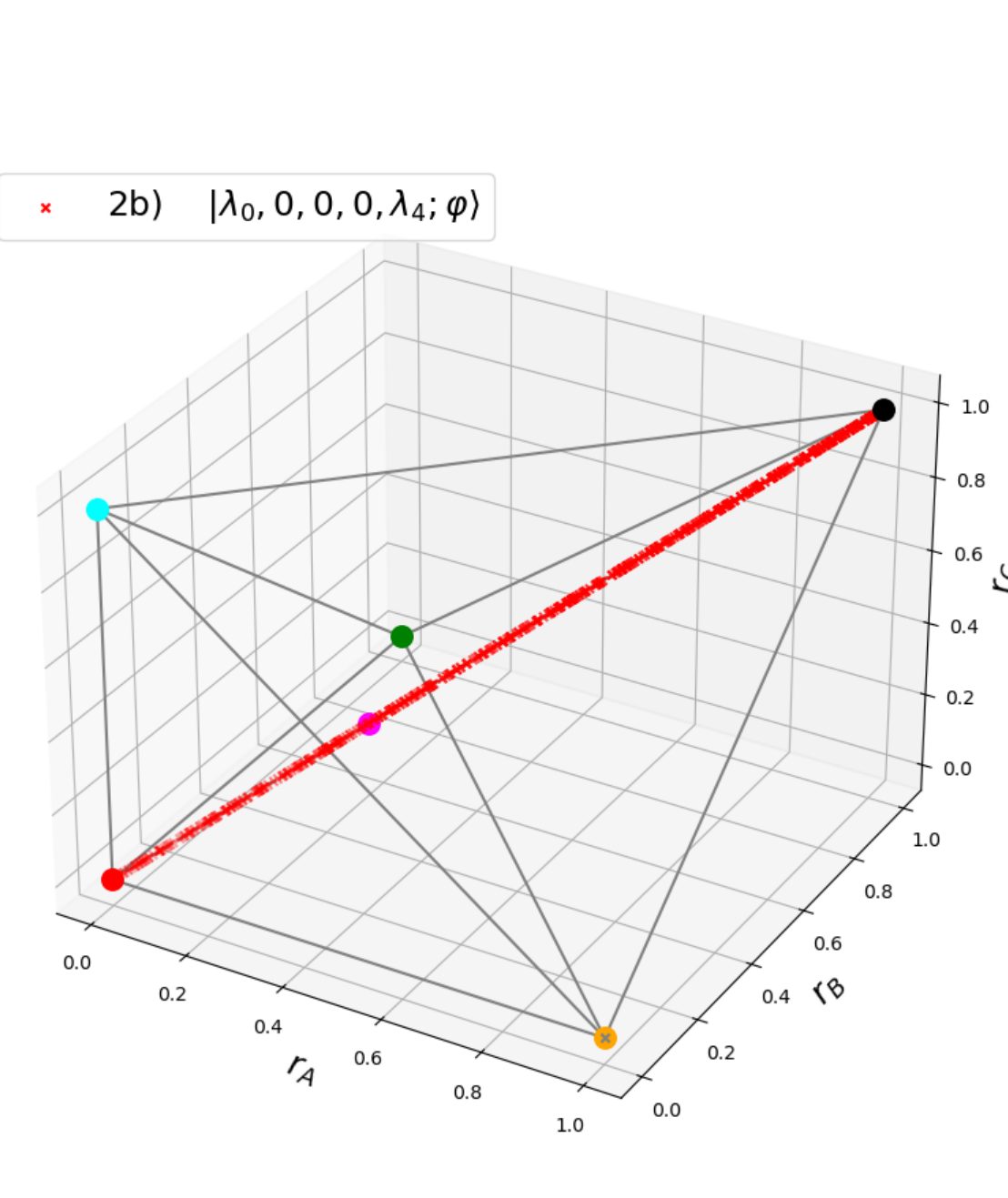}\label{tetra_2b_type__packed}} \\
	\subfloat[\textbf{Type 3b}.]{\includegraphics[trim=16cm 2cm 14cm 0, clip, scale=0.3]{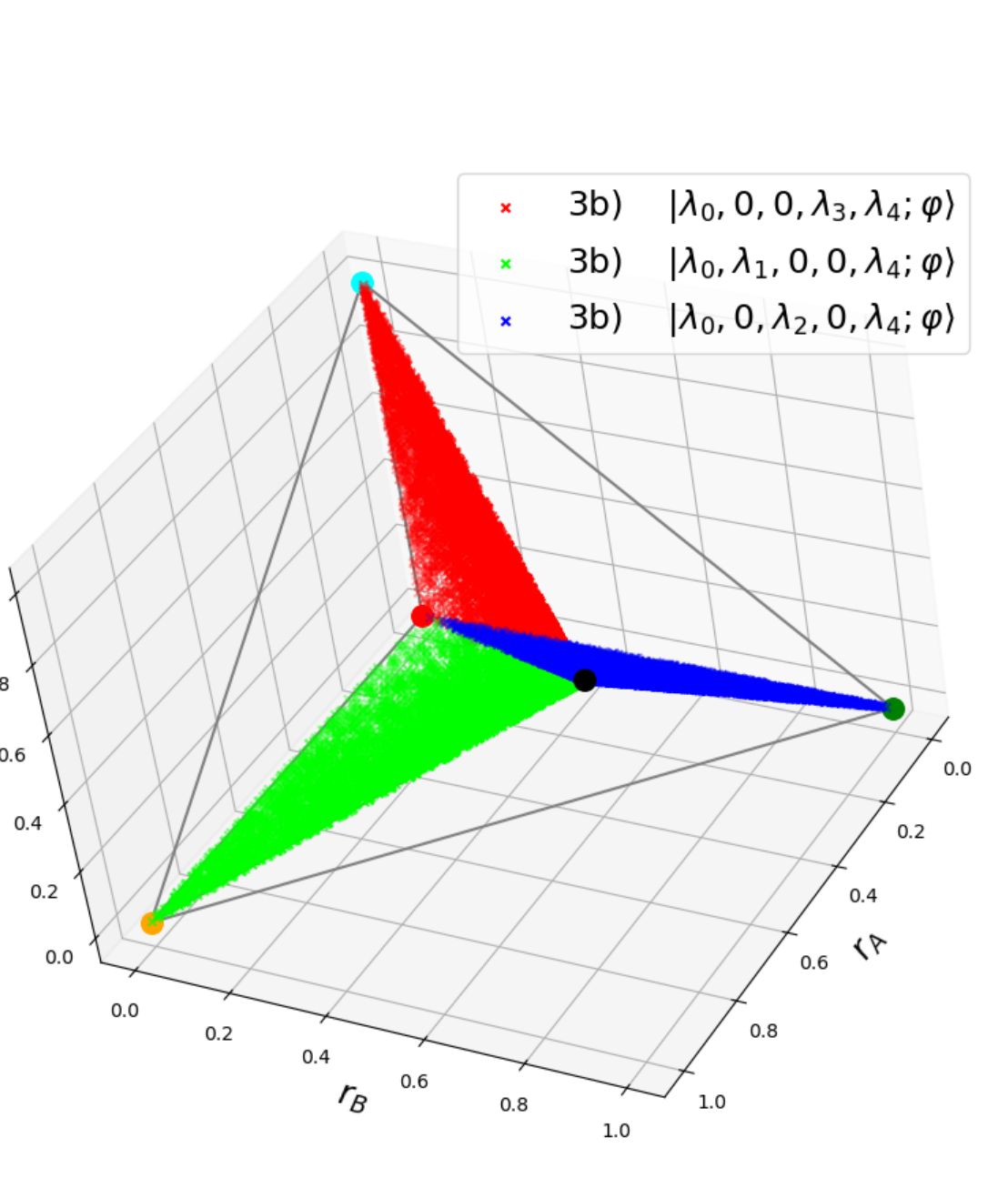}\label{tetra_3b_type__packed}} \\
	\subfloat[\textbf{Type 4b}.]{\includegraphics[trim=15cm 2cm 14cm 0, clip, scale=0.3]{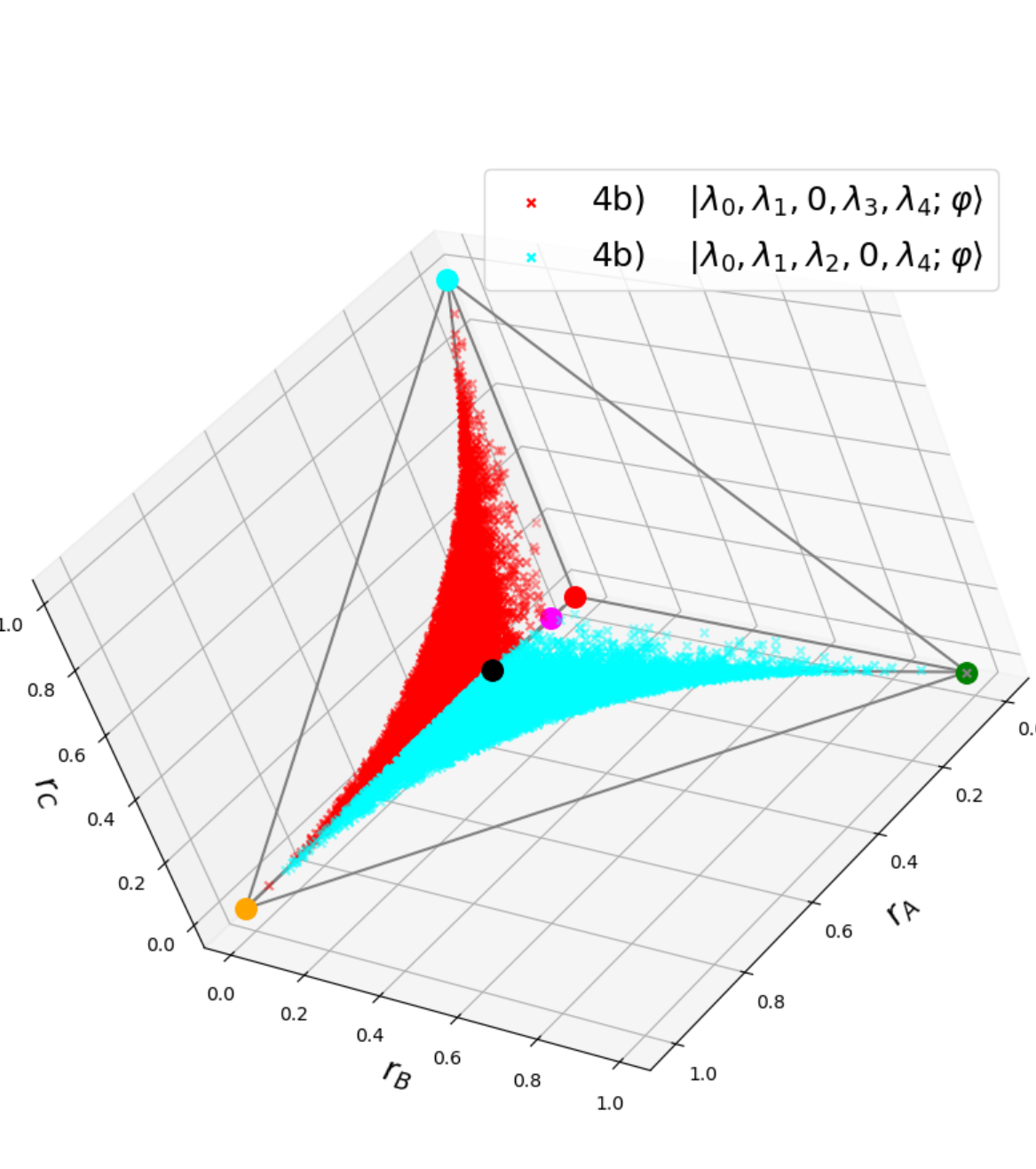}\label{tetra_4b_type__packed}} \\
	\caption{GHZ class.}
	\label{trial_classes_superFig}
\end{figure}

The proofs of the localization of the states in the polytope can be found in Appendix \ref{section_PolytopeProofs}. For related recent results, see \cite{natureDamageControl}. Observe that the states belonging to the GHZ class occupy any of the two tetrahedra, but the states not in this class are restricted to the upper tetrahedron. In particular, the state with the lowest value of $||(r_A,r_B,r_C)||$, with $\tau=0$, is the state $\tW$. \textcolor{black}{These observations points towards a connection between the tangle and the Bloch-norm geometrical picture.}

\section{Geometry of the tangle}\label{section_Tangle}


\multiComment{
\begin{itemize}
    \item \textbf{Type 2b}: $ 1-\tau= r_C^2 = r_A^2 = r_B^2$.

    \item \textbf{Type 3b}: $1-\tau=r_C^2 > r_A^2, r_B^2$ for \textit{type 1-2}, and similarly for \textit{2-3} and \textit{1-3}.

    \item \textbf{Type 4b}:
\end{itemize}
\begin{equation}
	\begin{matrix} 1-\tau \left( \lambda_2=0 \right)  =  r_C^2 - r_B^2 + r_A^2  \\ \\  1-\tau \left( \lambda_3=0 \right)  =  r_B^2 - r_C^2 + r_A^2  \end{matrix}
	\label{tangle_for_type_4b}
\end{equation}
}

In search of such a connection, we wish to study the relation between the tangle and $R$, where $R$ is defined as:
\begin{equation}
	R:= ||(r_A, r_B, r_C)|| = \sqrt{r_A^2 + r_B^2 + r_C^2}
	\label{norm_of_norms}
\end{equation}

\begin{figure}[hbpt!]
	\centering
	\subfloat[Type 2b.]{\includegraphics[trim=4cm 1cm 4cm 0, clip, scale=0.2]{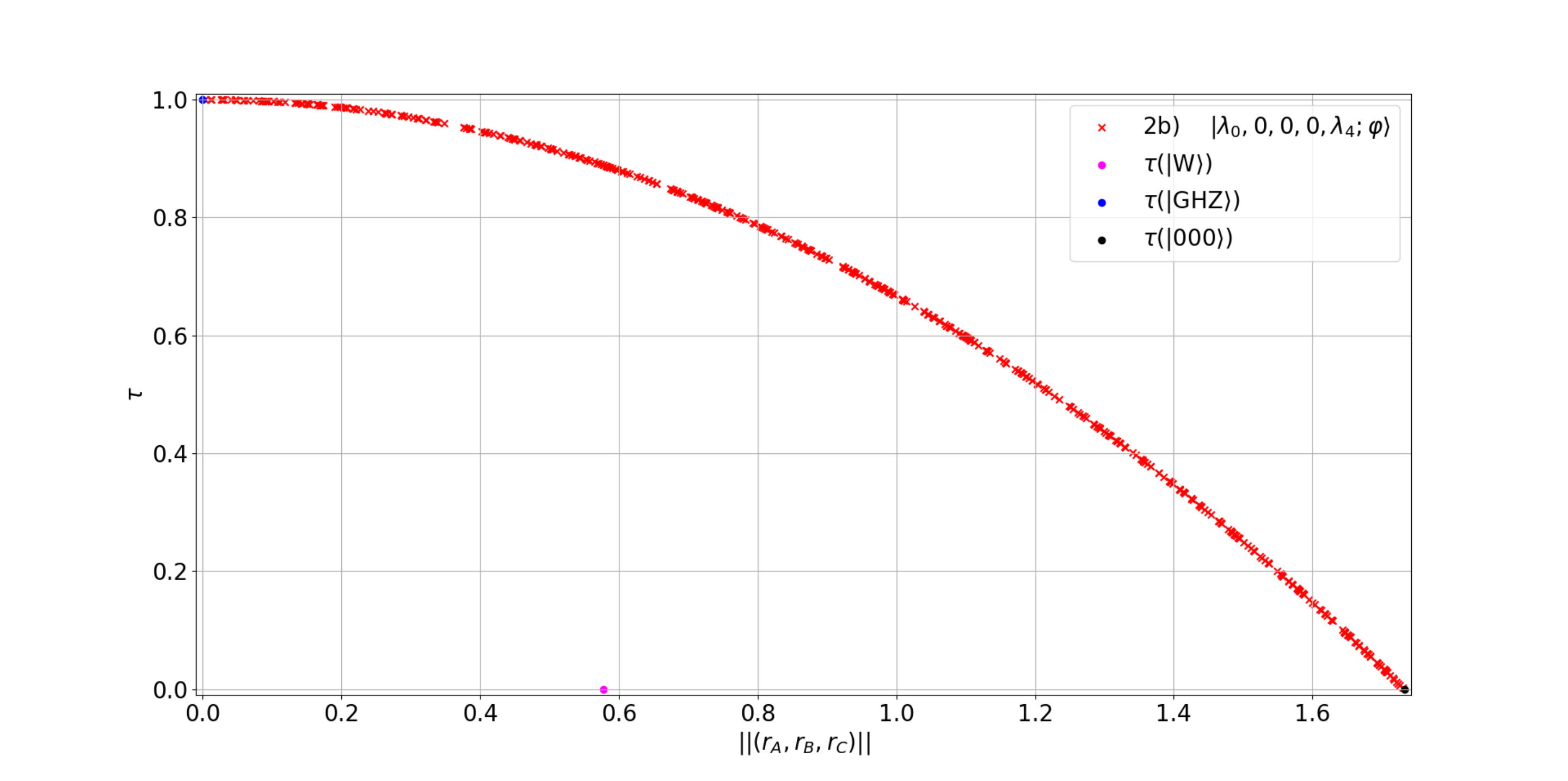}\label{different_R_tau_areas__a}} \\
	\subfloat[Type 3b.]{\includegraphics[trim=4cm 1cm 4cm 0, clip, scale=0.2]{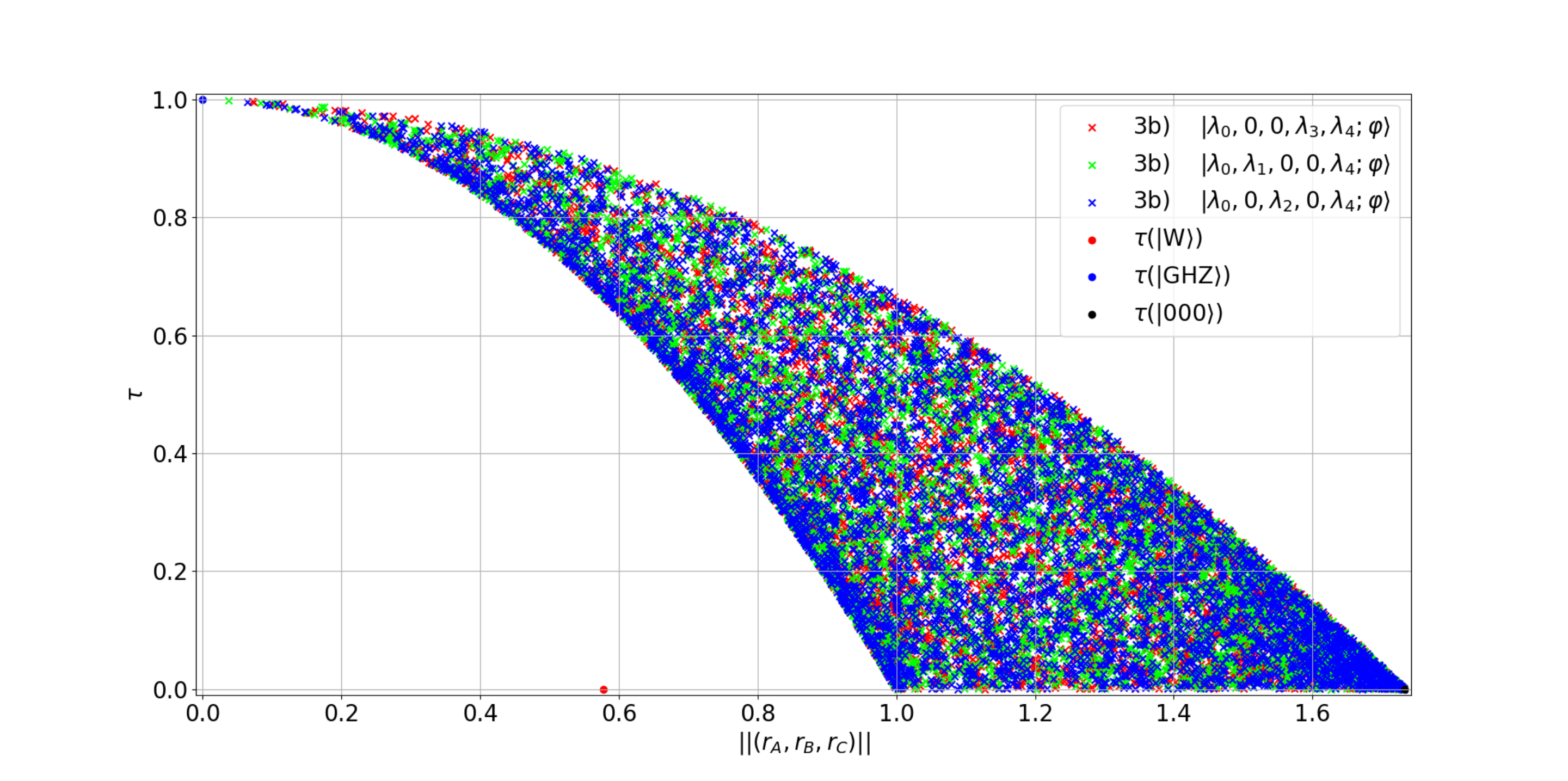}\label{different_R_tau_areas__b}} \\
	\subfloat[Type 4b.]{\includegraphics[trim=4cm 1cm 4cm 0, clip, scale=0.2]{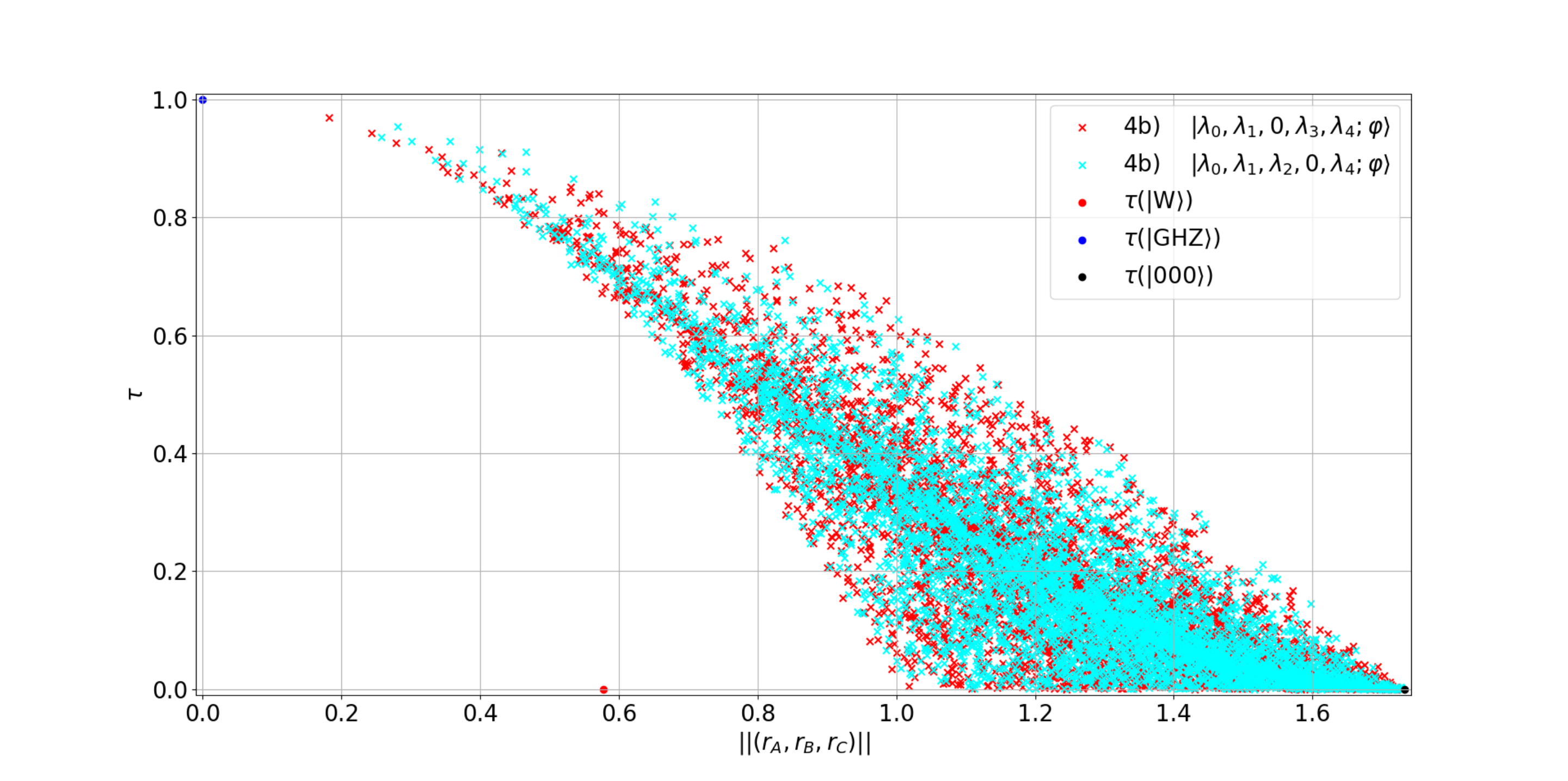}\label{different_R_tau_areas__c}} \\
	\subfloat[Type 4c.]{\includegraphics[trim=4cm 1cm 4cm 0, clip, scale=0.2]{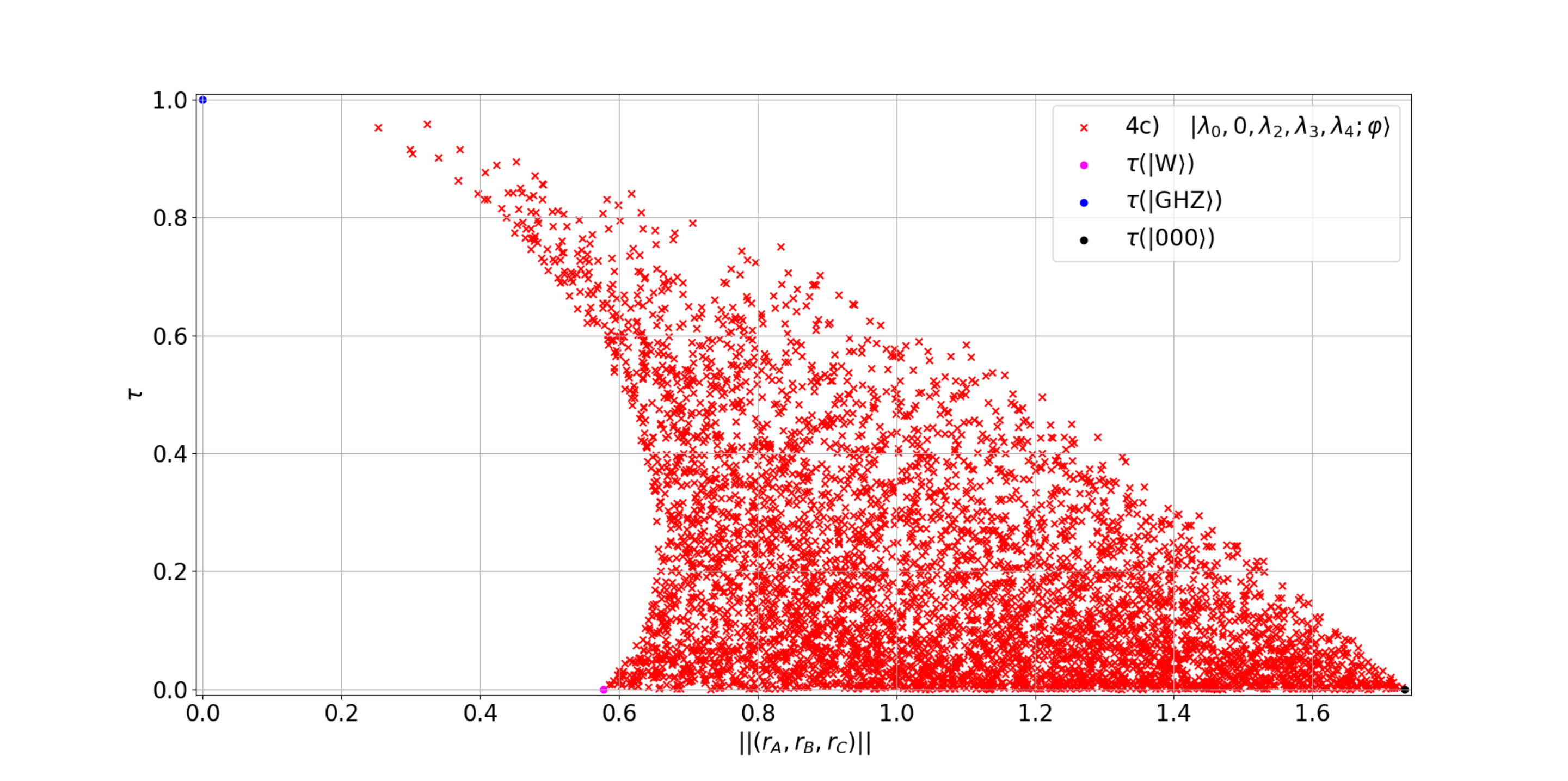}\label{different_R_tau_areas__d}}
	\caption{GHZ entanglement class states in the $(R,\tau)$ diagram.}
	\label{different_R_tau_areas}
\end{figure}

For a generic state, by means of the CD \eqref{cannonical_form}, one can compute:
\begin{align}
R^2 & = 3 \left( \lambda_0^4 + \lambda_1^4 +  \lambda_2^4 +  \lambda_3^4 +  \lambda_4^4   \right) \nonumber \\
 & + 6\lambda_1 \lambda_2 \lambda_3 \lambda_4 \cos{(\varphi)} \nonumber \\
& + 6 \left(  \lambda_1^2 \left[ +\lambda_0^2 + \lambda_2^2 + \lambda_3^2 \right]  + \lambda_4^2  \left[ -\lambda_0^2 + \lambda_2^2 + \lambda_3^2 \right]  \right) \nonumber \\
 & -2\left( \lambda_0^2 \left[ \lambda_2^2 + \lambda_3^2 \right] + \lambda_2^2 \lambda_3^2 - \lambda_1^2 \lambda_4^2 \right)
\label{R_norm_master_equation}
\end{align}

\noindent where we haven't used the normalization condition yet. On the other hand, the tangle can be computed as:
\begin{equation}
	\begin{matrix} \tangle{\psi} = 4 \left| \hdet{t_{ijk}} \right| = 4 \lambda_0^2 \lambda_4 ^2 \\ = 4 \lambda_0^2 \left( 1-\lambda_0^2 - \left| \vec{\lambda} \right|^2 \right)  \end{matrix}
	\label{tangle_master_eq}
\end{equation}
\noindent where Hdet is Cayley's hyperdeterminant \cite{cayley_paper, adrian_latorre_paper, latorre_cited_book, tangle_introduction}. Our aim is to substitute $\tau$ in \eqref{R_norm_master_equation} by using \eqref{tangle_master_eq} and then, obtain a function of the form $\tau=\tau(R; \{\lambda_j\})$. For this, we plot $(R,\tau)$ for the different types of GHZ states. We find that there are 3 zones where states lie (see Fig. \ref{different_R_tau_areas}):

\begin{itemize}
	\item Type 2a states occupy a curve $\upperTau$ (see Fig. \ref{different_R_tau_areas__a}), maximizing the value of $\tau$ for a given value of $R$.
	\item Type 3b and 4b occupy a common area (see Figures \ref{different_R_tau_areas__b} \& \ref{different_R_tau_areas__c}), bounded from above by $\upperTau$ and from below by $\middleTau$.
	\item Type 4c and 5 occupy a larger area than the previous cases (see Fig. \ref{different_R_tau_areas__d}), upper-bounded by $\upperTau$ and lower-bounded a curve with two branches: $\tau_{\uparrow}(R)$ and $\tau_{\downarrow}(R)$.
\end{itemize}
For type 2a, eq.\eqref{R_norm_master_equation} becomes:
\begin{equation}
	\upperTau =  1-R^2 /3
	\label{upperTau_solution}
\end{equation}
For types 3b and 4b, one can obtain (see Appendix \ref{section_DetailsBounds} ):
\begin{equation}
    \begin{matrix}
        \middleTau =  \left\{\begin{matrix} 5 \upperTau - 4\sqrt{\upperTau}  & \text{if } R\overset{<}{\sim} 0.56 \\   1-R^2 & \text{if } 0.56\overset{<}{\sim} R \le 1 \\ 0 & \text{ otherwise}  \end{matrix} \right. 
    \end{matrix}
    \label{second_case_R_tau_exploration__07__aligned__short}
\end{equation}
\noindent where $\middleTau \le \tau(R) \le \upperTau$. For types 4c and 5, the two branches are:
\begin{eqnarray}
\tau_{\uparrow}(R) & = \left( \frac{17}{49} - \frac{5}{21} R^2 \right) - \frac{32}{147} \sqrt{9-21 R^2} \quad \label{tau_up_branch} \\
\tau_{\downarrow}(R) & = \frac{1}{4} \left( 1-\sqrt{\frac{\left( R_{\tW}+R_{\star} \right) \left( R_{\star}-R \right)}{R_{\star}^2 - R_{\tW}^2}} \right) \label{tau_down_branch}
\end{eqnarray}
\noindent where $R_{\tW}=1/\sqrt{3}$, $R_{\star}=\sqrt{3/7}$ and with bounding conditions: $\tau \ge \tau_{\uparrow}(R)$ for $R\le R_{\tW}$; either $\tau \ge \tau_{\uparrow}(R)$ or $0 \le \tau \le \tau_{\downarrow}(R)$ when $R\in[R_{\tW},R_{\star}]$; and $\tau\in[0,\upperTau]$ when $R>R_{\star}$.\newline

Notice the following properties:
\begin{enumerate}
	\item States of type 2b maximize $\tau$ for a given $R$ (see Fig. \ref{different_R_tau_areas__a}). Moreover, they lie in the main diagonal of the polytope (i.e., $d( \vec{r}, \vec{V}_{\text{line}}  ) =0$, see Fig. \ref{tetra_2b_type__packed}).
	\item States of types 3b, 4b and 4c deviate from the main diagonal (i.e., $d( \vec{r}, \vec{V}_{\text{line}}  ) >0$) and have $\tau<\upperTau$.
\end{enumerate}

\noindent which leads us to the following geometrical ansatz for the tangle:
\begin{equation}
	\tau \left( \vec{r} \right)   =  1- \frac{\left| \vec{r}\right|^2}{3} - d\left( \vec{r}, V_{\text{line}}  \right) \cdot \mathcal{F}\left( \vec{r} \right)
	\label{geometry_for_tangle}
\end{equation}
\noindent where $\ket{\psi} \in  \text{GHZ}$ excluding type 5 and $\mathcal{F}(\vec{r}) \ge 0$. An example of this function for types \textit{3b-12} and \textit{4b-1} respectively:
\begin{equation}
	\begin{matrix} \mathcal{F} \left( \vec{r}\left( \lambda_1=\lambda_2=0 \right) \right) \simeq 2 r_C \sqrt{\frac{2}{3}} + O\left( r_C^2 \right)   \\  \mathcal{F} \left( \vec{r}\left( \lambda_2=0 \right) \right) \simeq   2\sqrt{3} r_B + O\left( r_B^2 \right) \end{matrix}
	\label{lowest_order_F_factor}
\end{equation}
\noindent with similar expressions for the different states in each type (see \eqref{lowest_order_F_factor_all_3b} and \eqref{lowest_order_F_factor_all_4b}).


\section{Study of the tangle in 3-qubit spin chains}\label{section_Chains}
\begin{figure}
	\centering
	\includegraphics[trim=3cm 0cm 4cm 0, clip, scale=0.19]{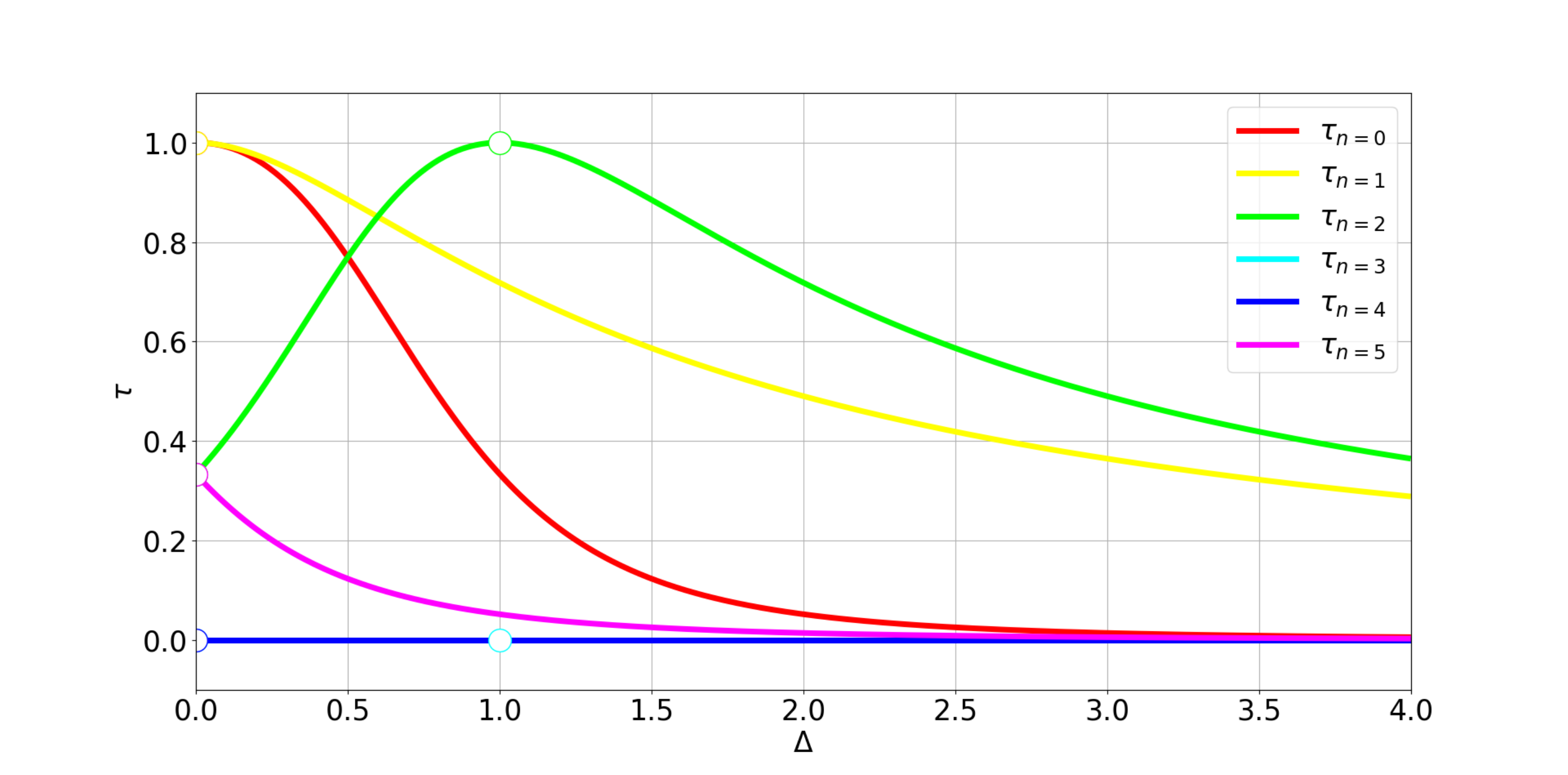}
	\caption{Tangle of TFIM levels \eqref{TFIM_eigenState_spectrum_nonDeg} \eqref{TFIM_eigenState_spectrum_deg}.}
	\label{chain_TFIM_tangles_plot_1}
\end{figure}

With these tools at hand, we now study tripartite entanglement present in energy levels of standard spin chain Hamiltonians with periodic boundary conditions (PBC). We begin by studying the Transverse Field Ising Model (TFIM) for 3 qubits:
\begin{equation}
	H_{\text{TFIM}} = -\displaystyle\sum_{j=0}^{2} \left( X_j X_{j+1}  \right)  -\Delta \displaystyle\sum_{j=0}^{2} Z_j
	\label{TFIM_hamiltonian}
\end{equation}
\noindent where $\Delta \ge 0$. The energy spectrum can be obtained exactly (see \eqref{TFIM_energy_spectrum} in Appendix \ref{section_DetailsChains_TFIM} and Fig. \ref{chain_TFIM_spectrum}). \textcolor{black}{Obtaining the eigenstates \eqref{TFIM_eigenState_spectrum_nonDeg} \eqref{TFIM_eigenState_spectrum_deg}) outside of \textit{level crossings} allows us to calculate the tangle at each energy level:}
\begin{equation}
	\tau_{n=0,2} = \frac{16 f_n}{g_n^4}; \quad \tau_{n=1,5} = \frac{48 f_n^3}{g_n^4}; \quad \tau_{n=3,4}=0;
	\label{TFIM_tangles_1}
\end{equation}
\noindent where $f_n$ and $g_n$ come from components of the eigenstates in the canonical basis \eqref{TFIM_eigenState_spectrum_nonDeg} \eqref{TFIM_eigenState_spectrum_deg} and depend on $\Delta$ \eqref{TFIM_eigenState_parameters}. The $(\Delta, \tau)$ plot is shown in Fig. \ref{chain_TFIM_tangles_plot_1}.\newline

\begin{figure}
	\centering
	\includegraphics[trim=4cm 0cm 4cm 0, clip, scale=0.20]{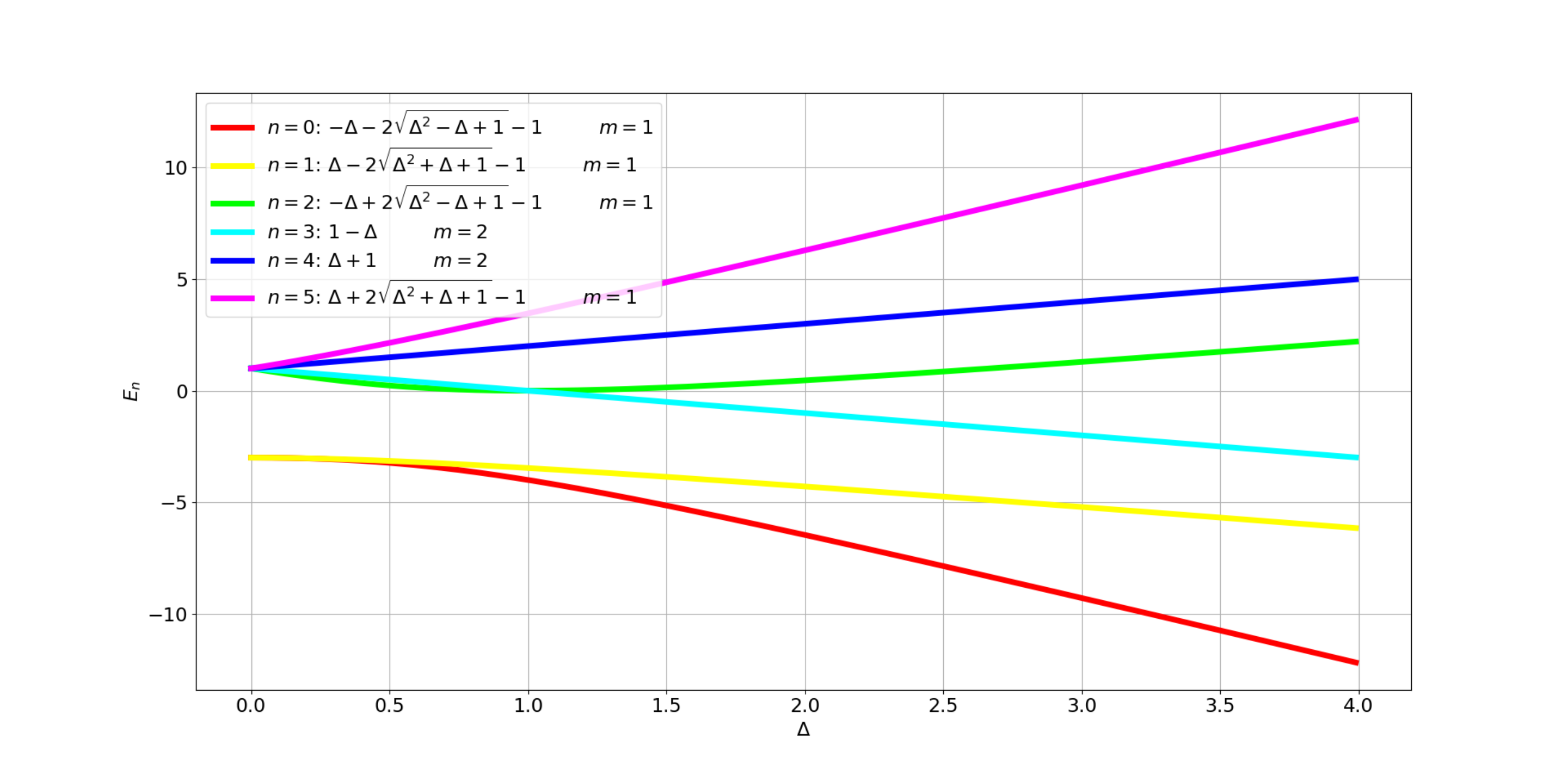}
    \caption{Energy spectrum of $H_{TFIM}$.}
	\label{chain_TFIM_spectrum}
\end{figure}



\begin{figure}[hbpt!]
	\centering
	\subfloat[$\ket{n=0}$]{\includegraphics[trim=18cm 2cm 7cm 2cm, clip, scale=0.23]{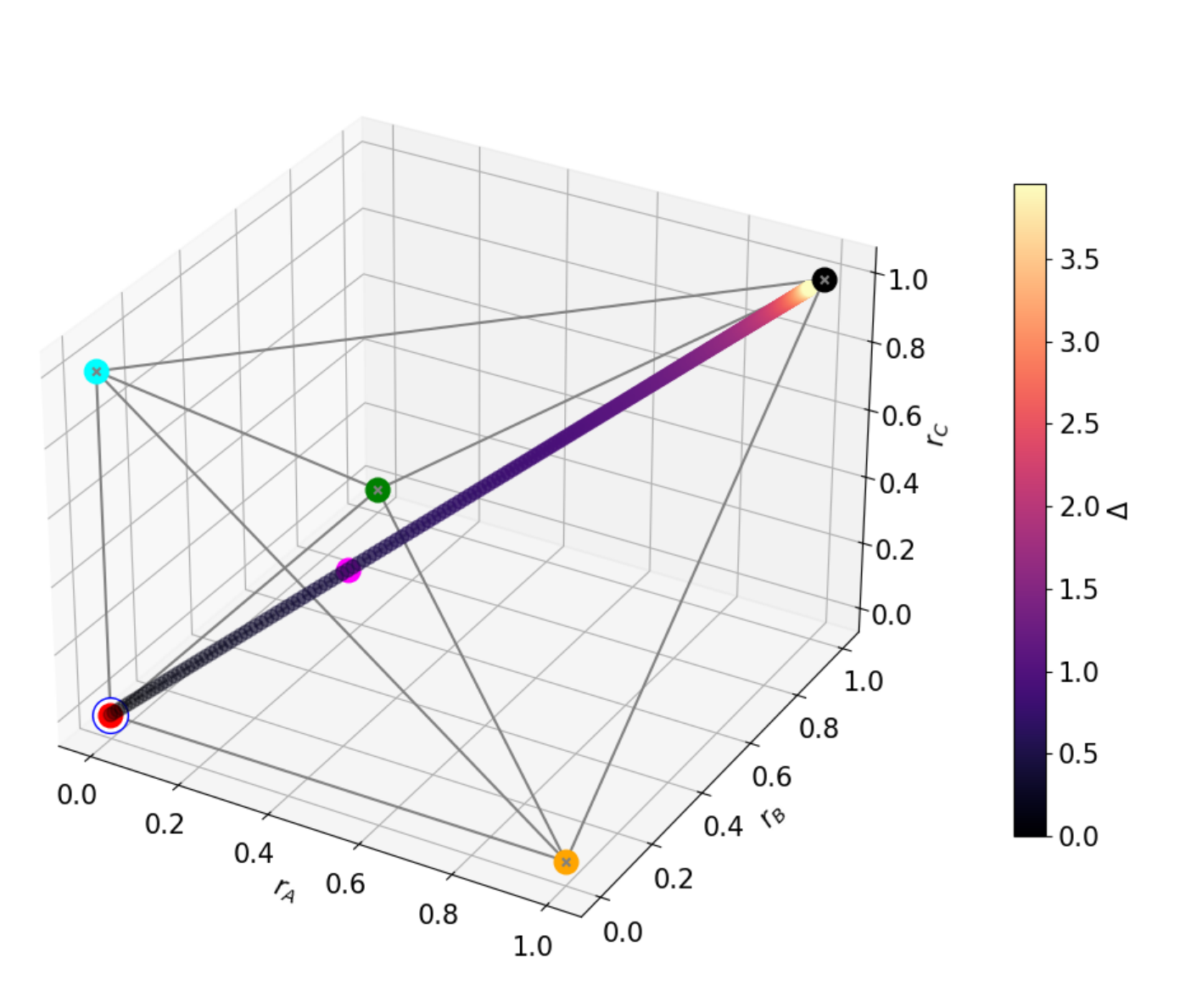}\label{chain_TFIM_trajectory_n0}} \\
	\subfloat[$\ket{n=1}$]{\includegraphics[trim=18cm 2cm 7cm 2cm, clip, scale=0.23]{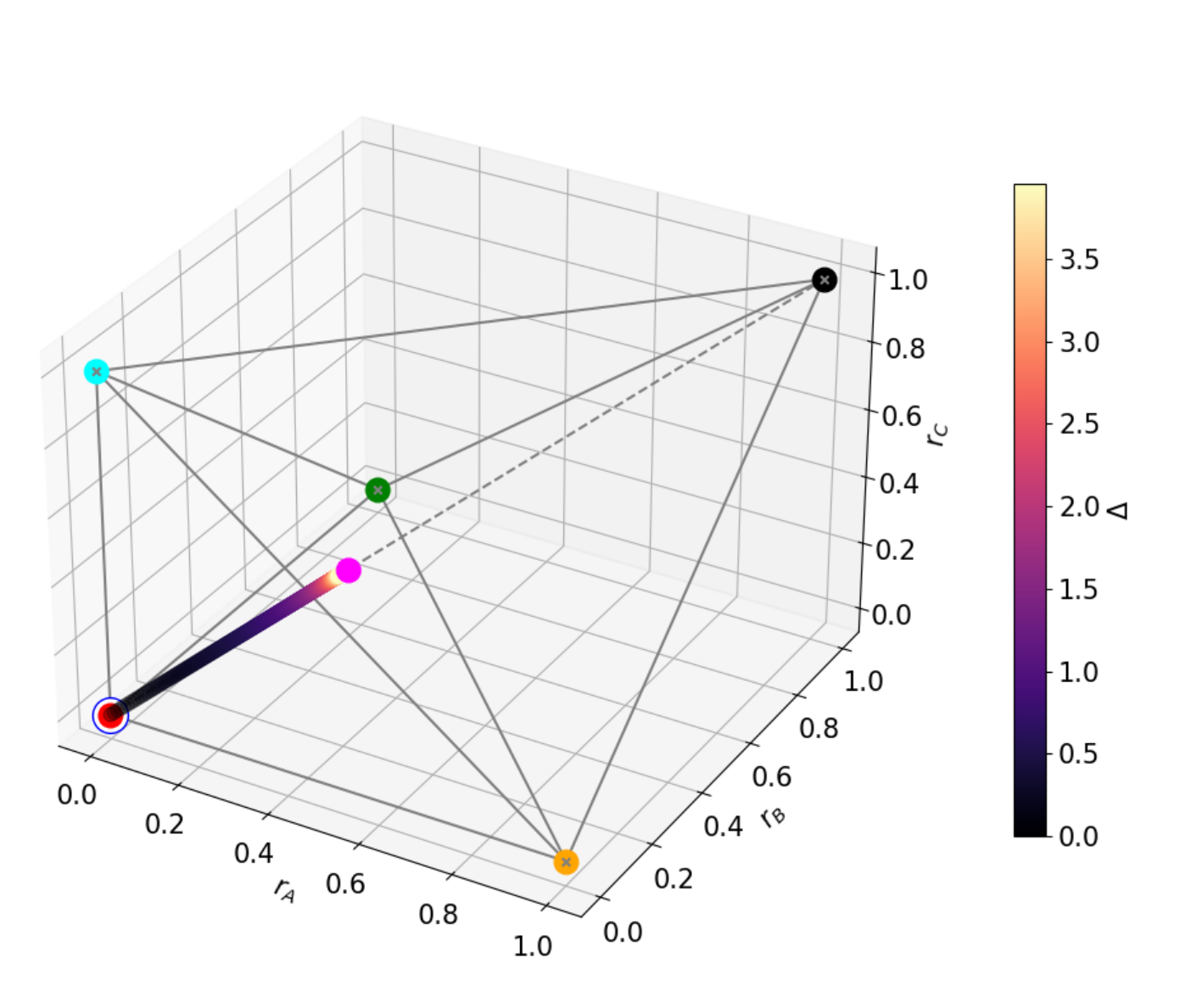}\label{chain_TFIM_trajectory_n1}} \\
	\subfloat[$\ket{n=5}$]{\includegraphics[trim=18cm 2cm 7cm 2cm, clip, scale=0.23]{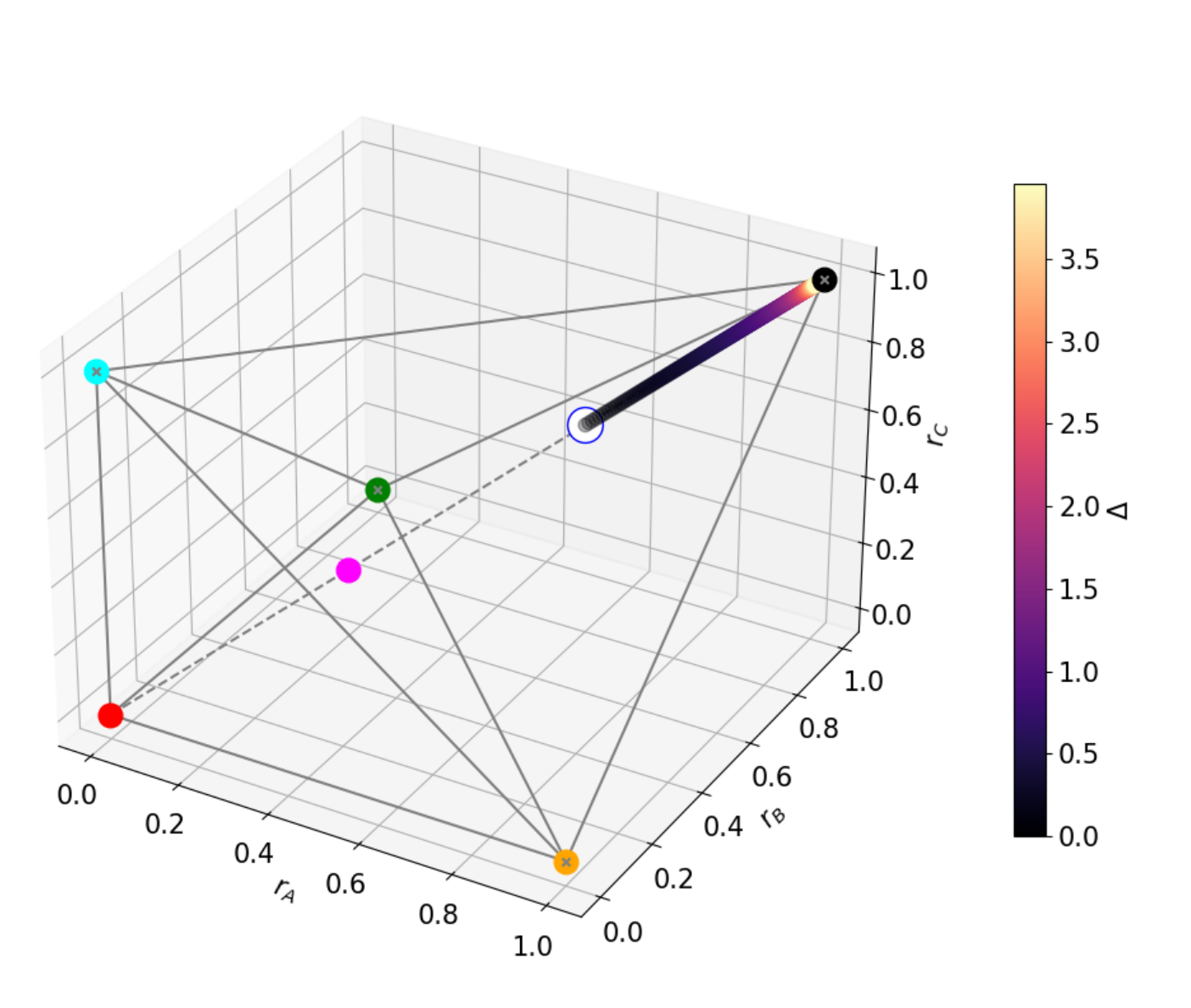}\label{chain_TFIM_trajectory_n5}} \\
        \subfloat[$n=3 \text{ } \& \text{ } 4$]{\includegraphics[trim=15cm 2cm 13cm 2cm, clip, scale=0.23]{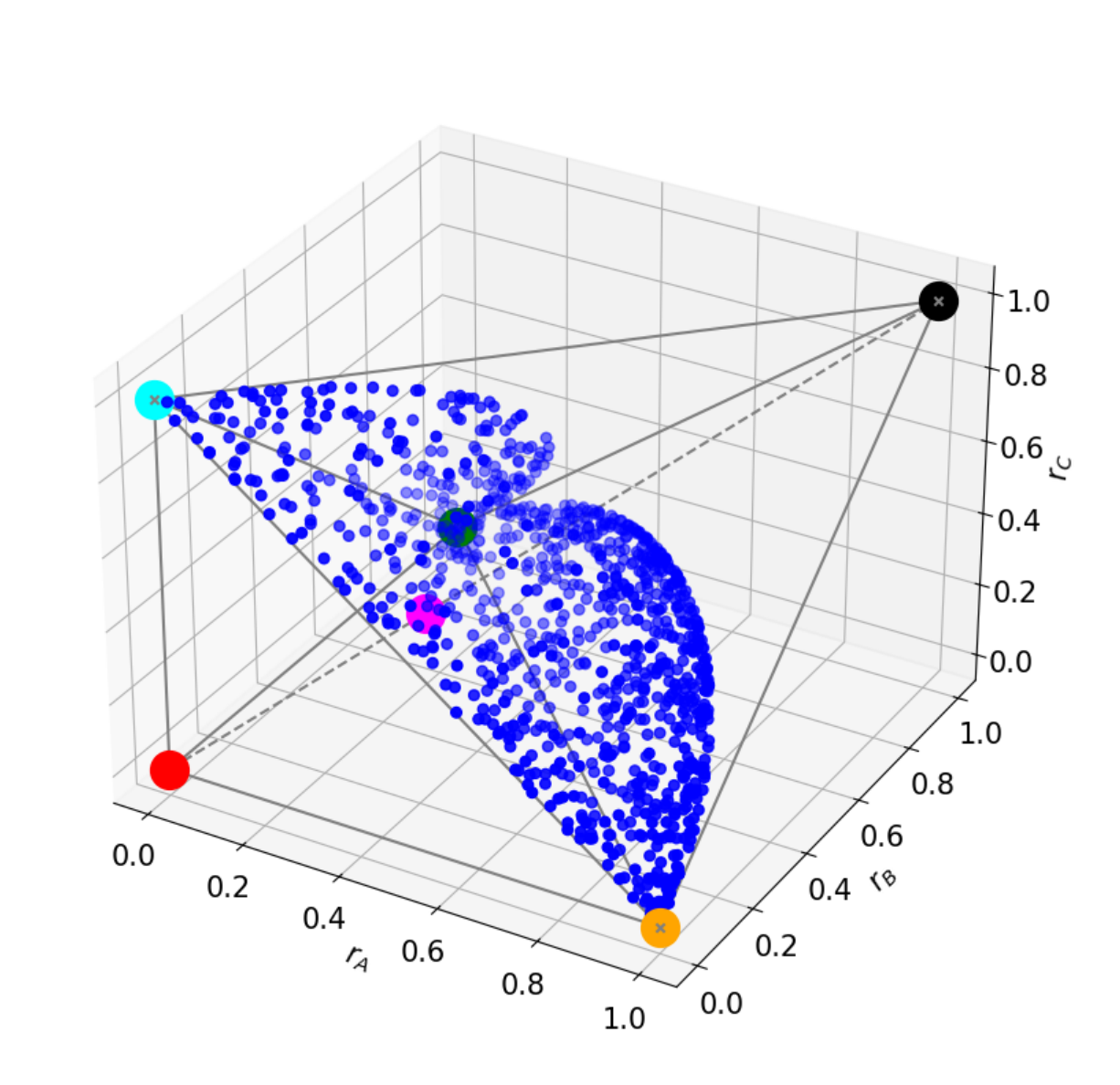}\label{chain_TFIM_trajectory_n3}}
	\caption{TFIM Bloch-norms out of level crossings.}
        \label{TFIM_trajectories_noCrossings}
\end{figure}

For non-degenerate subspaces, the Bloch-norms vector \eqref{TFIM_eigenState_spectrum_nonDeg_BlochNorms} defines a trajectory parametrized by $\Delta$ (see Figures \ref{chain_TFIM_trajectory_n0}, \ref{chain_TFIM_trajectory_n1}, \ref{chain_TFIM_trajectory_n5}) while for degenerate subspaces there is no $\Delta$-dependence and they will span a manifold of dimension greater than 1 (see Fig. \ref{chain_TFIM_trajectory_n3}). This is because the parameters controlling the Bloch-norm values are the weights of the allowed superposition. Notice that when increasing $\Delta$, levels $n=1,2$ loose their tangle slower than the other levels (see Fig. \ref{chain_TFIM_tangles_plot_1}). This shows that certain eigenstates have more \textit{robust} tangle than others under changes of the external field $\Delta$.\newline


\multiComment{
\begin{figure}[hbpt!]
	\centering
	\subfloat[$\ket{n=2}$ for the TFIM in $\Delta\in[0,1)$]{\includegraphics[trim=18cm 2cm 7cm 2cm, clip, scale=0.23]{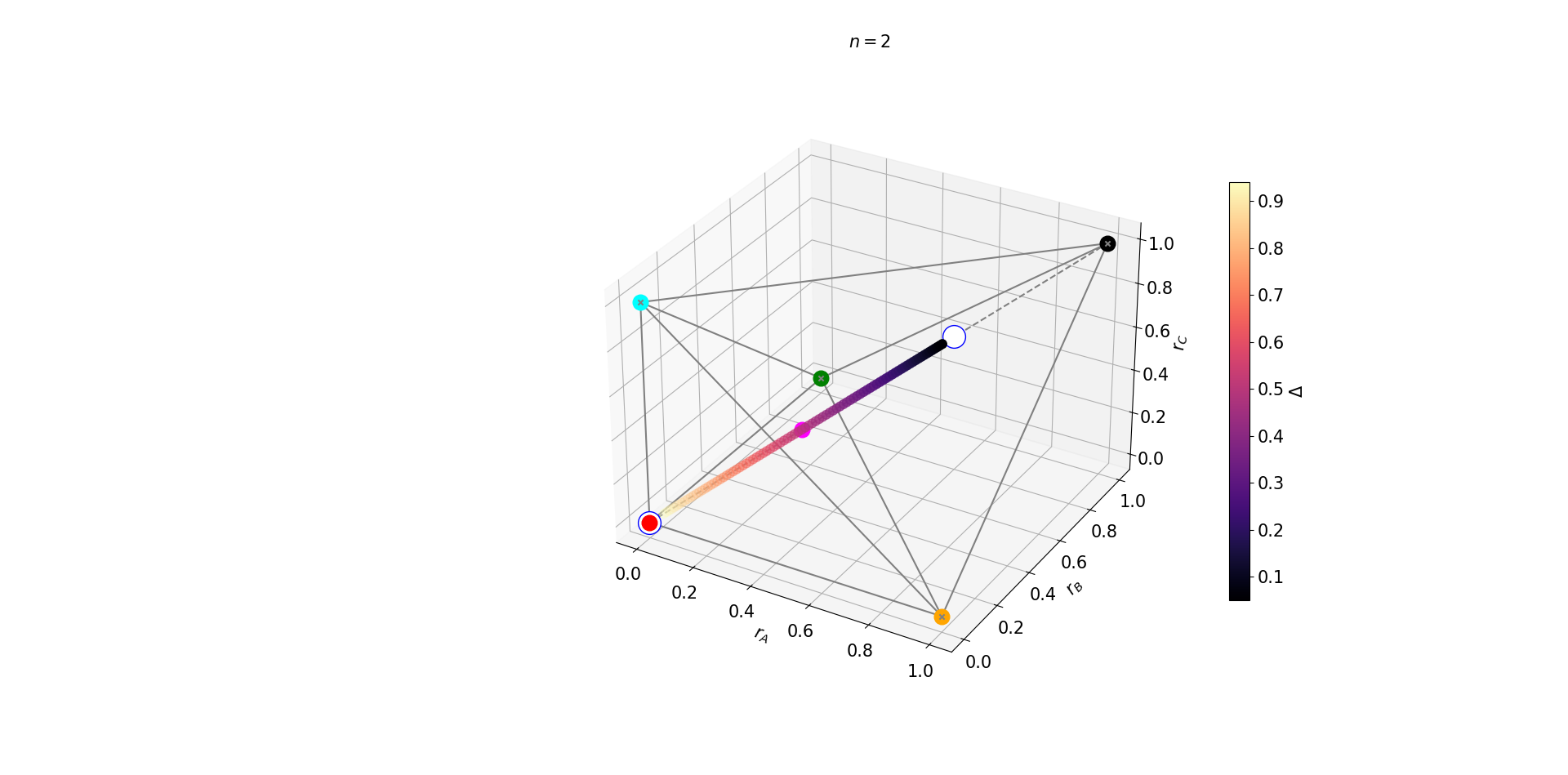}\label{chain_TFIM_trajectory_n2_pre}} \\
	\subfloat[$\ket{n=2}$ for the TFIM in $\Delta\in(1,4)$]{\includegraphics[trim=18cm 2cm 7cm 2cm, clip, scale=0.23]{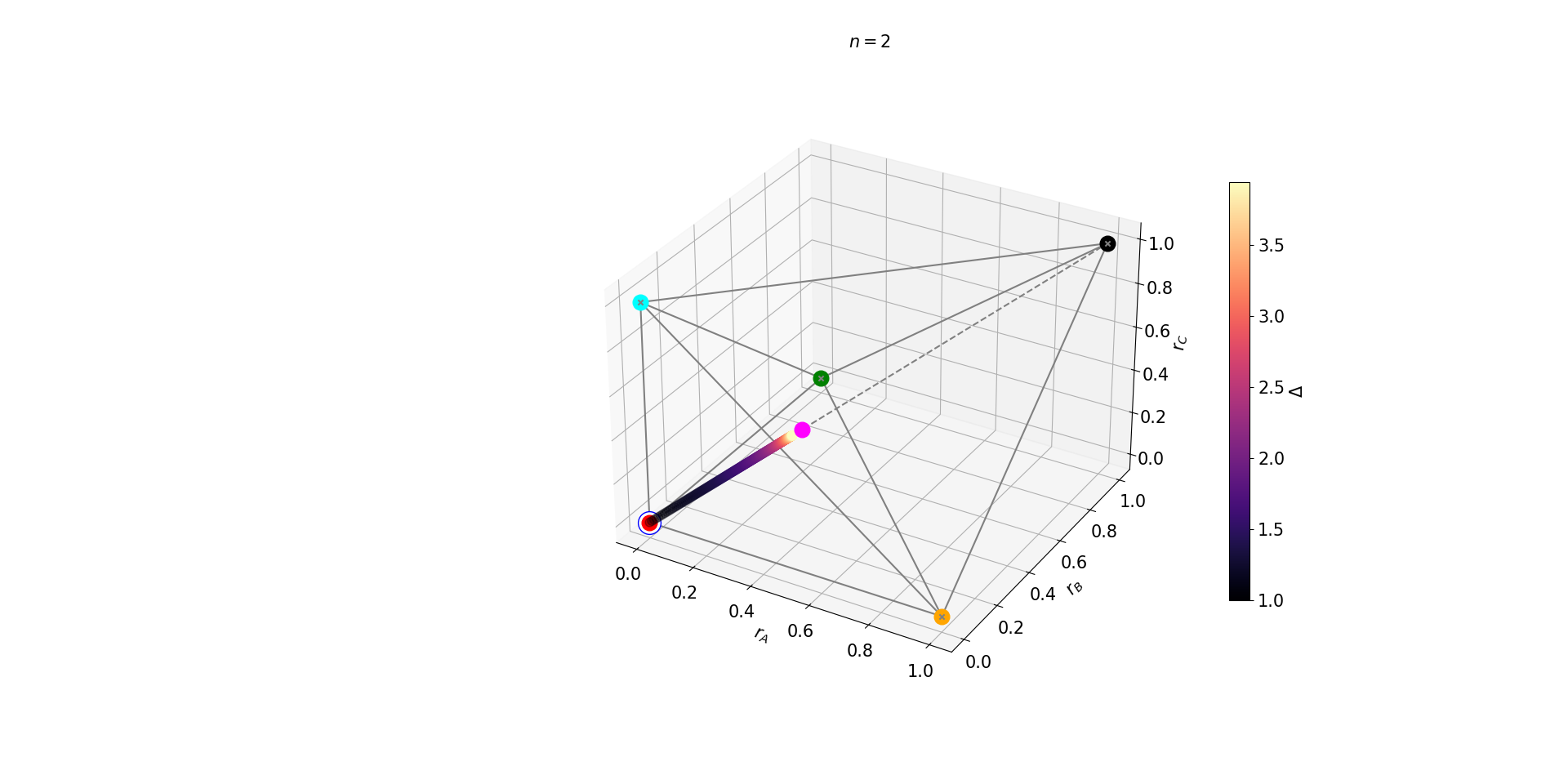}\label{chain_TFIM_trajectory_n2_post}} \\
	\caption{Trajectory for TFIM $\ket{n=2}$.}
\end{figure}
}

\begin{figure}[hbpt!]
	\centering
	\subfloat[$\ket{n=2}$ for the TFIM in $\Delta\in[0,1)$]{\includegraphics[trim=18cm 2cm 7cm 2cm, clip, scale=0.23]{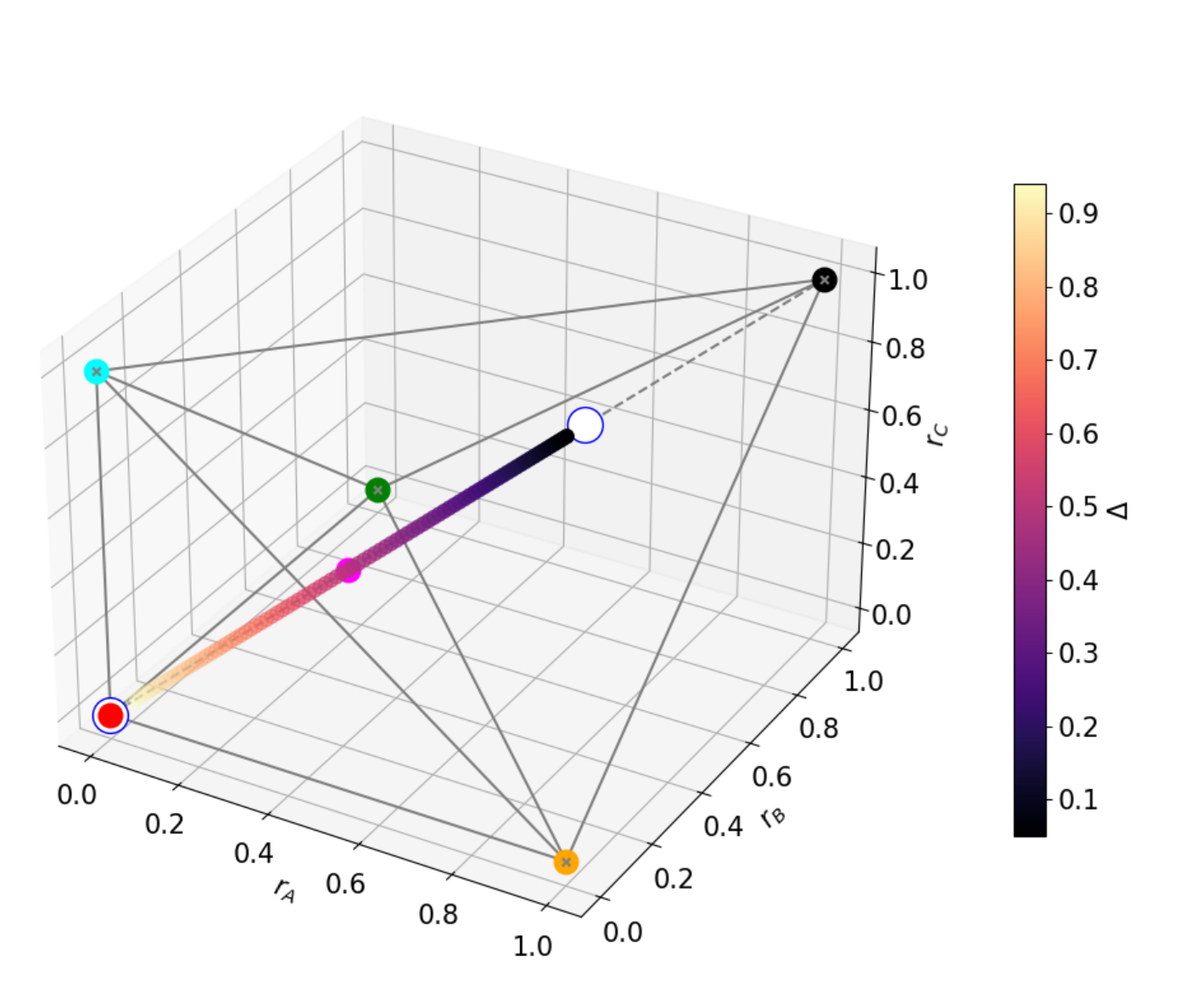}\label{chain_TFIM_trajectory_n2_pre}} \\
	\subfloat[$\ket{n=2}$ for the TFIM in $\Delta\in(1,4)$]{\includegraphics[trim=18cm 2cm 7cm 2cm, clip, scale=0.23]{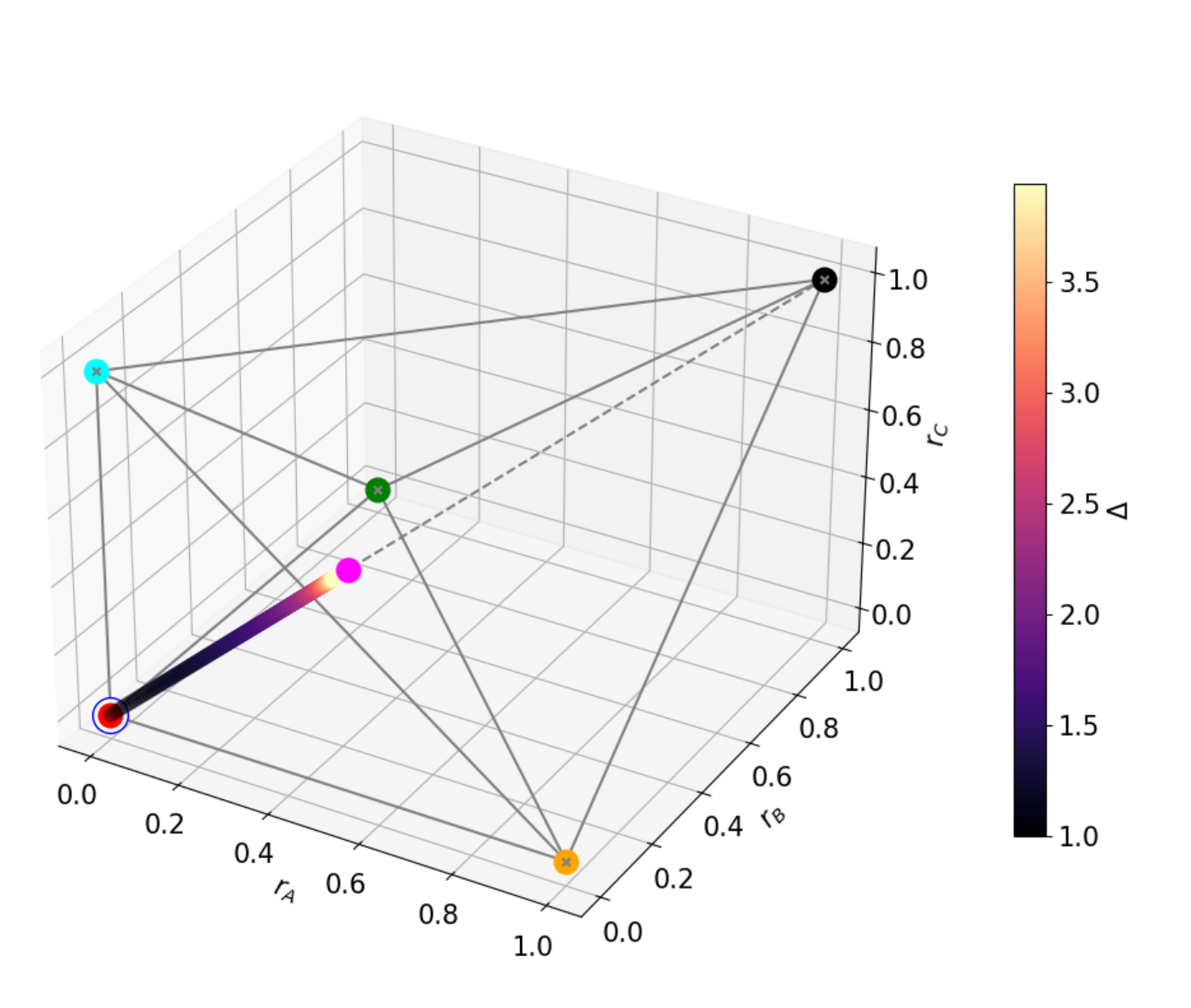}\label{chain_TFIM_trajectory_n2_post}} \\
	\caption{Trajectory for TFIM $\ket{n=2}$.}
\end{figure}

\multiComment{
\begin{figure}[hbpt!]
	\centering
	\subfloat[$n=0$ subspace at $\Delta=0$.]{\includegraphics[trim=15cm 2cm 15cm 2cm, clip, scale=0.23]{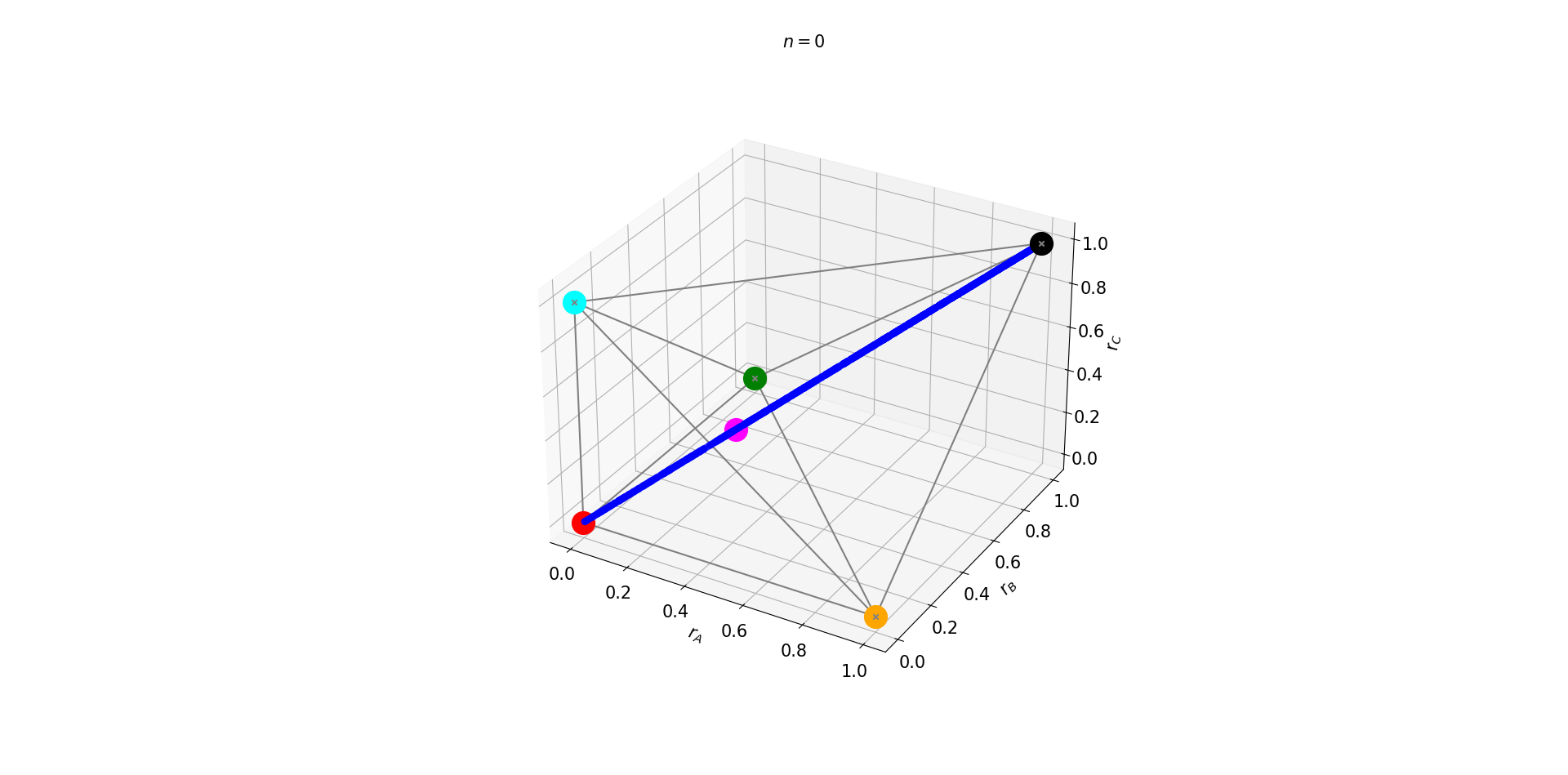}\label{chain_TFIM_trajectory_n0_delta0}} \\
	\subfloat[$n=2$ subspace at $\Delta=1$.]{\includegraphics[trim=15cm 2cm 15cm 2cm, clip, scale=0.23]{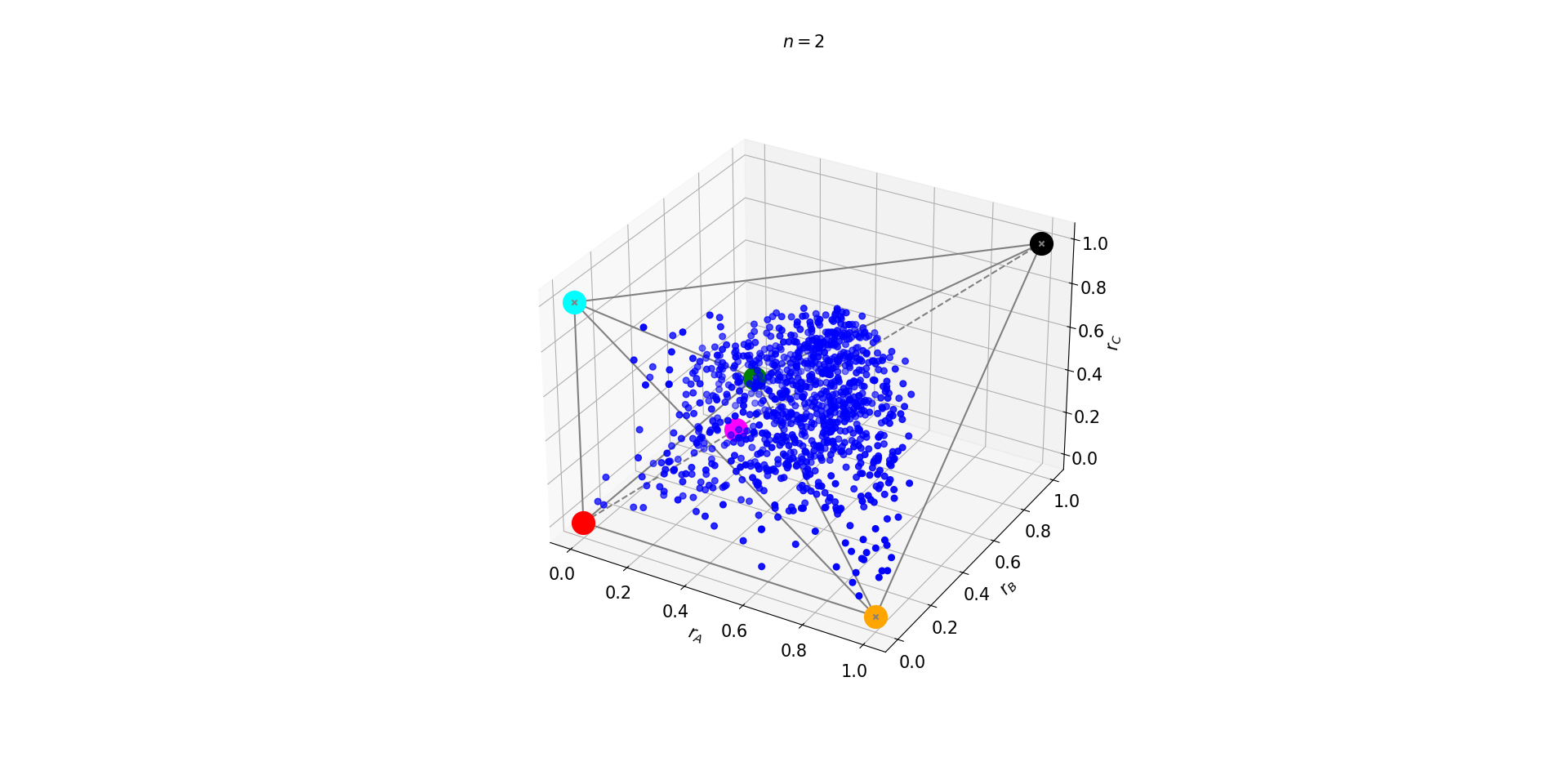}\label{chain_TFIM_trajectory_n2_delta1}} \\
	\caption{TFIM $n=0 \text{ } \& \text{ } 2$ at $\Delta=0$ and $1$ respectively.}
\end{figure}
}

\begin{figure}[hbpt!]
	\centering
	\includegraphics[trim=15cm 2cm 15cm 2cm, clip, scale=0.23]{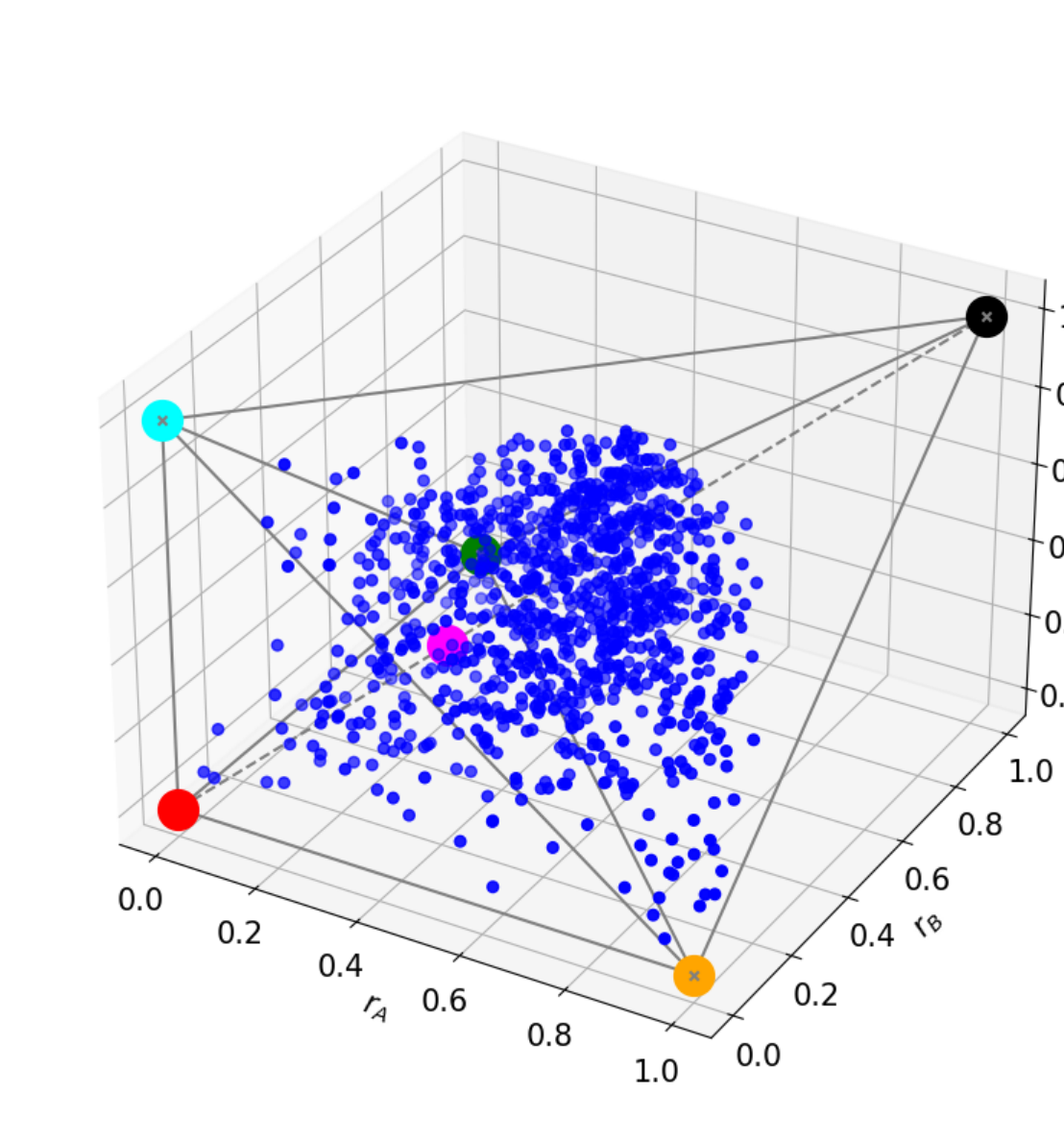}
    \caption{$n=2$ subspace at $\Delta=1$.}
    \label{chain_TFIM_trajectory_n2_delta1}
\end{figure}

We now turn to the level crossing points, specifically to $\Delta=1$ where subspaces $n=2$ and $n=3$ fuse, corresponding to a degeneracy of $m=3$ \eqref{TFIM_eigenState_n2_3_delta1} (see the Bloch-norms in Fig. \ref{chain_TFIM_trajectory_n2_delta1}, which now span a 3-dimensional manifold). The increase in degeneracy allows for new superpositions, changing the tangle of the energy level. This is in general the only observable change when considering level crossings: change of the tangle due to increase of degeneracy. It also makes it less likely that the Bloch-norms will maintain any geometrical pattern because superpositions the new superpositions might generate a state which is no longer translation-invariant.\newline

Consider the XX chain with a magnetic field $\Delta$:
\begin{eqnarray}
H_{\text{XX}} &= -\displaystyle\sum_{j=0}^{2} \left( X_j X_{j+1} + Y_j Y_{j+1} + \Delta Z_j \right) \quad\quad
\label{XX_hamiltonian}
\end{eqnarray}
\noindent where $\Delta \ge 0$, with exact energy spectrum \eqref{XX_energy_spectrum} shown in Fig. \ref{chain_XX_spectrum} (details in Appendix \ref{section_DetailsChains_XX}). The Bloch-norms of the non-degenerate levels are all either $1/3 \cdot (1,1,1)$ or $(1,1,1)$ \eqref{XX_eigenState_spectrum_nonDeg_BlochNorms}, while for both degenerate levels the shapes are the same as in Fig. \ref{chain_TFIM_trajectory_n3}.\newline

\begin{figure}
	\centering
	\includegraphics[trim=4cm 0cm 4cm 0, clip, scale=0.2]{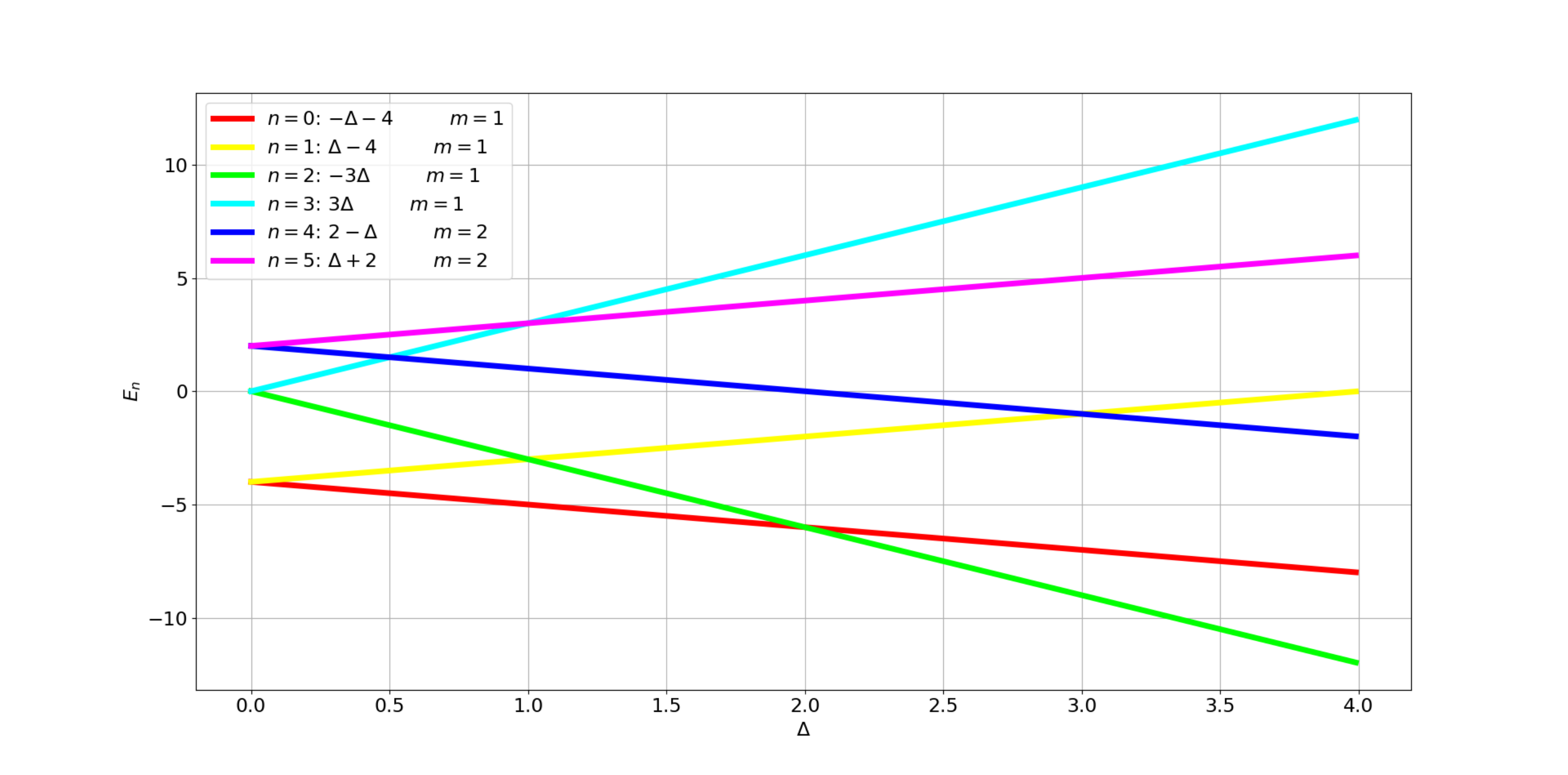}
	\caption{$H_{XX}$ energy spectrum.}
	\label{chain_XX_spectrum}
\end{figure}

\textcolor{black}{Notice that, while in the TFIM the tangle could be intuitively understood as coming from the competition between the 2-qubit and 1-qubit terms ($XX$ vs. $Z$ mediated by $\Delta$), in the XX chain this is no longer the case. This is because the $(XX+YY)$ terms commute with the $Z$ term, producing linear dependence with $\Delta$ of the energies and the independence of the eigenstates (and hence the tangle) from $\Delta$.}

\begin{figure}
	\centering
	\includegraphics[trim=3cm 0cm 4cm 0, clip, scale=0.19]{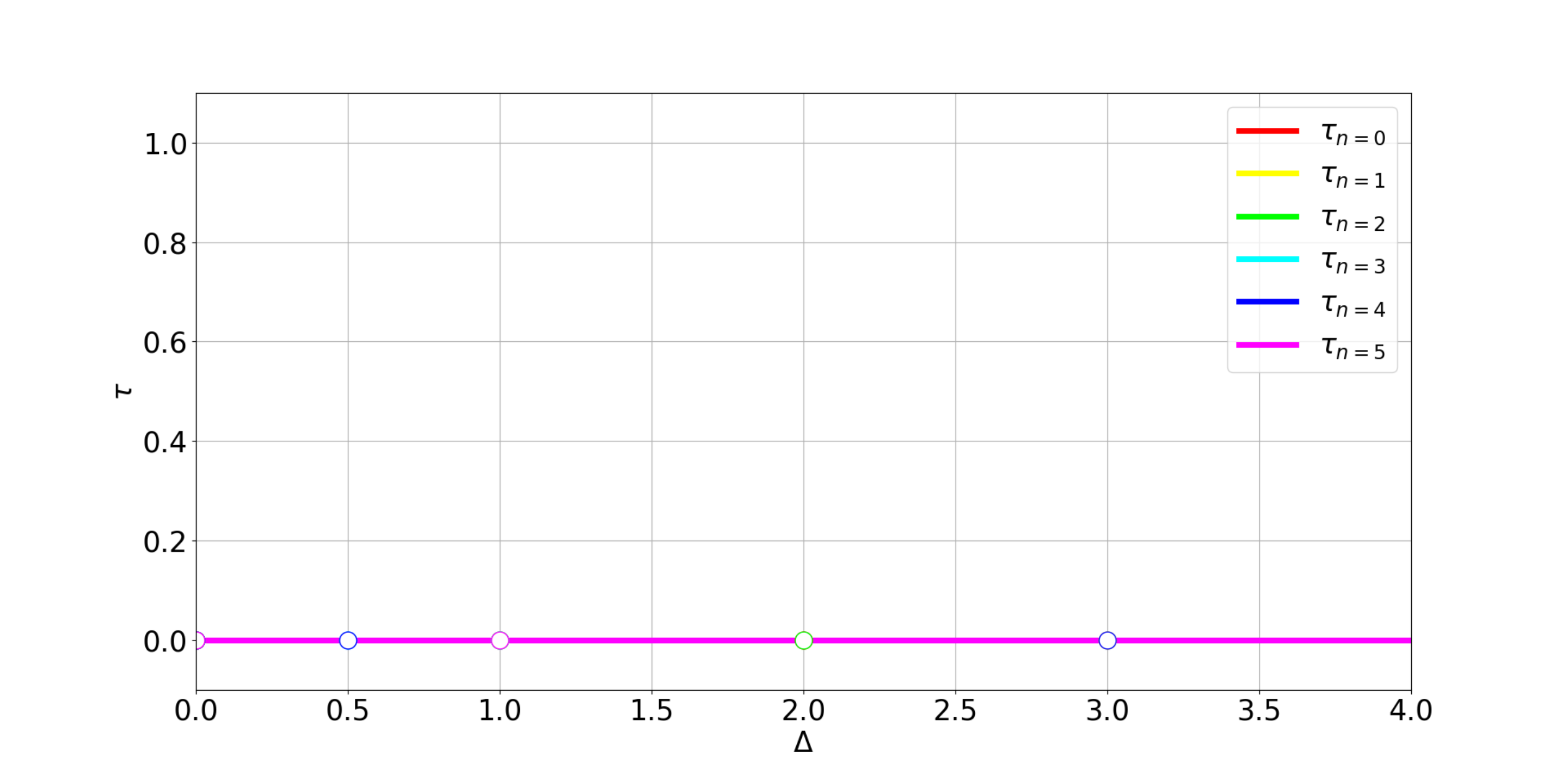}
    \caption{Tangle of XX levels\eqref{XX_eigenState_spectrum_yesDeg}} 
	\label{chain_XX_tangles_plot_1}
\end{figure}


\textcolor{black}{Furthermore, observe that in the XX chain the tangle of the energy levels is more \textit{fragile} to perturbations of the Hamiltonian than in the TFIM. The argument is as follows: adding a small perturbation (with parameter $\xi$) that breaks some symmetry will split the degenerate levels and all the tangle (which was due exclusively to these degenerate superpositions) will disappear. On the other hand, the tangle for the TFIM states is much more robust under the same procedure, since it does not come from degenerate superpositions. The same argument explains why the tangle generated at level crossings is also fragile: when introducing a small perturbation, the level crossings will generically disappear due to \textit{level repulsion} \cite{Haake2001_book, experimental_level_repulsion}\footnote{The phenomenon of level repulsion is known to arise even in classical systems \cite{classical_level_repulsion, classical_level_repulsion_2, classical_level_repulsion_3}}.}\newline



Consider the XXX chain, with Hamiltonian:
\begin{eqnarray}
H_{\text{XXX}} =& \displaystyle\sum_{j=0}^{2} \left( X_j X_{j+1} + Y_j Y_{j+1} +\Delta Z_j Z_{j+1} \right) \nonumber \\
\label{XXX_hamiltonian}
\end{eqnarray}
\begin{figure}
	\centering
	\includegraphics[trim=4cm 0cm 4cm 0, clip, scale=0.20]{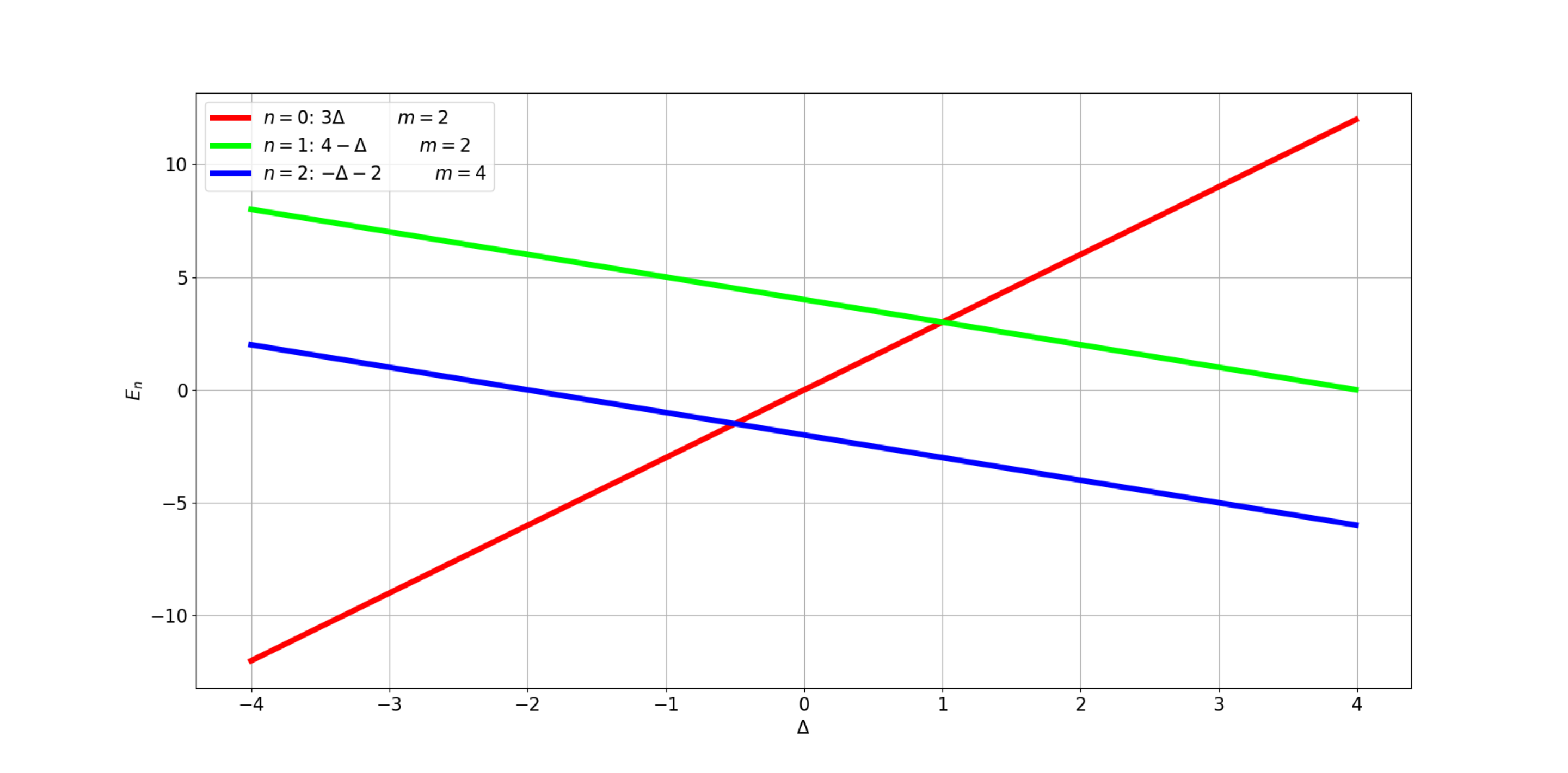}
	\caption{$H_{XXX}$ energy spectrum.}
	\label{chain_XXX_spectrum}
\end{figure}
\noindent where $\Delta\in\mathbb{R} $ and energy levels shown in Fig. \ref{chain_XXX_spectrum}. The key difference between the XX and XXX chains is the increase in degeneracies: all levels are degenerate (see \eqref{XXX_eigenState_spectrum_all} in Appendix \ref{section_DetailsChains_XXX}). We can find in general non-vanishing tangle:
\begin{eqnarray}
\tau_{n=0} &=& 4 \beta^{2} \left(1 - \beta^{2}\right) \nonumber \\
\tau_{n=1} &=& \frac{4}{3} \beta^2 \left( 1-\beta^2 \right)  \nonumber \\
\tau_{n=2} &=& \frac{4}{9} \Big| \left( S_{\alpha\beta}S_{\gamma\delta} - \left[ \eta_{\beta\alpha} \eta_{\delta\gamma} - \eta_{\alpha\beta} \eta_{\gamma\delta} \right] \right)^2 \nonumber \\
& & -4S_{\alpha\beta}S_{\gamma\delta}\eta_{\alpha\beta}\eta_{\gamma\delta} \Big|
\label{XXX_tangles_nonCrosses}
\end{eqnarray}
\noindent where $S_{\alpha\beta} = (\alpha+\beta)$ and $\eta_{\alpha\beta} =\alpha \exp{\left(i2\pi / 3 \right)} + \beta \exp{\left(-i2\pi / 3 \right)}$ and $\eta_{\beta\alpha} =\alpha \exp{\left(-i2\pi / 3 \right)} + \beta \exp{ \left( i2\pi /3 \right)}$. Notice that, just as in the XX chain, the tangle here will be fragile under any small perturbation of the Hamiltonian.\newline

\begin{figure}
	\centering
	\includegraphics[trim=3cm 0cm 4cm 0, clip, scale=0.20]{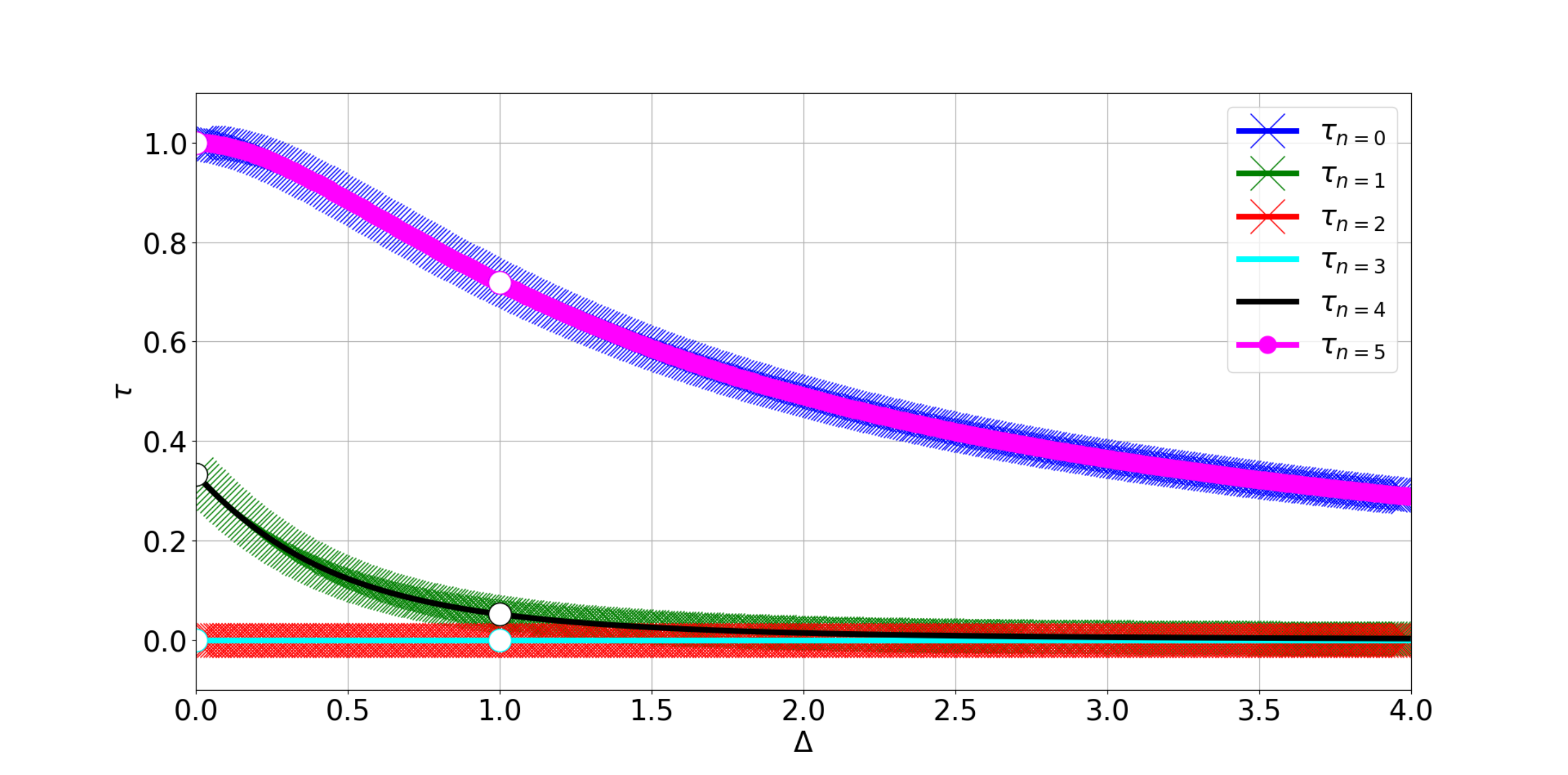}
	\caption{Tangle $H_{XZX}$ levels \eqref{XZX_eigenState_spectrum_nonDeg} \eqref{XZX_eigenState_spectrum_deg}.}
	\label{chain_XZX_tangles_plot_1}
\end{figure}

\begin{figure}[hbpt!]
	\centering
	\subfloat[$\ket{n=0}$]{\includegraphics[trim=18cm 2cm 7cm 2cm, clip, scale=0.23]{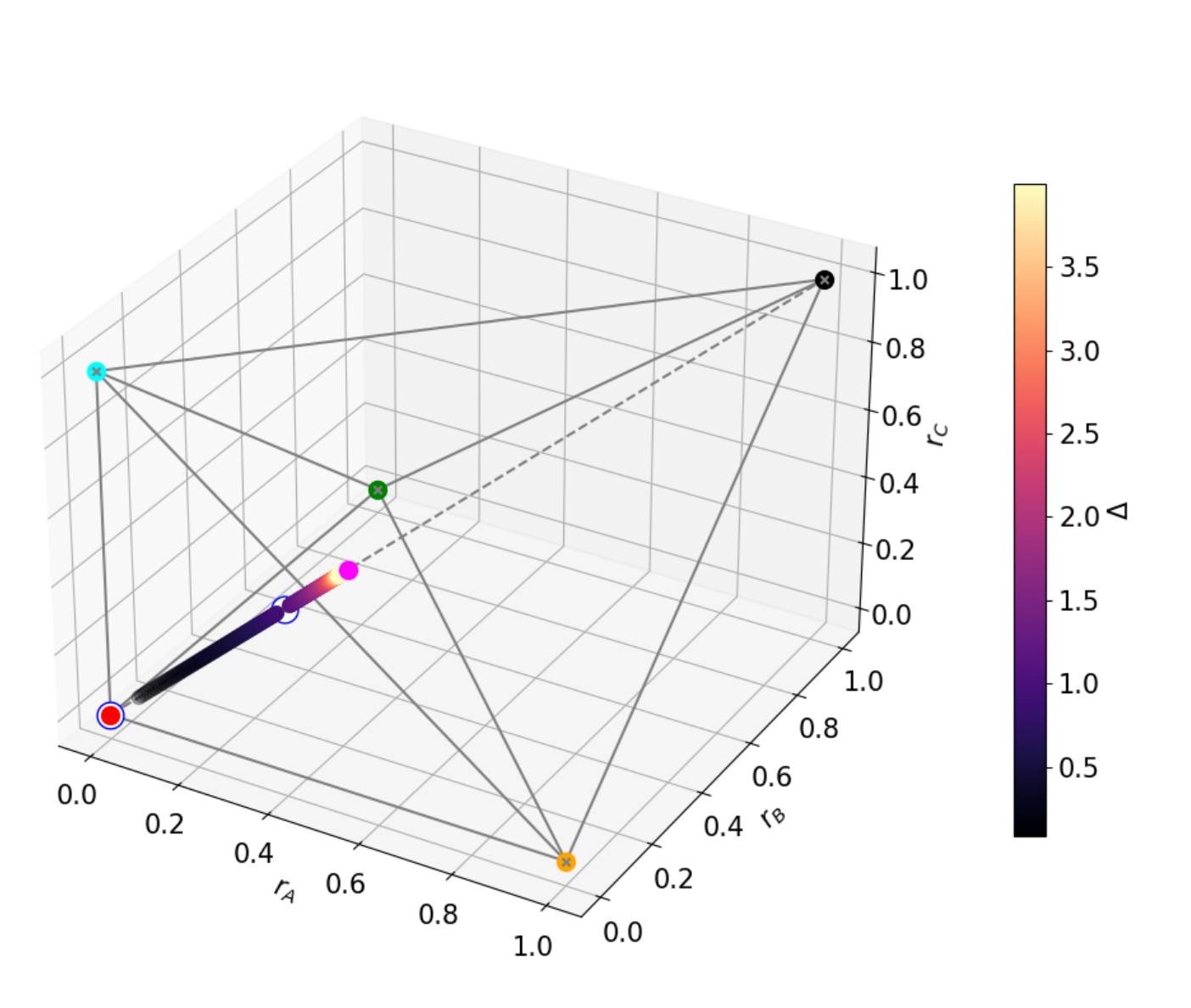}\label{chain_XZX_trajectory_n0}} \\
	\subfloat[$\ket{n=1}$]{\includegraphics[trim=18cm 2cm 7cm 2cm, clip, scale=0.23]{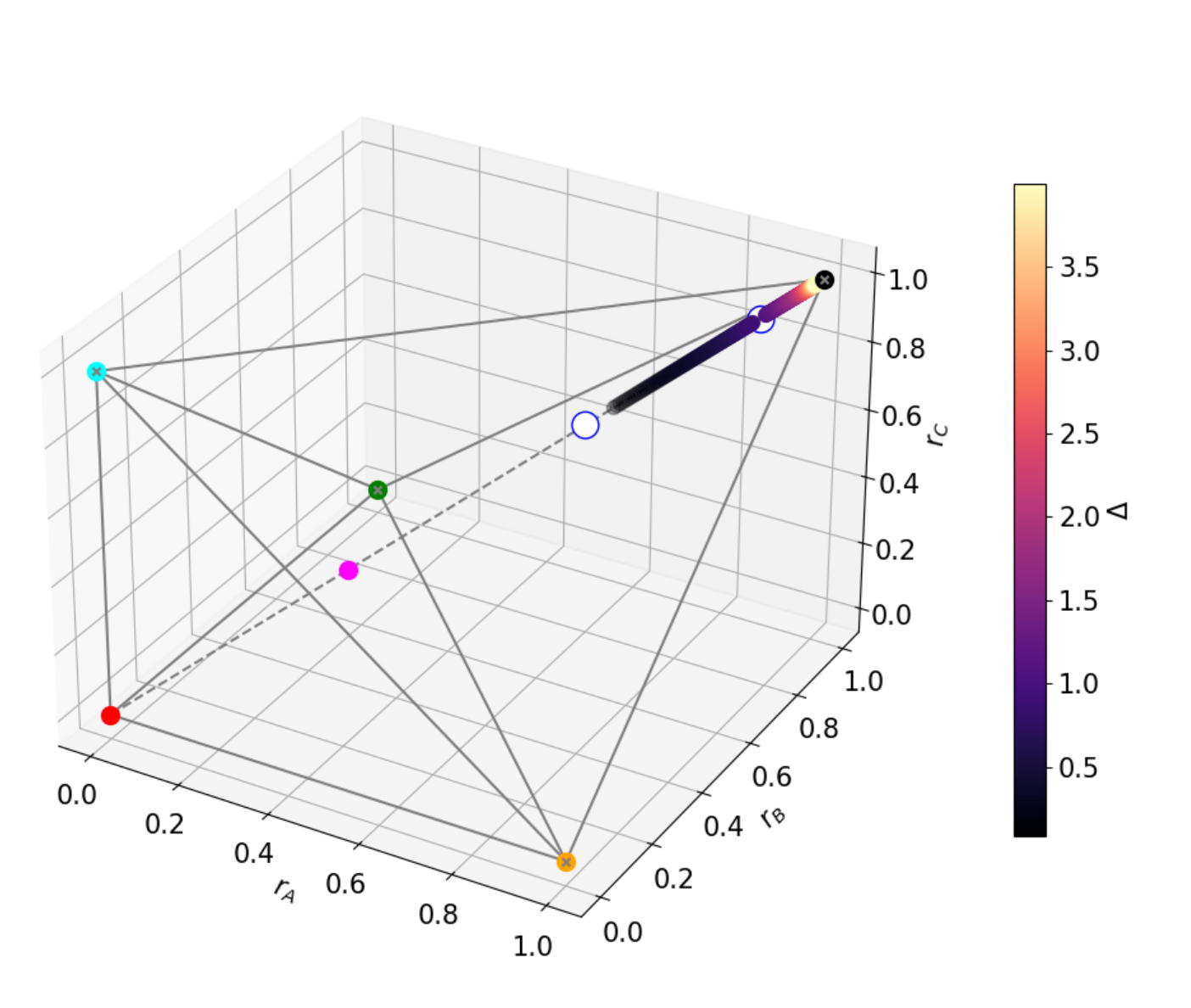}\label{chain_XZX_trajectory_n1}} \\
	\subfloat[$\ket{n=4}$]{\includegraphics[trim=18cm 2cm 7cm 2cm, clip, scale=0.23]{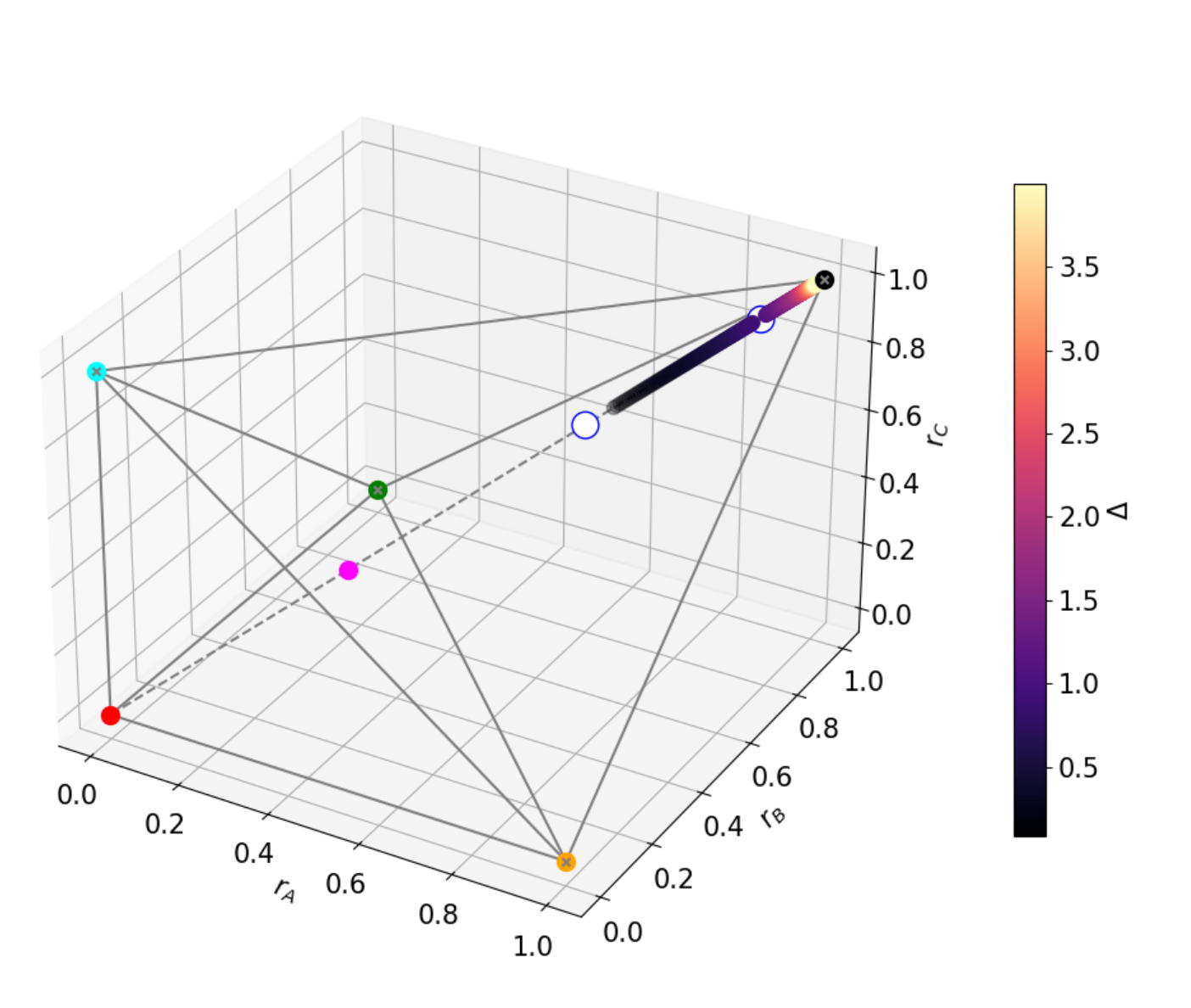}\label{chain_XZX_trajectory_n4}} \\
        \subfloat[$\ket{n=5}$]{\includegraphics[trim=18cm 2cm 7cm 2cm, clip, scale=0.23]{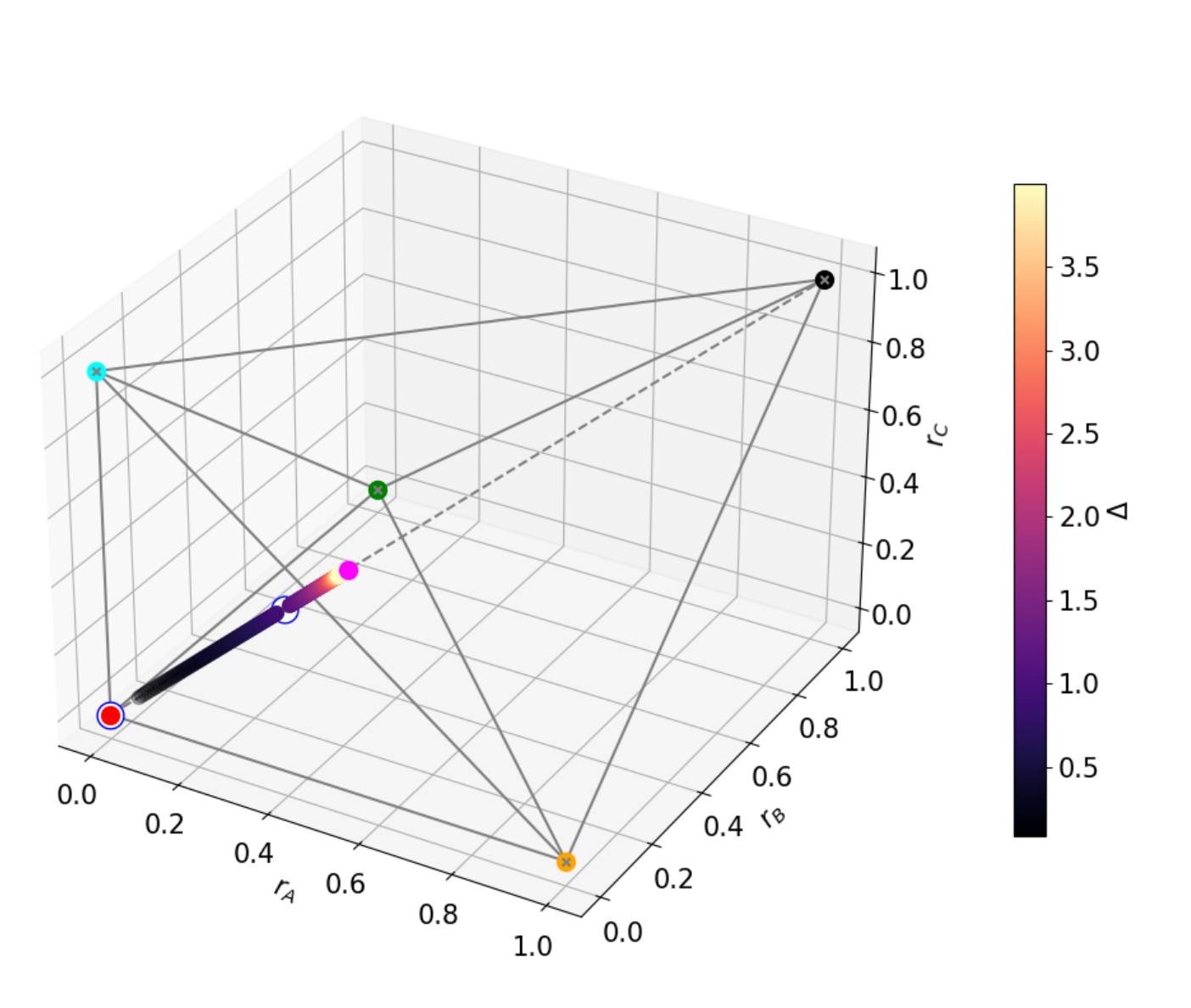}\label{chain_XZX_trajectory_n5}} \\
	\caption{XZX trajectories for$\Delta\in(0,1)\cup(1,4]$.}
    \label{XZX__non_deg_Trajectories}
\end{figure}

Finally, look at the XZX spin chain. This chain's Hamiltonian contains 3-body terms\footnote{Which are none other than the operators $K_j$ defining the cluster state \cite{cluster_state_paper} for a closed chain of 3 qubits: \begin{equation} \begin{matrix}K_j \ket{\phi_{\{\kappa\}}} = (-1)^{\kappa_j} \ket{\phi_{\{\kappa\}}} \\ \text{ where } K_j = Z_j \displaystyle\otimes_{l\in\text{nn}(j)} X_l \end{matrix} \label{cluster_state_definition}\end{equation}}  $X_{j} Z_{j+1} X_{j+2}$ competing against one-body terms $Z_j$:
\begin{equation}
	\begin{matrix} H_{\text{XZX}} = -\displaystyle\sum_{j=0}^{2} \left( X_j Z_{j+1} X_{j+2} 
 + \Delta Z_j \right)\end{matrix}
	\label{XZX_hamiltonian}
\end{equation}
\noindent where $\Delta\ge0$, with energy levels shown in Fig. \ref{chain_XZX_spectrum} (see \eqref{XZX_energy_spectrum} in Appendix \ref{section_DetailsChains_XZX}). Contrary to the XX and XXX models, the XZX chain presents a tangle \textit{robust} against small perturbations of its Hamiltonian:
\begin{eqnarray}
\tau_{n=0} &= & \frac{48 f_0^3}{g_0^4} = \tau_{n=5} = \frac{16 f_5}{g_5^4} \nonumber \\
\tau_{n=1} &= &\frac{48 f_0}{g_0^2} = \tau_{n=4} = \frac{48 f_4^3}{g_4^4} \nonumber \\
\tau_{n=2} &= &  \tau_{n=3} = 0 \nonumber \\
\label{XZX_tangles_1}
\end{eqnarray}
\noindent where $f_j,g_j$ are functions of $\Delta$ defined in \eqref{XZX_eigenState_parameters}. The $(\Delta, \tau)$ plot is shown in Fig. \ref{chain_XZX_tangles_plot_1}.  The Bloch-norm representation of the non-degenerate subspaces again forms a trajectory in the polytope restricted to subsets of the main diagonal (see Fig. \ref{XZX__non_deg_Trajectories}) and the shape of the degenerate subspaces is the same as the one shown in Fig. \ref{chain_TFIM_trajectory_n3}.
\begin{figure}
	\centering
	\includegraphics[trim=4cm 0cm 4cm 0, clip, scale=0.21]{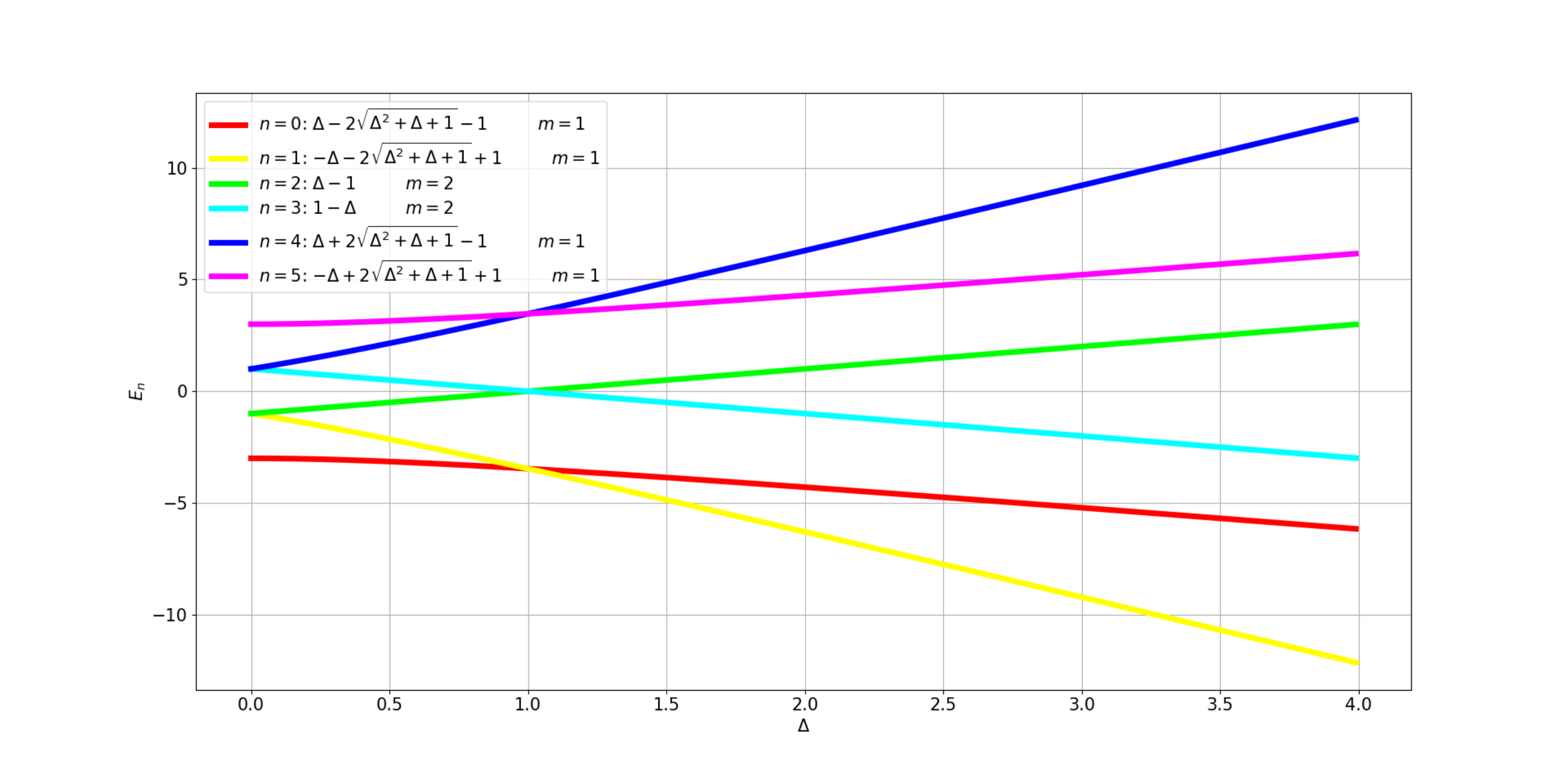}
	\caption{$H_{XZX}$ energy spectrum.}
	\label{chain_XZX_spectrum}
\end{figure}
\textcolor{black}{Furthermore, we can see the following emerging patterns:}
\begin{enumerate}
    \item \textcolor{black}{Eigenstates with \textit{robust} tangle $\in \vec{V}_{\text{line}}$. This is because translation invariance causes all 3 Bloch-norms to be equal. This also explains why the instances of Bloch-norms that are not in the main diagonal correspond to degenerate subspaces. Hence, the states with robust tangle belong to the GHZ class type 5 subset spanned by simultaneous eigenstates of the translation and parity operators.}
    \item \textcolor{black}{Out-of-level-crossing degenerate levels in $H_{TFIM}$ and $H_{XZX}$ have null tangle. This is because, when projected onto those subspaces, the \textit{kinetic} piece of the Hamiltonian will commute with the \textit{potential} term $\left[ \mathcal{P}_n H_0 \mathcal{P}_n, V \right]=0$, causing the tangle to take the same constant value $\forall \Delta$. Since for $\Delta \to \infty$ the resulting Hamiltonian cannot generate tangle (since it is a collection of 1-qubit operators), then necessarily $\tau=0$.}
\end{enumerate}


\section{Conclusions}\label{section_Conclusions}
In this work, we investigated the geometrical properties of genuine tripartite entanglement in 3-qubit states. We derived bounds for the tangle based solely on geometrical insights and arrived at a purely geometrical ansatz for Cayley's hyperdeterminant of non-generic GHZ states. We then explored how many of these states show up naturally in the energy eigenstate structure of usual spin-chain Hamiltonians. Due to translation and parity symmetries, only a very small subset of GHZ type 5 will ever appear in states that present \textit{robust} tangle. Moreover, we identified the necessary conditions for the tangle to be \textit{robust} and not disappear under realistic effects such as symmetry-breaking perturbations and level-repulsion.

A possible future direction for this work is the extension of these geometrical arguments to four-qubit systems, where entanglement classification schemes are already known \cite{verstraete_4qubits_classes, 4qubits_graphs}. However, this task becomes increasingly difficult as the number of qubits grows due to the calculation of the hyperdeterminant \cite{4_tangle_spinChains}. For 3-qubit states the canonical decomposition circumvented this issue. There are several proposals on how to expand the concept of a canonical decomposition for 4 qubits states \cite{Acin_4qubits} as well as for general $n$-qubit states \cite{n_qubits_CD}, however it is not yet clear if these simplify the computation in any way \cite{Acin_4qubits}.

\begin{acknowledgments}
We would like to thank Adrián Pérez-Salinas for insightful conversations and feedback. We acknowledge financial support from the Spanish MINECO grant PID2021-127726NB-I00, the CSIC Research Platform on Quantum Technologies PTI-001, and the QUANTUM ENIA project Quantum Spain funded through the RTRP-Next Generation program under the framework of the Digital Spain 2026 Agenda.
A.B. aknowledges support from the Spanish Ministry of Science, Innovation and Universities grant PRE2022-I0I93I.
G.S. also acknowledges partial support from NSF grant PHY-2309135 to the Kavli Institute for Theoretical Physics (KITP), as well as joint sponsorship from the Fulbright Program and the Spanish Ministry of Science, Innovation and Universities.

\end{acknowledgments}



\bibliographystyle{plain}  
\bibliography{main}  

\newpage
\appendix

\newpage
\onecolumngrid 

\section{Proofs of the geometrical properties of the entanglement types}\label{section_PolytopeProofs}

We now provide the aforementioned proofs for the polytope structure. Start by proving that all type 3a states lie in the faces of the upper tetrahedron: the CD of a state of such type is: $\type{3a} = \lambda_0 \ket{000} + \lambda_2 \ket{101} + \lambda_3 \ket{110}$. From it, we compute the Bloch vectors of each qubit by using \eqref{1_qubit_density_matrix_canonical_form}. We obtain that only the $z$ components are non-zero. Combining them with the normalization condition yields $z_A - z_B - z_C +1=0$ which is just the scalar equation of a plane. Introducing $s_{I} = \sign{z_{I}}$, $I\in\{A,B,C\}$:

\begin{equation}
	\begin{matrix} s_A r_A - s_B r_B -s_C r_C +1=0; & \text{ where } s_{\overline{I}}\in\{\pm 1\} \text{ and } \genNorm \in [0,1] \end{matrix}
	\label{type_3a_plane_parametric_equation_in_rs}
\end{equation}

\noindent which encodes 8 different plane equations (one for each different combination of signs). However, some of them will give a plane which is not present in the $[0,1]^3$ cube. These planes are redundant since they can be mapped to the correct ones (the ones present in the cube) via simple translations of the constant $+1$ in \eqref{type_3a_plane_parametric_equation_in_rs} to $-1$. The 4 configurations of signs for each face in the upper tetrahedron $\left( s_A, s_B, s_C \right)$ are:
\begin{table}[h!]
	\centering
	\begin{tabular}{c | c}
		\hline
		\hline
		$\left( s_A, s_B, s_C \right)$ & Vertices of the face \\
		\hline
		$(+1,+1,+1)$ & $(0,0,1),(1,1,1),(0,1,0)$ \\
		\hline
		$(-1,-1,+1)$ & $(0,0,1),(1,1,1),(1,0,0)$ \\
		\hline
		$(-1,+1,-1)$ & $(1,1,1),(1,0,0),(0,1,0)$ \\
		\hline
		$(-1,+1,+1)$ & $(0,0,1),(0,1,0), (1,0,0)$ \\
		\hline
		\hline
	\end{tabular}
	\caption{Values of the signs for each face of the upper tetrahedron.}
	\label{signs_for_upper_tetra_faces}
\end{table}
\noindent which completes the proof. Now, to show that type 2b states lie exclusively in the central diagonal $\vec{V}_{\text{line}}$, we just need a direct computation:
\begin{equation}
	z_A = z_B = z_C = \left(2\lambda_0^2 -1\right); \text{ } x_{\overline{I}} = y_{\overline{I}} = 0; \implies  r_A = r_B = r_C
	\label{z_components_type_2b}
\end{equation}
We now prove that type 3b states span the internal triangles of vertices $\{(0,0,0),(1,1,1),(0,0,1)\}$, $\{(0,0,0),(1,1,1),(1,0,0)\}$ and $\{(0,0,0),(1,1,1),(0,1,0)\}$ for kinds \textit{1-2}, \textit{2-3} and \textit{1-3} respectively. Start with  a state of the kind \textit{1-2}: $\type{3b-12} = \lambda_0 \ket{000} + \lambda_3 \ket{110} + \lambda_4 \ket{111}$ and compute the Bloch-vectors. From those, observe that $r_A=r_B$, so the point $(r_A, r_B, r_C)$ must lie in that plane. By using the normalization condition on the $\lambda$ parameters, one can check that:
\begin{equation}
	r_C = \sqrt{1-\tau} = \sqrt{ r_A^2 + 2\lambda_3^2 \left( 1+ s_A r_A\right) } > r_A
	\label{c_bloch_norm_type_3b_12}
\end{equation}
The relation in \eqref{c_bloch_norm_type_3b_12} cannot saturate to equality since that would require $\lambda_3=0$ redirecting us back to type 2b. We end up with the allowed region of values $r_C>r_A=r_B$, which is just the triangle of vertices $\{(1,1,1),(0,0,0),(0,0,1)\}$ (where the edge corresponding to the main diagonal $(0,0,0)-(1,1,1)$ is forbidden). This completes the proof for \textit{1-2}. For \textit{2-3}, $r_A$ and $r_C$ exchange roles resulting in the triangle of edges $\{(1,1,1),(0,0,0),(1,0,0)\}$ from $r_A>r_B=r_C$ and for kind \textit{1-3} $r_C$ and $r_B$ exchange roles so $r_B>r_A=r_C$ which gives the triangle with vertices $\{(1,1,1),(0,0,0),(0,1,0)\}$ (in all of these the central diagonal $(0,0,0)-(1,1,1)$ is forbidden). In summary:
\begin{equation}
	\begin{matrix} \type{3b-12} \to & r_C>r_A=r_B;   &   \type{3b-23} \to & r_A>r_C=r_B;  &   \type{3b-13} \to & r_B>r_A=r_C;  \end{matrix}
	\label{summary_type_3b}
\end{equation}
Finally, we show now where type 4b states lie. Start with with $\lambda_2=0$: $\type{4b,2} = \lambda_0 \ket{000} +\lambda_1 e^{i \varphi} \ket{110} + \lambda_3 \ket{110} + \lambda_4 \ket{111}$ and notice that it is just a normalized sum of a $\type{3b-12}$ state and a $\type{3b-23}$ state which one would naively expect to lie somewhere between both those states. To put this into more solid grounds, compute the Bloch-norms:
\begin{equation}
\begin{matrix}
    r_A^2 = \left(1-\tau \right) -4\lambda_0^2\lambda_3^2; & \quad r_B^2 = r_A^2 - 4\lambda_1^2 \lambda_4^2; & \quad r_C^2 & = \left(1-\tau \right) -4\lambda_1^2\lambda_4^2 
\end{matrix}
\label{r2_components_type_4b_2}
\end{equation}
\noindent where we have substituted $4\lambda_2^2 \lambda_4^2$ by $\tau$ as per the \textit{tangle master equation} \eqref{tangle_master_eq}. From these, we can obtain:
\begin{equation}
	\begin{matrix} r_B^2 < r_A^2 < \left( 1-\tau \right) \to r_B < r_A < \sqrt{1-\tau} \quad \\ r_B^2 < r_C^2 < \left( 1-\tau \right) \to r_B < r_C < \sqrt{1-\tau} \end{matrix}
	\label{bound_1_type_4b_2}
\end{equation}
\noindent where in the $\to$ step we have used that the square root is monotonous in the $[0,1]$ interval. Comparing with the cases \textit{1-2} and \textit{2-3} on \eqref{summary_type_3b}, we see that indeed this implies that the $\type{4b,2}$ states live in the space in between those two planes (with the planes themselves being forbidden). The same procedure can be repeated for the case in which $\lambda_3=0$ instead and one finds that the allowed zone is in between the planes of kinds \textit{2-3} and \textit{1-3}. This completes the proof.

\section{$(R,\tau)$ Surface fibration by $\vec{\lambda}$ curves}\label{section_DetailsBounds}

We illustrate for this one the general method one can follow to reproduce our results for the bounds presented in Sec. \ref{section_Tangle}. We start by looking at the case $\vec{\lambda} = \left( 0,0, \lambda_3\right)$. Substituting into \eqref{R_norm_master_equation} and then using the normalization condition to remove $\lambda_4$ gives:

\begin{equation}
	R^{2} = 12 \lambda_{0}^{4} + 4 \lambda_{0}^{2} \lambda_{3}^{2} - 12 \lambda_{0}^{2} + 3 \to \left( \lambda_0^2 \right)_{\pm} =  \frac{1}{2} - \frac{\lambda_{3}^{2}}{6} \pm \frac{\sqrt{3 R^{2} + \lambda_{3}^{4} - 6 \lambda_{3}^{2}}}{6}
	\label{begin_second_case_R_tau_exploration}
\end{equation}

Substituting $\left( \lambda_0^2 \right)_{\pm}$ into eq. \eqref{tangle_master_eq} will yield two solutions. The correct one:
\begin{equation}
    \tau \left( R, \lambda_3 \right)  = \upperTau  + \frac{4 \lambda_{3}^{2}}{9} \cdot \left( \lambda_{3}^{2} - 3 - \sqrt{3 R^{2} + \lambda_{3}^{2} \left( \lambda_{3}^{2} - 6\right) } \right)
    \label{TauRL3WannaRef}
\end{equation}
\noindent is the one that satisfies the consistency condition:
\begin{equation}
    \displaystyle\min_R \left\{ \displaystyle\argmin_{(R,\lambda_3)}\left[ \tau \left(R, \lambda_3 \right) \right] \right\} = 1
    \label{tau_0_R_min_is_1}
\end{equation}

Our next step will be to attempt to \textit{fibrate} the surface \eqref{TauRL3WannaRef} with curves $\lambda_3=\lambda_3(R)$ and then find the one curve that allows us to minimize the value of $\tau$ for a given value of $R$. Imposing now the reality condition $\tau\in\mathbb{R}$ yields the inequality:

\begin{equation}
	3 R^{2} + \lambda_{3}^{4} - 6 \lambda_{3}^{2} \ge 0 \text{ that saturates for: } \left( \lambda_{3,\text{sat}}^2 (R)\right)_{\pm} = 3 \pm \sqrt{9 - 3 R^{2}}
	\label{second_case_R_tau_exploration__04}
\end{equation}

We can now use a fact we learned from the analysis of the Bloch-norm diagram from the previous section: \textit{as $R\to 0$, the 3-qubit state $\to \GHZ$}. Imposing this constraint is equivalent to imposing $\lim_{R\to 0}\lambda_3(R) = 0 $, and it reveals that the correct solution is necessarily the $(-)$ one. With this, the $\tau$-reality condition is also telling us that $\lambda_3 \le \sqrt{ \left( \lambda_{3,\text{sat}}^2 \right)_{-}}$. Substituting the correct solution back into \eqref{TauRL3WannaRef} we arrive at:

\begin{equation}
	\begin{matrix} \tau \left( R, \lambda_3 = \sqrt{\left( \lambda_{3,\text{sat}}^2 (R)\right)_{-}} \right)   = 5 \upperTau - 4\sqrt{\upperTau}   \end{matrix}
	\label{second_case_R_tau_exploration__06}
\end{equation}

There exist values of $R$ for which $\left( \lambda_{3,\text{sat}}^2 (R)\right)_{-}$ will be greater than 1, which is forbidden since by construction $0<\lambda_3 < 1$. What this is telling us is that the solution has only a certain range of values of $R$ for which it is valid. This range is $R\le R_{o} := \sqrt{5/3} < \sqrt{3}$. Moreover, it does not fit the requirement observed in figures \ref{different_R_tau_areas__b}, \ref{different_R_tau_areas__c} that it makes the tangle vanish at $R=1$, meaning that this is not the curve we are looking for. Instead, we want a curve $\lambda_3^{(\star)}(R)$ such that $\tau_{-}\left( R, \lambda_3^{(\star)}(R) \right) = \middleTau$ which must fulfill the requirement $\tau_{-}\left( R=1, \lambda_3^{(\star)}(R=1) \right) =0 $. We can solve analytically for this, finding $\lambda_3^{(\star)}(R=1)=1/\sqrt{2} $. Moreover, in the small $R$ limit the curve that maximizes $\lambda_3$ will be the one minimizing $\tau$ so $\lambda_3^{(\star)}(R)$ must fulfill:
\begin{equation}
\begin{matrix}
    \displaystyle\lim_{R\to 0} \lambda_3^{(\star)}(R)   \simeq \sqrt{\left( \lambda_{3,\text{sat}}^2 (R) \right)_{-}} = \sqrt{3-\sqrt{9-3R^2}} & \quad\quad 
 \displaystyle\lim_{R\to 1} \lambda_3^{(\star)}(R)  \simeq \lambda_3^{(a)}(R) = \frac{R}{\sqrt{2}}
\end{matrix}
\label{second_case_R_tau_exploration__07}
\end{equation}
One can check numerically that the change from one curve to the other becomes appreciable around $R\simeq 0.56$. So, we finally arive at our approximated solution for the $\lambda_3^{(\star)}(R)$ curve such that $\tau \left( R, \lambda_3=\lambda_3^{(\star)}(R) \right) = \middleTau$: 

\begin{equation}
	\lambda_3^{(\star)}(R)  \simeq \begin{Bmatrix} \sqrt{3-\sqrt{9-3R^2}} & \text{ if } R\overset{<}{\sim} 0.56 \\  R/\sqrt{2}  & \text{ otherwise } \end{Bmatrix}
	\label{second_case_R_tau_exploration__07_01__aligned}
\end{equation}
\noindent which when substituted back into \eqref{TauRL3WannaRef} will yield \eqref{second_case_R_tau_exploration__07__aligned__short}.\newline

For the case of type 4c states the same procedure can be employed. The only difference being that in this case is that we have two curves to worry about, one for $\lambda_2$ and another for $\lambda_3$:

\multiComment{
\begin{widetext}
\begin{equation}
	\begin{matrix} \tau\left( R, \lambda_2, \lambda_3 \right)  = \upperTau - \frac{16 \lambda_2^2 \lambda_3^2}{9} + \frac{4\left( \lambda_2^4 + \lambda_3^4 \right)}{9} \\ \\ + \frac{4\left( \lambda_2^2 + \lambda_3^2 \right)}{9} \Big( -3 + \sqrt{ 3R^2 + \left( \lambda_2^4 + \lambda_3^4 \right) - 6\left( \lambda_2^2 + \lambda_3^2 \right) + 26\lambda_2^2 \lambda_3^2 } \Big)  \end{matrix}
	\label{third_case_R_tau_exploration__01}
\end{equation}
\end{widetext}
}

\begin{equation}
	\begin{matrix} \tau\left( R, \lambda_2, \lambda_3 \right)  = \upperTau - \frac{16 \lambda_2^2 \lambda_3^2}{9} + \frac{4\left( \lambda_2^4 + \lambda_3^4 \right)}{9} \\ \\ + \frac{4\left( \lambda_2^2 + \lambda_3^2 \right)}{9} \Big( -3 + \sqrt{ 3R^2 + \left( \lambda_2^4 + \lambda_3^4 \right) - 6\left( \lambda_2^2 + \lambda_3^2 \right) + 26\lambda_2^2 \lambda_3^2 } \Big)  \end{matrix}
	\label{third_case_R_tau_exploration__01}
\end{equation}

\noindent which makes the computations more cumbersome. The end results are \eqref{tau_up_branch} \eqref{tau_down_branch}.\newline

Finally, we now show where the \eqref{lowest_order_F_factor_all_3b} \eqref{lowest_order_F_factor_all_4b} results come from: start by considering an arbitrary point $\vec{r} \in [0,1]^3$, then the distance from that point to the straight line spanned by the main diagonal is the length of the vector connecting it to a point on the line such that this vector is perpendicular to the line. A simple trigonometric calculation gives:
\begin{equation}
	d\left( \vec{r}, \vec{V}_{\text{line}}  \right) = \sqrt{\frac{2}{3}} \sqrt{ R^2 - \displaystyle\sum_{i<j} \left( r_i r_j \right) }
	\label{distance_from_point_to_main_diagonal}
\end{equation}

Take a state $\type{3b-12}$ (such as the one we used for the type 3b geometry proof in Appendix \ref{section_PolytopeProofs}), then the distance to the main diagonal to lowest order is:
\begin{equation}
        \begin{matrix} \left[d\left( \vec{r}\left( \type{3b-12} \right), \vec{V}_{\text{line}}  \right)\right] ^2  & \simeq \frac{8 \lambda_0^4 \lambda_3^4}{3 r_C^2} \left( 1 + \frac{2\lambda_0^2 \lambda_3^2}{r_C^2} + O\left( \left[ \frac{\lambda_0^2 \lambda_3^2}{r_C^2} \right]^4 \right)  \right)   \end{matrix}
	\label{approx_distance_type_3b_12}
\end{equation}
\noindent so at lowest order:
\begin{equation}
	-\lambda_3^2 \left( \lambda_3^2 - 3 - \sqrt{3R^3 + \lambda_3^2 \left( \lambda_3^2 - 6 \right)} \right) \simeq -3 r_C \sqrt{\frac{3}{2}}  d\left( \vec{r}\left( \type{3b-12} \right), \vec{V}_{\text{line}}  \right)
	\label{lowest_order_smallF_factor}
\end{equation}
\noindent so by direct comparison to \eqref{TauRL3WannaRef}, one obtains \eqref{geometry_for_tangle}, where:
\begin{equation}
	\mathcal{F} \left( \vec{r}\left( \type{3b-12} \right) \right) \simeq 2 r_C \sqrt{\frac{2}{3}} + O\left( r_C^2 \right)
	\label{lowest_order_F_factor_12}
\end{equation}
For the other type 3b states, we can use the fact that the only difference in geometrical behavior is $r_C \leftrightarrow r_A$ and $r_C \leftrightarrow r_B$ for $\type{3b-23}$ and $\type{3b-13}$ respectively (see \eqref{summary_type_3b}) to find:
\begin{equation}
	\begin{matrix}  \mathcal{F} \left( \vec{r}\left( \type{3b-23} \right) \right) \simeq 2 r_A \sqrt{\frac{2}{3}} + O\left( r_A^2 \right)  \quad & \quad  \mathcal{F} \left( \vec{r}\left( \type{3b-13} \right) \right) \simeq 2 r_B \sqrt{\frac{2}{3}} + O\left( r_B^2 \right) \end{matrix}
	\label{lowest_order_F_factor_all_3b}
\end{equation}
\noindent which tells us that $\mathcal{F}$ somehow encodes the geometrical asymmetries of the Bloch-norm representations of the states.\newline

Furthermore, we can follow the same procedure for type 4b states as well:
\begin{equation}
	\begin{matrix} \mathcal{F} \left( \vec{r}\left( \type{4b-1} \right) \right) \simeq   2\sqrt{3} r_B + O\left( r_B^2 \right)  \quad & \quad  \mathcal{F} \left( \vec{r}\left( \type{4b-2} \right) \right) \simeq   2\sqrt{3} r_C + O\left( r_C^2 \right)  \end{matrix}
	\label{lowest_order_F_factor_all_4b}
\end{equation}


\section{Analytic details of 3-qubit spin chains}\label{section_DetailsChains}

It is possible to code a function that performs the CD procedure analytically for a simple enough input 3-qubit state. We have done so for the eigenstates of the chain Hamiltonians considered in Sect. \ref{section_Chains} and we show here the results.\newline

Such results depend in general on which one of the 2 possible solutions to the $\detZero$ equation of the CD procedure is chosen:

\begin{equation}
        \begin{matrix} U=\begin{pmatrix} z & w \\ -w^{*} & z^{*} \end{pmatrix} \Bigg| \left|z\right|^2 + \left| w \right|^2 = 1 & \text{ and } \detZero \text{ where } T_{i}' = \displaystyle\sum_{j} U_{ij}T_j   \end{matrix}
	\label{detZero_eq}
\end{equation}

\noindent which means one must specify which solution is picked each time. The exception to this is the tangle, which is the same for both solutions \cite{acin_decomp_paper}.\newline

We will label the energy levels by their integer ordering $n$ (with $n=0$ corresponding to the Ground State) and the degeneracies by the integer $m_n=\text{dim}(\text{span(\text{subspace }n)})$. We will also label the eigenstates by their eigenvalues under conserved quantities such as parity $p$ and momentum number $k$. For the XX and XXX models, the magnetization $m$ is conserved, so it will be used as well.

\subsection{TFIM analytics}\label{section_DetailsChains_TFIM}
Let's first consider the TFIM \eqref{TFIM_hamiltonian}. It can be solved exactly giving, an energy spectrum:

\begin{align}
E_0 & = - \Delta - 2 \sqrt{\Delta^{2} - \Delta + 1} - 1; & \quad m=1 \nonumber \\
E_1 & = \Delta - 2 \sqrt{\Delta^{2} + \Delta + 1} - 1; & \quad m=1 \nonumber \\
E_2 & = - \Delta + 2 \sqrt{\Delta^{2} - \Delta + 1} - 1; & \quad m=1 \nonumber \\
E_3 & = 1 - \Delta; & \quad m=2 \nonumber \\
E_4 & =\Delta + 1; & \quad m=2 \nonumber \\
E_5 & = \Delta + 2 \sqrt{\Delta^{2} + \Delta + 1} - 1; & \quad m=1
\label{TFIM_energy_spectrum}
\end{align}

\noindent where the degeneracy count remains valid outside of any level crossing point (like $\Delta=0$ which corresponds to a quantum critical point or $\Delta=1$). At any level crossing point, the energy expressions are still valid but the degeneracy count will change, so calculations need to be redone at these points for the relevant levels which are fusing together.

\subsubsection{TFIM outside level-crossing points}\label{section_DetailsChains_TFIM_NOcross}
We start by computing the energy eigenstates outside of these two level crossing points:
\begin{align}
\ket{n=0, p=+1,k=0} =& \frac{1}{g_0} \left[ f_0 \ket{000} + \sqrt{3} \hat{X}^{\otimes 3} \W \right]   \nonumber \\
\ket{n=1, p=-1, k=0} =& \frac{f_1}{g_1} \sqrt{3}\W + \frac{3}{g_1} \hat{X}^{\otimes 3}\ket{000}      \nonumber \\
\ket{n=2, p=+1, k=0} =& \frac{1}{g_2}\left[ -f_2\ket{000} +  \sqrt{3} \hat{X}^{\otimes 3} \W \right]   \nonumber \\
\ket{n=5, p=-1, k=0} =& \frac{-f_5}{g_5} \sqrt{3}\W + \frac{3}{g_5} \hat{X}^{\otimes 3} \ket{000}  
\label{TFIM_eigenState_spectrum_nonDeg}
\end{align}
\noindent for which we can calculate analytically the Bloch-norm vectors as well:
\begin{align}
\ket{n=0,2} \to & (r_A,r_B,r_C) =  \left( \frac{ \left| f_n^2 -1 \right| }{g_n^2} \right) \cdot \left( 1,1,1 \right) \nonumber \\
\ket{n=1,5} \to & (r_A,r_B,r_C) =  \left( \frac{\left| f_n^2 -9 \right|}{g_n^2} \right) \cdot \left( 1,1,1 \right)
\label{TFIM_eigenState_spectrum_nonDeg_BlochNorms}
\end{align}
For the degenerate states:
\begin{align}
\ket{n=3, p=-1, k=1} =& \wtrans{1} \nonumber \\
\ket{n=3, p=-1, k=2} =& \wtrans{2} \nonumber \\
\ket{n=3, p=-1, (\alpha, \beta)}  =& \alpha\ket{n=3, p=-1, k=1} + \beta\ket{n=3, p=-1, k=2};   \nonumber \\
\ket{n=4, p=+1, k=1} =& \hat{X}^{\otimes 3}\wtrans{1} \nonumber \\
\ket{n=4, p=+1, k=2} =& \hat{X}^{\otimes 3}\wtrans{2} \nonumber \\
\ket{n=4, p=-1, (\alpha, \beta)}  =& \alpha\ket{n=4, p=+1, k=1} + \beta\ket{n=4, p=+1, k=2}; 
\label{TFIM_eigenState_spectrum_deg}
\end{align}
\begin{eqnarray}
\ket{n=3, (\alpha,\beta)} & \to   (r_A,r_B,r_C) = \frac{1}{3} \Bigg( &  \left|1+2|\alpha|\beta \left[ \cos{\text{arg}(\alpha)} + \sqrt{3} \sin{\text{arg}(\alpha)} \right] \right|, \nonumber \\
& & \left|1+2|\alpha|\beta \left[ \cos{\text{arg}(\alpha)} - \sqrt{3} \sin{\text{arg}(\alpha)} \right] \right| ,   \left|1-4|\alpha|\beta \cos{\text{arg}(\alpha)}  \right|  \Bigg) \nonumber \\
\ket{n=4, (\alpha,\beta)} & \to \text{ same as for n=3} & 
\label{TFIM_eigenState_spectrum_yesDeg_BlochNorms}
\end{eqnarray}
\noindent where $\alpha \in \mathbb{C}$ and $\beta\in\mathbb{R}^+$ such that $|\alpha|^2 + \beta^2 = 1$ and we have already used the global $U(1)$ to remove the complex phase. Also, the states $\wtrans{1}$ and $\wtrans{2}$ are defined as:
\begin{eqnarray}
\wtrans{1} := \frac{1}{\sqrt{3}} \left( \ket{001} + e^{\left( i\frac{2\pi}{3} \right)} \ket{010} + e^{\left( i\frac{4\pi}{3} \right)} \ket{100} \right) \nonumber \\
\wtrans{2} := \frac{1}{\sqrt{3}} \left( \ket{001} + e^{\left( i\frac{4\pi}{3} \right)} \ket{010} + e^{\left( i\frac{2\pi}{3} \right)} \ket{100} \right)
\end{eqnarray}
The degeneracy allows for arbitrary superpositions characterized by the parameters $\alpha$ and $\beta$.\newline

For the non-degenerate subspaces, the $f_j,g_j$ parameters on \eqref{TFIM_eigenState_spectrum_nonDeg} are defined as:
\begin{align}
f_0 & =  -1+ 2\Delta + 2b(\Delta) \ge 0; & g_0 &= \sqrt{f_0^2 + 3}>0 \nonumber \\
f_{1} &= 2 \Delta + 2 a(\Delta) + 1 \ge 0; & g_1 &= \sqrt{3}\sqrt{f_1^2 + 3} >0 \nonumber \\
f_2 &=  \left[ -2\Delta +1 +2b(\Delta)\right] \ge 0; & g_2 &= \sqrt{f_2^2 + 3}>0 \nonumber \\
f_5 &= \left[ -2\Delta -1 +2a(\Delta)\right] \ge 0; & g_5 &= \sqrt{3}\sqrt{f_5^2 + 3}>0 
\label{TFIM_eigenState_parameters}
\end{align}
\begin{align}
        b(\Delta) &= \sqrt{1-\Delta + \Delta^2}; &  a(\Delta) &= \sqrt{1+\Delta + \Delta^2}; \label{a_and_b}
\end{align}
Now that we have the exact eigenstates at hand, we can begin computing the CD and the tangle for each level. For $n=0$, we have that the 2 solutions of the \ref{detZero_eq} equation are:
\begin{equation}
	w= \pm \frac{\sqrt{f_0}}{\sqrt{f_0 + 1}} \quad\quad   z= \frac{1}{\sqrt{f_0 + 1}}
	\label{TFIM_GS_sols}
\end{equation}
\noindent of which we pick the $+$ solution. The resulting $\lambda$ parameters are:
\begin{equation}
    \begin{matrix} \lambda_{0} = \frac{\sqrt{f_{0} + 1}}{g_{0}}; & \lambda_{1} = \frac{\sqrt{f_{0}} \left|{f_{0} - 1}\right|}{g_{0} \sqrt{f_{0} + 1}}; & \lambda_{2} = \lambda_3 = \frac{\left|{f_{0} - 1}\right|}{g_{0} \sqrt{f_{0} + 1}}; & \lambda_{4} = \frac{2 \sqrt{f_{0}}}{g_{0} \sqrt{f_{0} + 1}}; & \varphi = 0   \end{matrix}
    \label{CD_TFIM_GS}
\end{equation}
For the first excited state $\ket{n=1}$ the solutions to \eqref{detZero_eq} are:
\begin{equation}
	w= \frac{\pm\sqrt{f_1}}{\sqrt{f_1 + 3}} \quad \quad z=\frac{\sqrt{3}}{\sqrt{f_1 + 3}}
	\label{TFIM_n1_sols}
\end{equation}
\noindent of which we pick the $+$ solution. The resulting $\lambda$ parameters are:
\begin{equation}
    \begin{matrix}
        \lambda_{0} = \frac{\sqrt{f_{1}} \sqrt{f_{1} + 3}}{g_{1}} ; & \lambda_{1}  = \frac{\sqrt{3} \left|{f_{1} - 3}\right|}{g_{1} \sqrt{f_{1} + 3}}; \quad \lambda_{2} = \lambda_3 = \frac{\sqrt{f_{1}} \left|{f_{1} - 3}\right|}{g_{1} \sqrt{f_{1} + 3}}; \quad \lambda_{4} = \frac{2 \sqrt{3} f_{1}}{g_{1} \sqrt{f_{1} + 3}} ; \quad \varphi = 0
    \end{matrix}
\label{CD_TFIM_n1}
\end{equation}
For the second excited state $\ket{n=2}$ the solutions to \eqref{detZero_eq} are:
\begin{equation}
	w= \frac{\pm i\sqrt{f_2}}{\sqrt{f_2 + 1}} \quad \quad z=\frac{1}{\sqrt{f_2 + 1}}
	\label{TFIM_n2_sols}
\end{equation}
\noindent of which we pick the $+$ solution. The resulting $\lambda$ parameters are:
\begin{equation}
    \begin{matrix}
        \lambda_{0} = \frac{\sqrt{f_{2} + 1}}{g_{2}}  ; & \quad \lambda_{1} = \frac{\sqrt{f_{2}} \left|f_2 -1\right|}{g_{2} \sqrt{f_{2} + 1}}  ; & \quad \lambda_{2} = \lambda_3 = \frac{\left|{f_{2} - 1}\right|}{g_{2} \sqrt{f_{2} + 1}}  ; & \quad \lambda_{4} = \frac{2 \sqrt{f_{2}}}{g_{2} \sqrt{f_{2} + 1}}  ; & \quad \varphi = 0
    \end{matrix}
\label{CD_TFIM_n2}
\end{equation}
For the third excited subspace, we compute the CD for the arbitrary superposition $\ket{n=3; \left( \alpha, \beta \right)}$:
\begin{equation}
	\lambda_0 = \frac{\left| \alpha e^{-i\frac{2\pi}{3}} + \beta e^{+i\frac{2\pi}{3}} \right|}{\sqrt{3}}, \quad \lambda_1 = 0 = \lambda_4, \quad \lambda_2 = \left| \frac{\alpha+\beta}{\sqrt{3}} \right|, \quad \lambda_3 = \left| \frac{\alpha e^{-i\frac{2\pi}{3}}+\beta}{\sqrt{3}} \right|
	\label{CD_TFIM_n3_horizontal}
\end{equation}
For the fourth excited subspace, since $\ket{4, (\alpha,\beta)} = \hat{X}^{\otimes 3} \ket{3, (\alpha,\beta)}$ so the CD is the same as for $n=3$. For the fifth excited state $\ket{n=5}$ the solutions to \eqref{detZero_eq} are:
\begin{equation}
	w= \frac{\pm i\sqrt{f_5}}{\sqrt{f_5 + 3}} \quad \quad z=\frac{\sqrt{3}}{\sqrt{f_5 + 3}}
	\label{TFIM_n5_sols}
\end{equation}
\noindent of which we pick the $+$ solution. The resulting $\lambda$ parameters are:
\begin{equation}
    \begin{matrix}
        \lambda_{0} = \frac{\sqrt{f_{5}} \sqrt{f_{5} + 3}}{g_{5}}  ; & \quad \lambda_{1} = \lambda_3 = \frac{\sqrt{3} \left|{f_{5} - 3}\right|}{g_{5} \sqrt{f_{5} + 3}}  ; & \quad \lambda_{2} = \frac{\sqrt{f_{5}} \left|{f_{5}^{2} - 9}\right|}{g_{5} \left(f_{5} + 3\right)^{\frac{3}{2}}}  ; & \quad \lambda_{4} = \frac{2 \sqrt{3} f_{5}}{g_{5} \sqrt{f_{5} + 3}}  ; & \quad \varphi = 0
    \end{matrix}
\label{CD_TFIM_n5}
\end{equation}

\subsubsection{TFIM at level-crossing points}\label{section_DetailsChains_TFIM_cross}
Consider now the level-crossing point $\Delta=1$. Almost all levels are unchanged here, so the results for $\Delta$ outside the level-crossing still applies. The exception being levels $n=2$ and $n=3$ which fuse in this point. Diagonalizing the Hamiltonian at this value of $\Delta$, it is possible to obtain a new basis for the degenerate subspace:
\begin{align}
\ket{n=2, s=1} &= \frac{1}{2} \left[ -\ket{000} + \ket{011} + \ket{101} + \ket{110} \right] \nonumber \\
\ket{n=2, s=1} &= \wtrans{1} \nonumber \\
\ket{n=2, s=3} &= \wtrans{2} \nonumber \\
\ket{n=2, \left( \alpha,... \right)} & =\gamma \ket{2,1} + \alpha \ket{2,2} + \beta\ket{2,3}; &  
\label{TFIM_eigenState_n2_3_delta1}
\end{align}
\noindent with $\alpha,\gamma\in\mathbb{C} \text{ and } \beta\in\mathbb{R}^+$ the coefficients of an arbitrary superposition in this subspace. The hyperdeterminant formula allows to obtain directly the tangle:
\begin{equation}
    \tau_{n=2,\Delta=1} = |\gamma|^2 \left| \gamma^2 - \frac{1}{3} \left[ (\alpha+\beta) - (\xi-\eta) \right]^2 + \frac{4\eta}{3}(\alpha+\beta) \right|
	\label{TFIM_tau_n2_3_delta1}
\end{equation}
\noindent where $\xi = \left( \alpha e^{-i\frac{2\pi}{3}} + \beta e^{+i\frac{2\pi}{3}} \right)$ and $\eta =  \left( \alpha e^{+i\frac{2\pi}{3}} + \beta e^{-i\frac{2\pi}{3}} \right)$. We can see immediately that $\gamma=0 \implies \tau=0$, meaning that all the tangle in the superposition comes solely from the $\ket{n=2,s=1}$ state in the subspace.

\subsection{XX analytics}\label{section_DetailsChains_XX}
We now consider the XX chain \eqref{XX_hamiltonian}. The energy spectrum is:
\begin{align}
E_0 & = - \Delta - 4; & \quad m=1 \nonumber \\
E_1 & = + \Delta - 4; & \quad m=1 \nonumber \\
E_2 & = -3\Delta; & \quad m=1 \nonumber \\
E_3 & = +3\Delta; & \quad m=1 \nonumber \\
E_4 & = 2-\Delta; & \quad m=2 \nonumber \\
E_5 & = 2+\Delta; & \quad m=2 
\label{XX_energy_spectrum}
\end{align}

\subsubsection{XX outside level-crossing points}\label{section_DetailsChains_XX_NOcross}
We start by computing the energy eigenstates outside of level crossing points:
\begin{align}
\ket{n=0, k=0, m=+1, p=-1} &= \W \nonumber \\
\ket{n=1, k=0, m=-1, p=+1} &= \hat{X}^{\otimes 3} \W \nonumber \\
\ket{n=2, k=0, m=+3, p=+1} &= \ket{000} \nonumber \\
\ket{n=3, k=0, m=-3, p=-1} &= \ket{111} \nonumber \\
\ket{n=4, k=1, m=+1, p=-1} &= \wtrans{1} \nonumber \\
\ket{n=4, k=2, m=+1, p=-1} &= \wtrans{2} \nonumber \\
\ket{n=4; m=+1, p=-1, (\alpha, \beta)} & = \beta\ket{4, k=1} + \alpha\ket{4, k=2};  \nonumber \\
\ket{n=5, k=1, m=-1, p=+1} &= \hat{X}^{\otimes 3}\wtrans{1} \nonumber \\
\ket{n=5, k=2, m=-1, p=+1} &= \hat{X}^{\otimes 3}\wtrans{2} \nonumber \\
\ket{n=5; m=-1, p=+1, (\alpha, \beta)} & = \beta\ket{5, k=1} + \alpha\ket{5, k=2}; 
\label{XX_eigenState_spectrum_yesDeg}
\end{align}
\noindent where $\alpha$ and $\beta$ are defined as in \eqref{TFIM_eigenState_spectrum_deg}. We can calculate analytically the Bloch-norms for the non-degenerate levels:
\begin{equation}
\ket{n=0 \text{ \& }1} \to  \frac{1}{3} \left( 1,1,1 \right)   \quad \quad  \ket{n=2 \text{ \& }3} \to   \left( 1,1,1 \right)
\label{XX_eigenState_spectrum_nonDeg_BlochNorms}
\end{equation}
We can now compute the CD for each level: For $n=0$, since it is plainly the $\tW$ state we have:
\begin{equation}
	\lambda_0 = \lambda_2 = \lambda_3 = \frac{1}{\sqrt{3}}; \quad \lambda_1 = \lambda_4 = 0;
	\label{CD_XX_n0_horizontal}
\end{equation}
For the $n=1$ subspace, the result of the CD is the same. For $n=3$ and $n=4$ one has $\lambda_0=1$ and $\lambda_{n>0}=0$ since they are product states. For both $n=4$ and $n=5$ one has that $\lambda_1 = \lambda_4 = 0$ while the non-zero $\lambda$ are the same as we saw for levels $n=3$ and 4 for the TFIM \eqref{CD_TFIM_n3_horizontal}.

\subsection{XXX analytics}\label{section_DetailsChains_XXX}
We now consider the XX chain \eqref{XXX_hamiltonian}. The energy spectrum is:
\begin{align}
E_0 & = +3\Delta; & \quad m=2 \nonumber \\
E_1 & = 4-\Delta; & \quad m=2 \nonumber \\
E_2 & = -\Delta - 2; & \quad m=4 \nonumber \\
\label{XXX_energy_spectrum}
\end{align}
\noindent for $\Delta\in\mathbb{R}$. At $\Delta=-1/2$ there is a level crossing, where the role of the GS changes from $n=0$ to $n=1$. We will still keep the labels used in $\Delta\in(-\infty, -1/2)$ for consistency.

\subsubsection{XXX outside level-crossing points}\label{section_DetailsChains_XXX_NOcross}
We start by computing the energy eigenstates outside of level crossing points:
\begin{align}
\ket{n=0, k=0, m=+3, p=+1} &= \ket{000} \nonumber \\
\ket{n=0, k=0, m=-3, p=-1} &= \ket{111} \nonumber \\
\ket{n=0, k=0, \left( \alpha,\beta \right)} & = \beta\ket{000} + \alpha\ket{111} \nonumber \\
\ket{n=1, k=0, m=-1, p=+1} &= \hat{X}^{\otimes 3} \W \nonumber \\
\ket{n=1, k=0, m=+1, p=-1} &=  \W \nonumber \\
\ket{n=1, k=0, m=+1, \text{ } \left( \alpha,\beta \right)} & = \beta \hat{X}^{\otimes 3} \W + \alpha\W \nonumber \\
\ket{n=2, k=1, m=+1, p=-1} &= \wtrans{1} \nonumber \\
\ket{n=2, k=1, m=-1, p=+1} &= \hat{X}^{\otimes 3}\wtrans{1} \nonumber \\
\ket{n=2, k=2, m=+1, p=-1} &= \wtrans{2} \nonumber \\
\ket{n=2, k=2, m=-1, p=+1} &= \hat{X}^{\otimes 3}\wtrans{2} \nonumber \\
\ket{n=2, \left( \alpha, ... \delta \right)} & = \alpha \ket{2, 1} + \beta \ket{2,2} + \gamma \ket{2,3} + \delta \ket{2,4} 
\label{XXX_eigenState_spectrum_all}
\end{align}
We can compute the CD for the $n=0$ subspace, obtaining:
\begin{equation}
	\lambda_0 = \beta; \text{ } \lambda_4 = |\alpha|; \text{ } \lambda_j = 0 \text{ for } j=1,2,3;
	\label{XXX_n0_CD}
\end{equation}
For $n=1$, the CD is:

\begin{equation}
    \begin{matrix}
        \lambda_0 = \frac{1}{\sqrt{3}}; & \lambda_2 = \lambda_3 = \sqrt{\frac{1}{3} + \beta^2 \left( \beta^2 -1 \right)}; & \lambda_4 = \beta \sqrt{1-\beta^2}; & \varphi = \frac{\pi}{6} -2\text{arg}(\alpha) + \text{atan} \left( \frac{-\beta^2 \sqrt{3}}{2-3\beta^2} \right);
    \end{matrix}
    \label{XXX_n1_CD}
\end{equation}

\noindent but for level $n=2$ the degeneracy is so large that computing the CD is not possible due to a timeout error from the automated solver, but one can still obtain the tangle from the Hyperdeterminant formula \eqref{XXX_tangles_nonCrosses}.

\subsection{XZX analytics}\label{section_DetailsChains_XZX}
The energy spectrum of the XZX chain \eqref{XZX_hamiltonian} reads:
\begin{align}
E_0 & = + \Delta - 2a(\Delta) - 1; & \quad m=1 \nonumber \\
E_1 & = - \Delta - 2a(\Delta) + 1; & \quad m=1 \nonumber \\
E_2 & = -1+\Delta; & \quad m=2 \nonumber \\
E_3 & = +1-\Delta; & \quad m=2 \nonumber \\
E_4 & = + \Delta + 2a(\Delta) - 1; & \quad m=1 \nonumber \\
E_5 & = - \Delta + 2a(\Delta) + 1; & \quad m=1 \nonumber \\
\label{XZX_energy_spectrum}
\end{align}
\noindent where $a(\Delta)$ is defined as in \eqref{a_and_b}.

\subsubsection{XZX outside level-crossing points}\label{section_DetailsChains_XZX_NOcross}
The energy eigenstates outside of level crossing points are:
\begin{align}
\ket{n=0, p=-1, k=0} &= \frac{1}{g_0} \left[ 3\ket{111} + -f_0 \sqrt{3}\W \right] \nonumber \\
\ket{n=1, p=+1, k=0} &= \frac{1}{g_0} \left[ \sqrt{3}f_0 \ket{000} + 3 \hat{X}^{\otimes 3} \W \right] \nonumber \\
\ket{n=4, p=-1, k=0} &= \frac{1}{g_4} \left[ 3 \ket{111} + \sqrt{3} f_4  \W \right] \nonumber \\
\ket{n=5, p=+1, k=0} &= \frac{1}{g_5} \left[ -f_5 \ket{000} + \sqrt{3} \hat{X}^{\otimes 3} \W \right] 
\label{XZX_eigenState_spectrum_nonDeg}
\end{align}
\noindent for the non-degenerate levels, with the $f_j,g_j$ defined as:
\begin{align}
f_0 &= 2 \Delta + 2 a(\Delta) + 1 \ge 0; & g_0 &= \sqrt{3} \sqrt{f_0^2 + 3} \nonumber \\
f_4 &= - 2 \Delta + 2 a - 1 \ge 0 & g_4 &= \sqrt{3} \sqrt{f_{4}^{2} + 3} \nonumber \\
f_5 &= f_4 \ge 0 & g_5 &=  \sqrt{f_{5}^{2} + 3}
\label{XZX_eigenState_parameters}
\end{align}
\noindent while for the degenerate levels:
\begin{align}
\ket{n=2, p=+1, k=1} &= \hat{X}^{\otimes 3} \wtrans{1} \nonumber \\
\ket{n=2, p=+1, k=2} &= \hat{X}^{\otimes 3} \wtrans{2} \nonumber \\
\ket{n=2, p=+1, (\alpha, \beta)} & = \alpha\ket{2, 1} + \beta\ket{2, 2};   \nonumber \\
\ket{n=3, p=-1, k=1} &= \wtrans{1} \nonumber \\
\ket{n=3, p=-1, k=2} &= \wtrans{2} \nonumber \\
\ket{n=3, p=-1, (\alpha, \beta)} & = \beta\ket{3, 1} + \alpha\ket{3, 2};
\label{XZX_eigenState_spectrum_deg}
\end{align}
We can now compute the CD of the eigenstates. Starting with $n=0$, the solutions to \eqref{detZero_eq} are:
\begin{equation}
	w=  \frac{ \pm i\sqrt{f_0}}{\sqrt{f_0 + 3}}; \quad  z= \frac{\sqrt{3}}{\sqrt{f_0 + 3}}
	\label{XZX_GS_sols}
\end{equation}
\noindent and picking the $(+)$ solution:
\begin{equation}
    \begin{matrix}
        \lambda_{0} = \frac{\sqrt{f_{0}} \sqrt{f_{0} + 3}}{g_{0}}  ; & \quad \lambda_{1} = \frac{\sqrt{3} \left|{f_{0} - 3}\right|}{g_{0} \sqrt{f_{0} + 3}}  ; & \quad \lambda_{2} = \lambda_3 = \frac{\sqrt{f_{0}} \left|{f_{0} - 3}\right|}{g_{0} \sqrt{f_{0} + 3}}  ; & \quad \lambda_4 = \frac{2 \sqrt{3} f_{0}}{g_{0} \sqrt{f_{0} + 3}}  ; & \quad \varphi = 0
    \end{matrix}
\label{CD_XZX_GS}
\end{equation}
For $n=1$, the solutions are:
\begin{equation}
	w=  \frac{ \pm \sqrt{f_0}}{\sqrt{f_0 + 1}}; \quad  z= \frac{1}{\sqrt{f_0 + 1}}
	\label{XZX_level_1_sols}
\end{equation}
\noindent and we pick the $(+)$ one:
\begin{equation}
    \begin{matrix}
        \lambda_{0} = \frac{\sqrt{f_{0} + 1}}{\sqrt{f_{0}^{2} + 3}}  ; & \quad \lambda_{1} = \frac{\sqrt{f_{0}} \left|{f_{0} - 1}\right|}{\sqrt{f_{0} + 1} \sqrt{f_{0}^{2} + 3}}  ; & \quad \lambda_{2} = \lambda_3 = \frac{\left|{f_{0} - 1}\right|}{\sqrt{f_{0} + 1} \sqrt{f_{0}^{2} + 3}}  ; & \quad \lambda_4 = \frac{2 \sqrt{f_{0}}}{\sqrt{f_{0} + 1} \sqrt{f_{0}^{2} + 3}}  ; & \quad \varphi = 0
    \end{matrix}
\label{CD_XZX_level_1}
\end{equation}
For $n=2$ and $n=3$ we can reuse the results from levels $n=3$ and 4 of the TFIM. For $n=4$, the solutions are:
\begin{equation}
	w=  \frac{ \pm \sqrt{f_4}}{\sqrt{f_4 + 3}}; \quad  z= \frac{\sqrt{3}}{\sqrt{f_4 + 3}}
	\label{XZX_level_4_sols}
\end{equation}
\noindent and we pick the $(+)$ one:
\begin{equation}
    \begin{matrix}
        \lambda_{0} = \frac{\sqrt{f_{4}} \sqrt{f_{4} + 3}}{g_{4}}  ; & \quad \lambda_{1} = \frac{\sqrt{3} \left|{f_{4} - 3}\right|}{g_{4} \sqrt{f_{4} + 3}}  ; & \quad \lambda_{2} =\lambda_3 = \frac{\sqrt{f_{4}} \left|{f_{4} - 3}\right|}{g_{4} \sqrt{f_{4} + 3}}  ; & \quad \lambda_4 =  \frac{2 \sqrt{3} f_{4}}{g_{4} \sqrt{f_{4} + 3}}  ; & \quad \varphi = 0
    \end{matrix}
\label{CD_XZX_level_4}
\end{equation}
Finally, for $n=5$:
\begin{equation}
	w=  \frac{ \pm i \sqrt{f_5}}{\sqrt{f_5 + 1}}; \quad  z= \frac{1}{\sqrt{f_5 + 1}}
	\label{XZX_level_5_sols}
\end{equation}
\begin{equation}
    \begin{matrix}
        \lambda_{0} = \frac{\sqrt{f_{5} + 1}}{g_{5}}  ; & \quad \lambda_1 =  \frac{\sqrt{f_{5}} \left|{f_{5} - 1}\right|}{g_{5} \sqrt{f_{5} + 1}}  ; & \quad \lambda_{2} =\lambda_3= \frac{\left|{f_{5} - 1}\right|}{g_{5} \sqrt{f_{5} + 1}}  ; & \quad \lambda_4 = \frac{2 \sqrt{f_{5}}}{g_{5} \sqrt{f_{5} + 1}}   ; & \quad \varphi = 0
    \end{matrix}
\label{CD_XZX_level_5}
\end{equation}


\end{document}